\documentclass[11pt]{uhthesis}

\newif\ifpdf
\ifx\pdfoutput\undefined
  \pdffalse               
\else
  \pdfoutput=1            
  \pdftrue
\fi

\ifpdf
  \usepackage[pdftex]{graphicx}
\else
  \usepackage{graphicx}
\fi

\usepackage{natbib}
\usepackage{chapterbib}
\usepackage{uhaas}
\usepackage{rotating}
\usepackage{lscape}

\begin{document}
\frontmatter
%

\title{FROM LAB TESTING TO SCIENCE: APPLYING SAPHIRA HGCDTE L-APD DETECTORS TO ADAPTIVE OPTICS}
\author{Sean B. Goebel}
\date{August 2018}
\chairperson{Donald Hall}
\memberA{Olivier Guyon}
\memberB{Mark Chun}
\memberC{Andrew Howard}
\memberD{Michael Connelley}
\memberE{John Learned}
\maketitle
\makesig

\makecopyright{2018}{Sean B. Goebel}


\makeacknowledgements
Sean gratefully thanks his advisors, Donald Hall and Olivier Guyon, for the time and 
expertise they have shared with him and the innumerable questions they have answered.

Sean greatly appreciates the contributions to the papers presented in the following pages
by his co-authors. He thanks Shane Jacobson, without whom none of the cameras and equipment
would work. He expresses his gratitude to Charles Lockhart and Eric Warmbier for tirelessly
working on the Pizza Box readout electronics outside their work duties and troubleshooting
the late-night crises of his observing runs.

Sean particularly thanks his parents for putting up with him as a weird child, supporting his university education, and continuing to care for him as a (still weird) adult.

\makeabstract
Due to their high frame rates, high sensitivity, low noise, and low dark current, SAPHIRA detectors provide new capabilities for astronomical observations. The SAPHIRA detector is a 320$\times$256@24 $\micron$ pixel HgCdTe linear avalanche photodiode array manufactured by Leonardo. It is sensitive to $0.8-2.5$ $\micron$ light. Unlike other near-infrared arrays, SAPHIRA features a user-adjustable avalanche gain, which multiplies the photon signal but has minimal impact on the read noise. This enables the equivalent of sub-electron read noise and therefore photon-counting performance, which has not previously been achieved with astronomical near-infrared arrays. SAPHIRA is intended for high clocking speeds, and we developed a new readout controller to utilize this capability and thereby enable the high frame rates ($\sim 400$ Hz for the full frame or $\sim 1.7$ kHz for a 128$\times$128 pixel subarray). Beginning with the first science-grade SAPHIRA detectors and continuing with later improved devices, we deployed SAPHIRAs to the SCExAO instrument at Subaru Telescope. SCExAO is an extreme adaptive optics instrument intended for observations of high-contrast objects such as debris disks and extrasolar planets. While at SCExAO, we demonstrated the ability of SAPHIRA to function as a focal-plane wavefront sensor, and we performed extensive studies of speckle evolution. Our demonstration of SAPHIRA’s ability to wavefront sense behind pyramid optics contributed to the decision to select a SAPHIRA detector and pyramid optics for the facility-class Keck Planet Imager. Additionally, we utilized the high Strehl provided by SCExAO to characterize the morphology of the HIP 79977 debris disk. Due largely to our characterization of the performance of SAPHIRA detectors and our demonstration of their capabilities, numerous facilities throughout the world have recently proposed to use them in instruments currently in development.

\tableofcontents
\listoftables
\listoffigures

\mainmatter


\chapter{Introduction}\label{chapter:intro}

\section*{Abstract}
This chapter contextualizes the work described in the rest of the dissertation. 
It provides general information about adaptive optics and the SAPHIRA detector 
not contained following chapters. It briefly describes the work carried out 
during the course of Sean Goebel's PhD, the reasons for those actions, and
the ongoing efforts that are building upon those accomplishments. Finally,
it provides a road map to the four papers reproduced in Chapters 2-5.

\section{Adaptive Optics}
Although an optical/infrared ground-based 8-m-class telescope collects far more light than a 
hobby-level 30 cm backyard telescope, it is unable to resolve greater detail in the object 
being observed. This is because the resolutions of both telescopes are limited by atmospheric
turbulence, not diffraction from the telescope pupil. At even the best astronomical sites in
the world, such as Mauna Kea, atmospheric turbulence prevents the resolving of structures 
with angular scales smaller than several tenths of an arcsec at visible wavelengths.

Adaptive optics (AO) is a technology which reduces the blurring effects of the atmosphere in order to
regain the diffraction limit of the telescope. If the aberrated wavefronts of light from the 
object being observed can be measured, it is possible to reflatten them by reflecting the light
off a deformable mirror (DM) with the appropriate shape. Measuring the wavefront, calculating
the correction that should be placed on the DM, and driving the mirror to this shape before the
turbulence substantially changes is a major challenge; adaptive optics was first proposed 
by~\citet{Babcock1953}, but due to computational and funding limitations, it was not
actually implemented at civilian telescopes until the 1980s (see Table 1 
in~\citet{Beckers1993} for a summary of these early efforts). 

\begin{figure}
\begin{centering}
\includegraphics[width=0.7\textwidth]{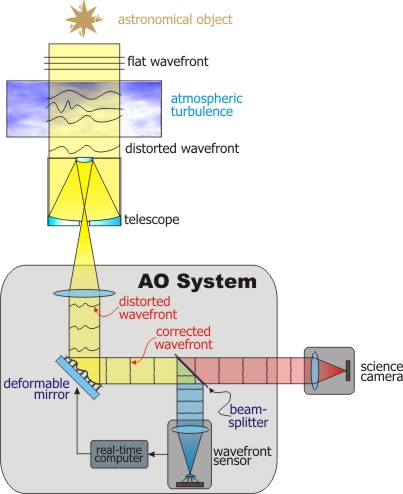}
\caption[Schematic diagram of an adaptive optics system]
{Schematic diagram of an adaptive optics system. Because astronomical objects are far away, their spherical
 wavefronts are effectively planar when they reach the Earth. The Earth's atmosphere distorts the wavefronts, thereby
 aberrating images of the object. A deformable mirror whose updates are calculated by a real-time computer's analysis
 of wavefront sensor data can significantly (though never perfectly) correct the wavefront. In this diagram, the
 wavefront sensor uses blue light and the science camera uses the red light; this is the typical setup. However,
 some astronomical objects (such as late-type stars or stars embedded in dust) have little optical 
 emission. For these sorts of objects, wavefront sensing at near-infrared wavelengths would be ideal. SAPHIRA
 detectors are a technology that enables this. This figure is reproduced with permission from  http://slittlefair.staff.shef.ac.uk/teaching/phy217/lectures/telescopes/L10.}
\label{fig:aoschematic}
\end{centering}
\end{figure}

The basic design of an AO system is shown in Figure~\ref{fig:aoschematic}. After the light
reflects from the deformable mirror, it is split (most commonly using a dichroic mirror which
reflects certain wavelengths and is transparent to other wavelengths) and sent on different
paths to the wavefront sensor and the science camera. The ``science camera" is most commonly 
a focal-plane imager or spectrograph. The wavefront sensor consists of optical elements and a 
detector to receive the light from them. A real-time computer then converts the images
from the detector to voltages applied to the DM. Several optical setups have been used for 
wavefront sensing;
in decreasing order of prevalence, the most common setups used in adaptive optics are 
Shack-Hartmann sensors~\citep{Shack1971, Allen1988}, curvature sensors~\citep{Roddier1988}, 
and pyramid sensors~\citep{Ragazzoni1996, Esposito2000}.

Shack-Hartmann sensors use an array
of microlenses to divide the pupil into subapertures and create a spot corresponding to each.
A perfect wavefront would cause the spots to have the same spacing as the microlenses; 
spots produced by aberrated
wavefronts shift according to the local slope of the wavefront at each microlens. By measuring
the shift of the spots from their unaberrated positions, the wavefront can be reconstructed.
Shack-Hartmann wavefront sensors have a large linear response range, so they are well-suited
for sensing highly aberrated wavefronts. However, the spot size is determined by the 
diffraction limit of each subaperture instead of the significantly smaller telescope aperture diffraction limit, so Shack-Hartmanns do not make the optimal use of the availble light. Curvature wavefront sensors also use 
microlenses, but instead of looking at the shifts of spots caused by aberrations, they measure
changes in spot intensity in front of and behind the focal plane. Finally, pyramid wavefront
sensors are discussed at length in this dissertation. Pyramid wavefront sensors focus the
point spread function (PSF) onto the tip of a pyramid-shaped prism\footnote{Two rooftop prisms
placed tip to tip with perpendicular orientations can also be used.}. With the addition of 
refocusing optics, this produces four pupil images. Aberrations can be measured from the 
relative illuminations of the four pupil images; a perfect PSF would produce identical
pupil images. Because these images are defined by the diffraction limit of the entire
telescope pupil (instead of individual subapertures), pyramid sensors have the potential to 
be more sensitive than Shack-Hartmann sensors. However, pyramid sensors have a narrow range of 
wavefront error amplitude over which their response is linear, so they are best-suited for use
in woofer-tweeter AO setups in which preliminary wavefront corrections have already been
made. The range of linear response can be extended by modulating the PSF around the tip
of the pyramid, but this reduces its sensitivity~\citep{Esposito2001}.

In addition to the optical elements, the second component of an effective wavefront sensor
is a sensitive and high-speed detector. Because AO systems have update rates in the range
of $\sim 100$ Hz to $\sim$ 2 kHz, the detector needs to produce images at least this 
quickly. Because astronomical targets tend to be dim\footnote{If they were bright, someone else
would have already observed them when the technology was less mature.} and the frame
rates necessary for adaptive optics wavefront sensing are high, relatively low signal
levels are available per exposure. For this reason, detectors for wavefront sensing should
have low read noise and high quantum efficiency. 
The detector should also have a low
latency between the production of photoelectrons by incoming photons
and the production of a digital signal from them, but in most cases this is not a problem.

Wavefront sensing has traditionally be done at visible wavelengths with CCD detectors, or 
in recent years, electron-multiplying CCDs (EMCCDs). These have high speed and quantum
efficiency, and in the case of EMCCDs, can detect individual photons. Other AO instruments,
such as AO188 at Subaru Telescope~\citep{minowa10}, have used large numbers of 
single-pixel photodiodes. These are fast and highly sensitive, but tend to be cumbersome
and expensive. In both situations, the wavefront sensing is done at optical wavelengths, 
and the science instrument uses longer wavelengths. Wavefront sensing at optical and
carrying out the science observations at near-infrared (NIR) or infrared (IR) is the most
common setup for two reasons. First, the amplitude (in meters) of wavefront errors is approximately 
constant with wavelength, and the reduction in image quality due to aberrations
scales as the wavefront error divided by the wavelength. Therefore, at longer wavelengths,
a given amount of wavefront error results in a smaller impact on image quality. Because
it is desirable to have as good of image quality as possible, one is motivated to place the
science instrument at longer wavelengths. Second, until the research described in this 
dissertation was carried out, there have not existed high sensitivity, low noise, high 
speed, reasonable cost detector arrays sensitive to NIR wavelengths.

However, there are compelling reasons to wavefront sense at NIR wavelengths. Many
targets have more emission at NIR wavelengths than at optical wavelengths, so better
wavefront sensing (and therefore AO corrections) are possible if the wavefront sensing
detector is
sensitive to these wavelengths. Such targets include late-type stars or stars obscured
by dust. If one is trying to directly image extrasolar planets in the habitable zones
of their host stars, the lowest contrast and therefore most feasible targets are planets
around M dwarfs~\citep{guyon2012}, which are extremely red stars and have low optical emission. 
Second, NIR wavefront sensing extends the performance of natural guide star AO to redder 
targets and expands the sky coverage available for laser guide star AO~\citep{Wizinowich2016}.
Finally, characterization
of young stars and the dusty disks around them is an exciting and emerging area
of research (e.g.~Chapter~\ref{chapter:hip79977}). The optical light from the youngest 
of these systems is attenuated by the dust, but wavefront sensing at longer 
wavelengths could be possible.

\section{The SAPHIRA Detector}
\begin{figure}
\begin{centering}
\includegraphics[width=0.9\textwidth]{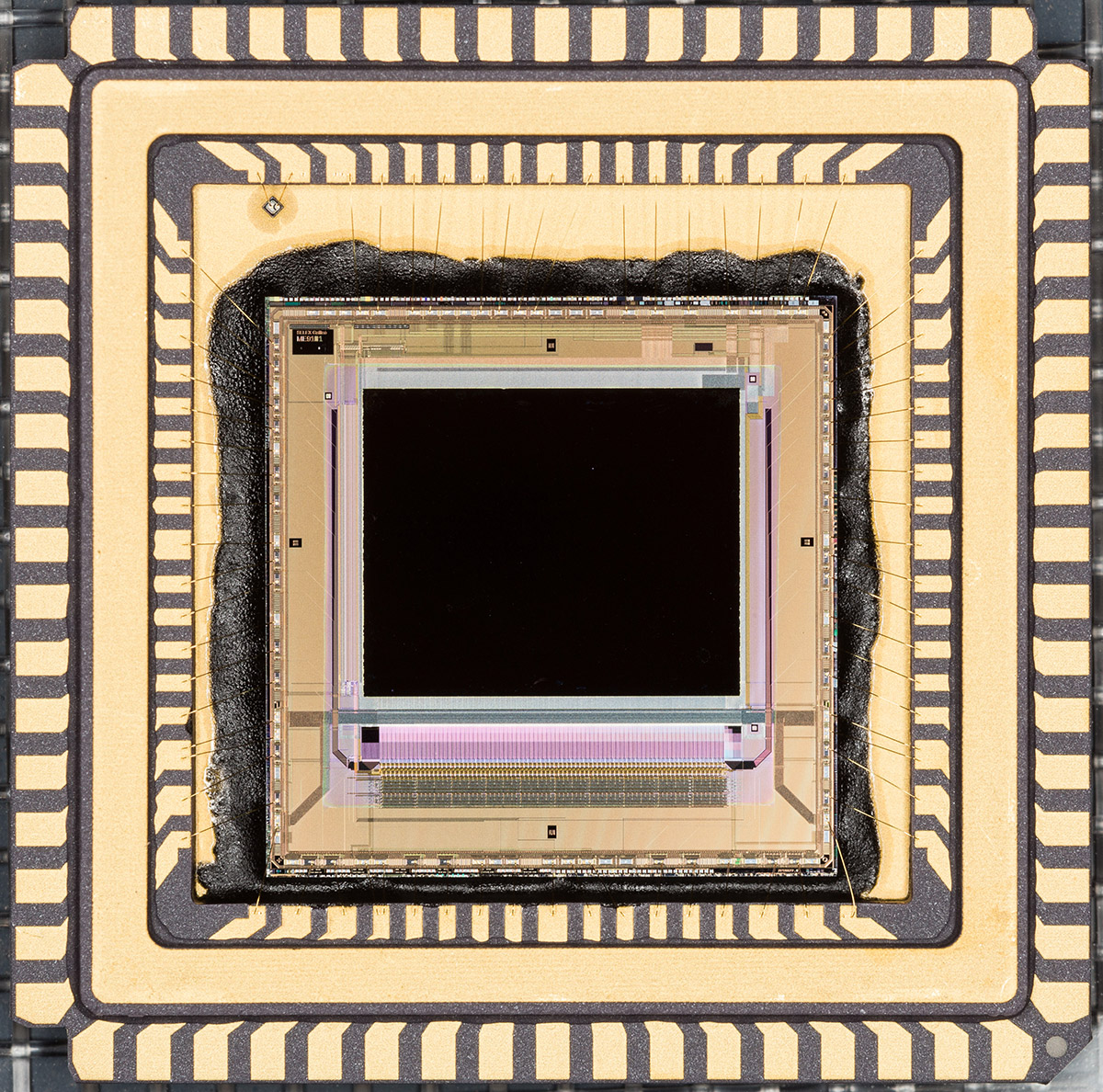}
\caption[Photograph of the SAPHIRA detector]
{The SAPHIRA detector. The black area in the center is the HgCdTe material. Surrounding it is the
read out integrated circuit (ROIC), which applies the voltages to pixels and reads and resets them.
The ROIC is epoxied (black foamy material) to the leadless chip carrier (LCC) and the appropriate 
connections are wire bonded. The entire package in the image measures 24 mm per side.}
\label{fig:saphiraphoto}
\end{centering}
\end{figure}

The Selex Avalanche Photodiode for HgCdTe InfraRed Array (SAPHIRA) detector is the first
high speed, low noise, reasonable cost array sensitive to NIR 
wavelengths. Therefore, it provides the potential for wavefront sensing at longer wavelengths
than are possible for EMCCDs. SAPHIRA is a 320$\times$256 pixel mercury cadmium telluride 
(HgCdTe) array with a 24 $\micron$ pitch and is
manufactured by Leonardo (formerly SELEX). The detector is pictured in Figure~\ref{fig:saphiraphoto}.

\begin{figure}
  \begin{centering}
  \includegraphics[width=0.8\textwidth]{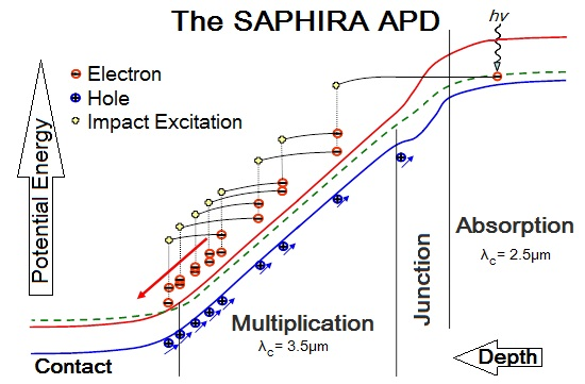}
  \caption[Schematic diagram of the SAPHIRA avalanche multiplication process]{Photons deposit 
  photoelectrons in the top right of the diagram. These diffuse to the junction and are accelerated
  into the multiplication region. There they are accelerated by the bias voltage and can set off
  avalanches of electrons. Depending
  on the bias voltage applied, one photoelectron can ionize up to several hundred additional
  electrons. These gather at the contact with the ROIC, where their total charge is read. In 
  HgCdTe, the holes created in the avalanche process have far lower mobility than the 
  electrons, so the avalanche process is effectively single carrier, making it extremely low 
  noise. Since the read noise is determined by the ROIC and readout electronics, it is 
  independent of the avalanche process. By multiplying the signal of an incoming photon and 
  keeping the read noise fixed, it is possible to obtain the equivalent of sub-electron 
  read noise. Photons with wavelengths of $0.8-2.5 \; \micron$ are absorbed in the absorption
  region (far right of diagram) and therefore cross the full multiplication region. 
  However, in some cases photons with wavelengths of $2.5-3.5 \; \micron$ can be absorbed in 
  the multiplication region and cause a gain-dependent and spurious signal, so it is important
  to use a cold filter that blocks this wavelength range of photons.
  This figure is courtesy of Leonardo.}
  \label{fig:saphiradepthvspe}
  \end{centering}
\end{figure}

\begin{figure}
  \begin{centering}
  \includegraphics[width=0.8\textwidth]{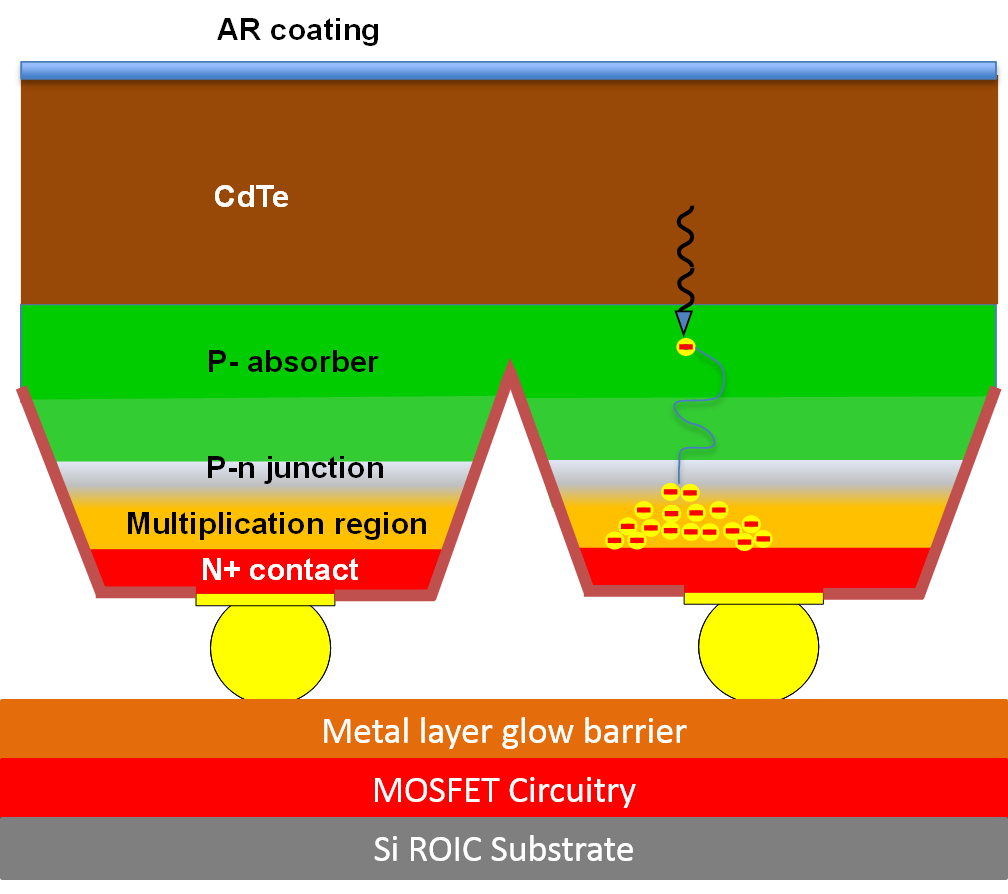}
  \caption[SAPHIRA cutaway diagram]{A cutaway diagram of the structure of two SAPHIRA pixels. The CdTe
  layer defines the 0.8 $\micron$ wavelength cutoff of SAPHIRA. The yellow dots are the indium bump
  bonds by which the HgCdTe structure is hybridized to the ROIC. The mesa structure (inverted pyramid
  shape) of the pixels is designed to prevent photoelectrons from diffusing into neighboring pixels, since
  there is no field in the absorption layer. It also prevents photons with wavelengths of 
  $\sim 2.5 \; \micron$ from reflecting out of the pixels after entering.
  This figure is courtesy of Leonardo.}
  \label{fig:saphirastructure}
  \end{centering}
\end{figure}

The pixels of SAPHIRA are linear mode avalanche photodiodes (APDs), and the basic principle of their
operation is illustrated in Figure~\ref{fig:saphiradepthvspe}. Photons shorter than 0.8 $\micron$
are unable to penetrate the CdTe substrate on the surface of the detector (a cross-section cutaway
of the detector is illustrated in Figure~\ref{fig:saphirastructure}). Photons with wavelengths of
$0.8-2.5 \; \micron$ are absorbed in the absorption layer. Their photoelectrons diffuse
to the junction and then are accelerated in the multiplication region by the bias voltage and 
collisionally set off avalanches
of electrons. The number of avalanches that occur depends on the bias voltage applied across the 
multiplication region. At 1.5 V bias, there are no avalanches and SAPHIRA operates like a 
traditional HgCdTe array. At bias voltages of $\sim20$ V, nearly 1000 electrons are produced by 
each detected photon.
Unlike Geiger mode APDs, which need to be reset after each photon arrival~\citep{Renker2006}, 
SAPHIRA's output is linear with the number of incoming photons (within the detector's $\sim 10^5$ e$^-$ 
dynamic range). Therefore, unlike Geiger mode APDs, there is no dead time between photon arrivals, and
the detector is useful in high-flux regimes. Also, the avalanches in Geiger mode APD arrays typically 
emit significant amounts of glow, which effectively causes high dark current in pixels adjacent to the
one experiencing the avalanche.

The avalanche process in SAPHIRA is effectively
single-carrier because the holes have far lower mobility than the electrons. In other words,
the electrons set off additional avalanches during their path across the multiplication region, but the
holes do not cause additional avalanches. For this reason, the excess noise factor $F$ is very near 1.
$F$ is the ratio of the signal to noise of photoelectrons (before avalanche multiplication) to the
signal to noise of avalanche-multiplied electrons. An ideal APD would have $F=1$, and it has been 
measured to be in the range of $1-1.3$~\citep{Finger2016} for SAPHIRA. On the other hand, both the holes and electrons contribute to the avalanche process in Si APDs, so they have a much larger excess noise
factor than is the case for SAPHIRA.

Because the avalanche multiplication is essentially noise free and occurs prior to the reading of charge
in the ROIC and digitization
of the signal in the readout electronics, it does not affect the read noise. The avalanche multiplication
increases the signal but has no impact on the read noise. Therefore, one can divide the read noise
by the avalanche gain in order to obtain the equivalent noise without the avalanche gain; gain-corrected
read noises below 1 e$^-$ are easily achievable~\citep{Finger2016}. This sub-electron read 
noise and the low excess noise factor $F$ makes SAPHIRA well-suited for photon-counting operation,
which previously has only been achieved at $\sim 1 \; \micron$ wavelengths using Geiger mode APDs 
and photomultiplier tubes. Both of these technologies have limited applications because they need to 
be reset after each photon arrival and therefore cannot accommodate high flux levels.
The photon-counting capability of SAPHIRA was demonstrated by~\citet{Atkinson2018}.

The quantum efficiency, which is the percent of incoming photons of a given wavelength that are
detected, averages $60-80\%$ over $0.8-2.5 \; \micron$ and therefore is competitive with other 
detectors~\citep{Finger2016}. 
As stated previously, photons
The dark current of SAPHIRA detectors is at least as good as the best other HgCdTe devices.
\citet{Atkinson2017} measured the dark current as a function of temperature and detector 
version and provided an in-depth analysis of this. In short, the gain-corrected dark current
for bias voltages of $2.5-8$ V is about $0.02-0.03 \; \rm{e}^- \rm{s}^{-1} \rm{pix}^{-1}$
at 62.5 K for Mk.~3, 13, 14, and 19 detectors. However, this is primarily due to the ROIC glow, not
intrinsic dark current in the HgCdTe material. They placed a $2\sigma$ upper bound on an intrinsic
dark current of $0.0015\; \rm{e}^- \rm{s}^{-1} \rm{pix}^{-1}$ at 62.5 K, and a similar value
was reported by~\citet{Hall2018}. This makes SAPHIRA a highly promising detector for 
low-background observations, but that capability has not been explored yet. The dark current
is not a major consideration for AO applications because the individual exposures are on the
order of a millisecond, so even a 1000 e$^-$/s/pix dark current would only produce about an 
electron per pixel per exposure. This is negligible compared to the signal level.

\section{Lab Testing and Deployments of SAPHIRA}\label{sec:whatwedid}
A wide variety of work with SAPHIRAs not explicitly described in the following chapters was carried out 
during the course of this dissertation. First, we tested tested many generations of SAPHIRA detectors.
Leonardo iterated the design of the detectors by varying parameters such as the bandgap structure, 
doping, and anneal time. Particularly notable are the Mk.~3 and 13/14 detectors; the mark number refers
to the generation number of metal organic vapour phase epitaxy (MOVPE) used to produce the detector
structure. Higher numbers were later designs. The Mk.~3 detectors were
the first science-grade SAPHIRAs and were delivered in late 2013; Sean joined Donald Hall's group in 
mid 2014. Mk.~13 and 14 detectors (these were very similar designs) were delivered in 2015 and 
presently remain the detectors of choice for wavefront sensing due to their excellent cosmetics and
consistent behavior. A list of the SAPHIRA detectors tested by UH is given in Table 1.1
of the dissertation of~\citet{Atkinson2018dissertation}. 
During the course of the iterations with Leonardo, the short wavelength
cutoff of SAPHIRA was extended from 1.4 $\micron$ to 0.8 $\micron$, the incidence of hot pixels
was reduced, the persistence was reduced (this was noticed in a Mk.~3 array and not in any later
ones),  the readout integrated circuit (ROIC) glow was reduced, 
and the dark current was  reduced. This process is ongoing; Leonardo continues to refine and 
improve their detector designs.

Second, we assisted with the commissioning of three SAPHIRA ROICs. The initial SAPHIRA detectors
came with ME911 ROICs, which only permitted global resets. Therefore, we could only do sample
up the ramp (conceptually similar to Fowler sampling) reads. This is ideal for low flux 
observations because
the detector is read many times between each reset and the reads can be combined to minimize read
noise, but it is poorly suited for fast observations such as those needed for wavefront sensing.
The first detectors with ME1000 ROICs were delivered in late 2015. This ROIC was developed by
ESO and permitted line-by-line resets, which enabled new readout modes better suited for wavefront
sensing applications. These modes are discussed in detail in Chapter~\ref{chapter:overview}. A gate 
was unintentionally left floating in the ME911 and ME1000 ROICs, and this caused a large amount of ROIC glow. 
This was noticed after the first ME1000s had been produced. The glow inhibited the
performance of SAPHIRA for low-background observations and prevented measurement of its intrinsic
dark current. The ME1001 ROIC was functionally identical to the ME1000, but it fixed the floating
gate issue and re-stacked the metal layers above the glowing output amplifiers. This
greatly reduced the glow seen by the pixels.

Third, we introduced, tested, and refined new readout electronics for operating the SAPHIRA
detectors. For most lab testing and the early observatory deployments, SAPHIRA was operated using a
Gen.~III ARC Leach Controller~\citep{Leach2000}. However, this had a fixed 265 kHz clocking rate, which
enabled a $\sim 100$ Hz frame rate for full frame SAPHIRA reads. This was inadequate for
some of our SAPHIRA usages, and the SAPHIRA was designed for clocking speeds of up to 10 MHz.
Additionally,
the Leach controller was not compatible with line-by-line resets and therefore could not be used
for read modes other than sample-up-the-ramp. Due to these limitations, the IFA developed a 
new, field programmable gate array (FPGA)-based readout controller affectionately called the
Pizza Box. It featured clocking speeds of up to 2 MHz, supported all the SAPHIRA read modes, and
utilized a USB 3.0 interface instead of the Leach controller's proprietary PCI card interface.
Developing and testing the Pizza Box was a several-year effort, but they are now used with 
SAPHIRA detectors at SCExAO at Subaru, the Keck Planet Imager at Keck Observatory, and 
ROBO-AO at the Kitt Peak 1.5-m telescope, and operate a HAWAII-2RG array at the CryoNIRSP
instrument at the Daniel K.~Inouye Solar Telescope on Maui.
The Pizza Box is also increasingly used for laboratory detector testing in Hilo, and other
groups have expressed interest in deploying them.

Fourth, we solved noise problems that occurred during observatory deployments. SAPHIRA deployments
prior to spring 2018 to Subaru and Keck were plagued by noise many times greater than that experienced
in the lab at Hilo. This is largely because they had poor grounding (the volcanic earth of Mauna Kea 
is an especially poor conductor), high magnetic fields, and copious amounts of radio frequency
emissions from nearby instruments. The output amplifiers of SAPHIRA struggled to drive signals
stably over the $\sim 1$ m of cables to the readout electronics, resulting in ringing and noise pickup
in the cables. We collaborated with Australia National University to adapt cryogenic preamplifiers
they had developed to amplify the detector's signals prior to exiting the camera. These reduced the 
read noise in the laboratory by nearly half and caused the noise at SCExAO to finally match that of 
Hilo. This effort is described in Chapter~\ref{chapter:preamps}.

\begin{figure}
\begin{centering}
\includegraphics[width=0.8\textwidth]{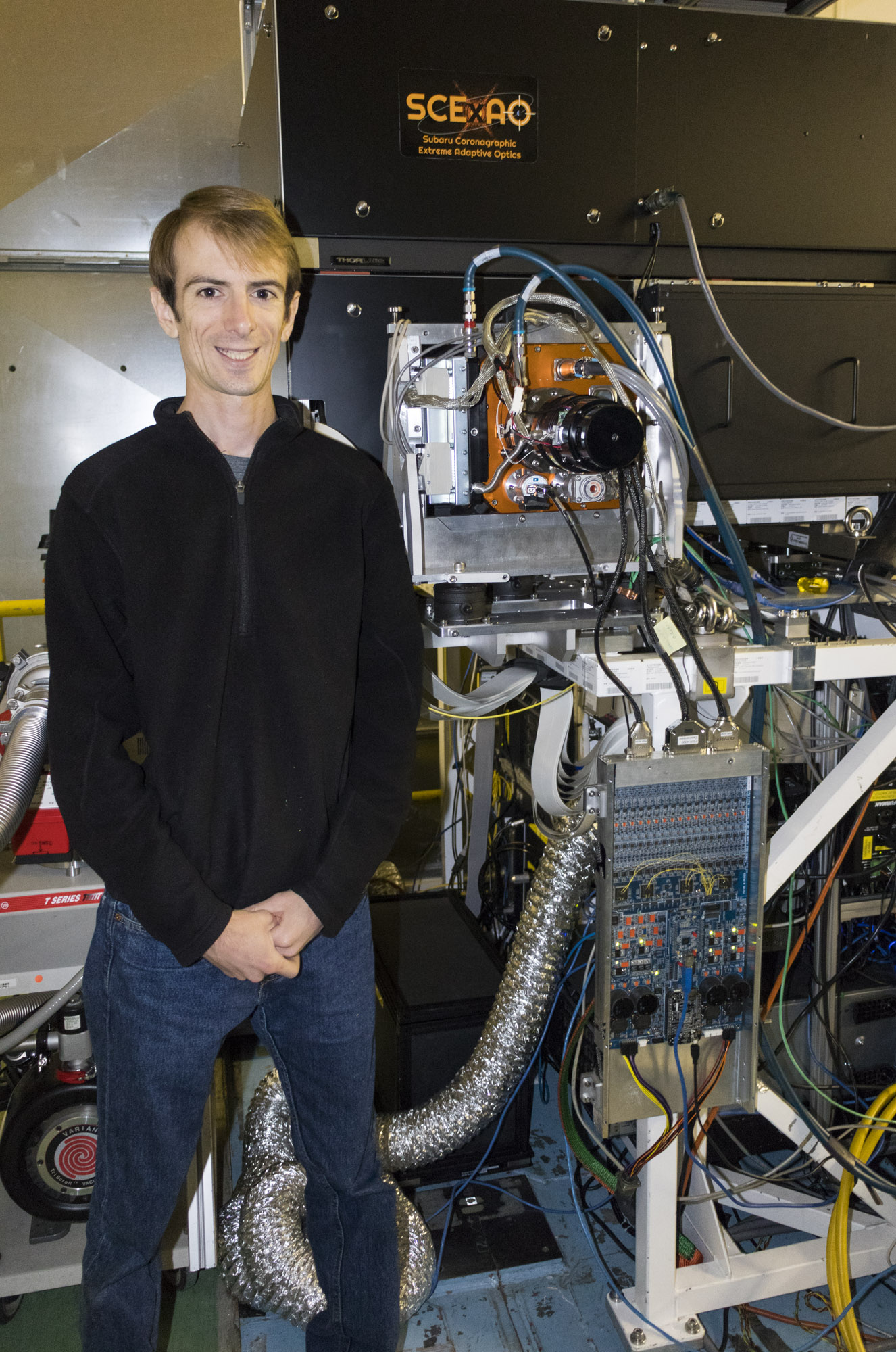}
\caption[The author and SAPHIRA camera at SCExAO]{The author with the SAPHIRA system at the SCExAO instrument
at Subaru Telescope. The SAPHIRA is housed inside the orange Sterling-cooled cryostat (SCC), and it is read
by the FPGA-based ``Pizza Box'' readout electronics below it. Normally the Pizza Box is covered for 
protection.}
\label{fig:authoratscexao}
\end{centering}
\end{figure}

Fifth, and perhaps most importantly, we demonstrated the detectors on-sky. SAPHIRAs are fundamentally
designed for astronomical work, so this was particularly fulfilling. \citet{Atkinson2014} performed 
the first-ever on-sky deployment of
SAPHIRA at the NASA Infrared Telescope Facility, where it was used for lucky imaging. Sean Goebel
dedicated most of his time to the SAPHIRA deployment at the Subaru Coronagraphic Extreme Adaptive
Optics (SCExAO) instrument at Subaru Telescope (the system at that deployment is pictured in 
Figure~\ref{fig:authoratscexao}). The IFA also deployed a SAPHIRA system to the ROBO-AO instrument
at Palomar and then later at Kitt Peak, where it was used for tip/tilt tracking~\citep{Salama2016,Baranec2015,Jensen-Clem2018} 
and produced its first science publication~\citep{Han2017}. We also performed a brief deployment to the Keck II
telescope in preparation for a deployment in the Keck Planet Imager upgrade to NIRC2.

Sixth, we demonstrated new technologies and techniques. We performed the first-ever demonstration of 
pyramid wavefront sensing at NIR wavelengths (Lozi et al., in preparation). We measured speckle
evolution with three orders of magnitude higher temporal resolution than previously had been done
at NIR wavelengths and developed a new method to quantify speckle lifetimes 
(Chapter~\ref{chapter:specklelives}). We demonstrated the usefulness of SAPHIRA for focal-plane
wavefront sensing by connecting it to the speckle nulling loop originally developed by~\citet{Martinache2014}. 
SAPHIRA provided high frame rate images during the testing of a new vector apodizing phase plate
coronagraph developed by~\citet{Otten2017}. 
Finally, as previously stated, we commissioned the Pizza Box readout electronics and ANU cryogenic 
preamplifiers.

\section{Future Work Building upon Our Foundation}
Our efforts described in Section~\ref{sec:whatwedid} and the following chapters laid the foundation
for future work utilizing the new capabilities of SAPHIRA detectors. Due to our demonstrations of
SAPHIRA's ability to perform focal plane and pyramid wavefront sensing and our characterization of
its speed and noise performance, Keck has chosen to use the detector in the Keck Planet Imager
upgrade to NIRC2~\citep{Mawet2016}. The instrument is currently undergoing lab testing in Hilo 
and will be deployed to Mauna Kea in fall 2018~\citep{Mawet2018}.

There are two parallel efforts to develop new SAPHIRA formats. Due to its extremely low dark current,
the detector offers great promise for low-background imaging. The NASA ROSES ARPA has funded the 
development of a 1024$\times$1024@15 $\micron$ pixel SAPHIRA 
array and new ROIC to accommodate it~\citep{Hall2016, Hall2018}. This array will incorporate
reference pixels for improved common-mode noise reduction. Reference pixels have proven extremely
useful for reducing noise in HAWAII devices~\citep{Rauscher2011}, but have not been present on 
SAPHIRA detectors tested so far. The first
of these larger format SAPHIRA devices will be delivered to the IFA in late 2018. These will have
16 outputs (half that of the current, smaller SAPHIRAs) in order to make them three-side close 
buttable. These detectors prioritize low noise and dark current at the expense of speed. Second,
a consortium of ESO, Max Planck Institute, and Herzberg (Canada) is pursuing the development of
a 512$\times$512@24 $\micron$ pixel SAPHIRA detector with 64 outputs (Ian Baker, private
communication). This detector is intended for high speed operation, and the ordering of the 
channels will be 
optimized for pyramid wavefront sensing. The 320$\times$256 pixel SAPHIRAs are well-sized
for wavefront sensing on current 8-10-m-class telescopes, but the 512$\times$512 size is ideal
for the upcoming 30-m-class telescopes.

A number of instruments currently in development have proposed to employ SAPHIRA detectors.
The Magdalena Ridge Observatory Interferometer in New Mexico has proposed to replace their PICNIC
detector with a SAPHIRA for fringe tracking because SAPHIRA's improved sensitivity reduces
the opto-mechanical requirements of the instrument~\citep{Ligon2018}. The Giant Magellan Telescope will use a SAPHIRA
in its Acquisition, Guiding, and Wavefront Sensing instrument in order to phase the mirrors to
50 nm accuracy~\citep{Kopon2018}. 
A commercially-available C-RED ONE SAPHIRA camera~\citep{Greffe2016}, produced by First Light 
Imaging, has been purchased for the CHARA array at Mt.~Wilson, CA. This will be used in a $K$-band 
interferometer which combines the beams from six telescopes for high-resolution imaging of 
disks around young stars~\citep{Lanthermann2018}. The ESO Extremely Large Telescope will use
at least one SAPHIRA for low- and high-order wavefront sensing with natural guide stars in the 
inFraRED cAmera (FREDA) module\footnote{This dissertation grants FREDA the award for the most 
forced acronym contained within its pages.}~\citep{Downing2018}. These instruments have published
in the SPIE proceedings their intent to use SAPHIRA detectors; there likely are additional proposals.

\section{From Lab Testing to Science: a Road Map to the Chapters Ahead}
The ensuing chapters are ordered such that their topics cross the spectrum from engineering
to science. Chapter~\ref{chapter:overview} provides an overview of the SAPHIRA detector and
discusses a variety of considerations that must be taken into account when the detector
is employed for wavefront sensing.
Chapter~\ref{chapter:preamps} describes efforts to reduce
the noise and increase the clocking speeds in SAPHIRA arrays. These characteristics were
particularly limiting at the SCExAO instrument.
Chapter~\ref{chapter:specklelives} describes measurements of speckle temporal evolution 
in extreme AO images and their implications for
observations. Chapter~\ref{chapter:hip79977} contains characterization of the
morphology of the HIP 79977 debris disk that was made possible by the extreme adaptive optics
corrections of SCExAO. These four chapters also appear as journal publications. Lastly, Appendix~\ref{app:codes} links to an online database of the 
codes used to produce the results contained in this dissertation and contains brief 
descriptions of them. 

\bibliographystyle{apj}
\bibliography{ch1bib}

%

\chapter{Overview of the SAPHIRA Detector for AO Applications}\label{chapter:overview}

Note: this chapter originally appeared as \citet{Goebel2018},
with co-authors Donald N.B. Hall, Olivier Guyon, Eric Warmbier, and Shane M. Jacobson.

\section*{Abstract}
We discuss some of the unique details of the operation and behavior of Leonardo SAPHIRA detectors,
particularly in relation to their usage for adaptive optics wavefront sensing. SAPHIRA 
detectors are 320$\times$256@24 $\mu$m pixel HgCdTe linear avalanche photodiode arrays and are sensitive to
0.8-2.5 $\mu$m light. SAPHIRA arrays
permit global or line-by-line resets, of the entire detector or just subarrays of it, and the order
in which pixels are reset and read enable several readout schemes. We discuss three readout modes,
the benefits, drawbacks, and noise sources of each, and the observational modes for which each is 
optimal. We describe the ability of the detector to read subarrays for increased frame rates, and
finally clarify the differences between the avalanche gain (which is user-adjustable) and the 
charge gain (which is not). 

\section{Introduction}
Due to their ability to detect individual photons with both high temporal and spatial resolutions, 
electron-multiplying CCDs (EMCCDs) have greatly improved the sensitivity limits of adaptive optics (AO) 
systems. 
However, EMCCDs are sensitive to optical wavelengths, and in most AO implementations, the
science instrument operates at near-infrared wavelengths. This wavelength sensitivity difference 
between the wavefront sensor and science instrument occurs for two reasons. First, the 
path length differences introduced by atmospheric turbulence are approximately independent of
wavelength, so the aberrations as a fraction of phase are reduced at longer wavelengths. 
Therefore, better image quality is obtained at longer wavelengths, and so the science
instruments are designed to take advantage of this. Second,
until now there have not existed high frame rate, low noise,
reasonable cost infrared detector arrays, so it was not feasible to do high-order wavefront 
sensing at
near-infrared wavelengths. The Selex Avalanche Photodiode for HgCdTe InfraRed Array (SAPHIRA)
detector is the first such technology that enables this.

Operating the wavefront sensor at near-infrared wavelengths provides several 
benefits~\citep{Wizinowich2016}. It has the potential to expand the sky coverage available for 
natural guide
star sensing because it enables the observation of targets that have minimal emission at optical 
wavelengths, such as late-type stars or obscured objects. In particular, there is significant 
interest in observing nearby
M dwarfs because they have favorable contrasts for directly imaging reflected-light extrasolar 
planets located in their habitable zones.
Second, if both the science module and wavefront sensor are sensitive to similar
wavelengths, they can share a greater fraction of optical elements inside the instrument and thereby 
minimize noncommon path errors. For these reasons, SAPHIRA detectors enable exciting new AO 
observing modes.

SAPHIRA arrays are well-suited for wavefront sensing because they provide two advantages over other
infrared devices. First, they have a user-adjustable avalanche gain. This is
discussed at length in Section~\ref{sec:gains}, but in short, it amplifies the signal from photons
but not read noise sources. At high gains, the signal of an individual photon is greater than the read
noise. If the noise is divided by the avalanche gain, it can be equivalent to
$<1 e^-$. The second advantage of SAPHIRA is that all outputs are still used when reading 
subarrays, so greatly increased frame rates can be achieved by reading less than the full detector
(Section~\ref{sec:subwindows}). This is in contrast to (for example) Teledyne HAWAII detectors, in 
which each output reads a contiguous stripe of the detector and only a single output is used
when in sub-array mode.

SAPHIRA detectors are 320$\times$256 pixel mercury cadmium telluride arrays and have a 24 $\mu$m
pixel pitch~\citep{Baker2016}. They are manufactured by Leonardo (formerly called Selex)
and have 32 outputs. The absorption layer (where photons are intended to be absorbed) has 0.8-2.5 
$\mu$m sensitivity
and is transparent at longer wavelengths. These longer wavelengths must be filtered out to avoid 
spurious signal from
photons absorbed in the multiplication layer. The multiplication layer, which is where the avalanche
multiplications occur, lies below the absorption layer and has sensitivity to 3.5 
$\mu$m~\citep{Finger2016}.
Because SAPHIRAs are a developmental program, they are assigned 
mark numbers which refer to the generation of metalorganic vapour phase epitaxy (MOVPE)
layer architecture; higher numbers are newer designs. Mk.~3 SAPHIRA arrays were the first
science-grade ones and were delivered in 2013. The Mk.~13 and 14 arrays (which are indistinguishable 
for the purposes of wavefront sensing) were delivered in 2015 and (as of the time of writing) remain 
the best arrays for telescope deployments due to their uniform cosmetic qualities. Later generations
of arrays have improved in some areas (e.g. dark current) but had problems in other areas (e.g.
dead pixels or operational reliability). There are plans to produce larger format SAPHIRA arrays
for low-background imaging and high-order wavefront sensing on thirty-meter-class 
telescopes~\citep{Hall2016A}.
SAPHIRA devices produced until late 2015 utilized ME911 read out integrated circuits (ROICs), 
which only permitted global resets and therefore were limited to up-the-ramp readout mode 
(Section~\ref{sec:utr}). The ME1000 ROIC enabled row-by-row resets, which made 
possible read-reset-read and read-reset modes (Sections~\ref{sec:rrr} and~\ref{sec:rr}).
The ME1001 ROIC is functionally identical to the ME1000 ROIC, but it features reduced
glow during operation.

While the operation of the SAPHIRA APD array is very similar to that of conventional HgCdTe arrays such
as the Teledyne HAWAII series~\citep{Beletic2008} and the Raytheon Virgo~\citep{Starr2016}, there are some
important differences. The conventional arrays are normally operated at a bias of a few hundred mV, 
occasionally up to a volt, and the full bias voltage between the substrate and the node is within the 
dynamic range of the readout circuit. After reset, the array discharges to the substrate voltage. The 
depletion is modest and varies with
bias voltage, and that results in a highly bias-dependent diode capacitance.
In contrast, the SAPHIRA typically operated at bias voltages of $2.5-18$ V, way beyond the dynamic range of 
the readout circuit, and the node voltage is referenced to ground in the ROIC. At these bias voltages the 
photodiode is largely or completely depleted of electrical charge, and the response is highly linear over the dynamic range. As 
a result of the depletion, there is no settling of charge due to the change in electric field when the bias voltage is changed. In conventional detectors or
when SAPHIRA is operated at very low bias voltage, the detector needs to be allowed
to settle after changing the bias voltage.

SAPHIRA arrays are beginning to enter widespread deployments. The University of Hawaii Institute for
Astronomy has extensively tested the detectors in the lab and at telescopes. Over the last several
years, we have continuously deployed a SAPHIRA system to the SCExAO extreme AO instrument at Subaru 
Telescope, where it is primarily used for focal-plane wavefront 
sensing~\citep{Goebel2016}. It has also been deployed short-term to the NASA 
Infrared Telescope Facility (IRTF)~\citep{Atkinson2014}, where it obtained diffraction-limited images of binary star systems using
the lucky imaging\citep{Fried1978} technique. Also, a SAPHIRA system is currently being integrated into
the Keck II AO system in order to enable near-infrared 
pyramid wavefront sensing~\citep{Mawet2016}. SAPHIRA is being used for fast NIR imaging on the Robo-AO instrument, which was initially deployed to the 1.5 m telescope at 
Palomar~\citep{Baranec2015} and then 
later to the 2.1 m telescope at Kitt Peak~\citep{Jensen-Clem2018}. While at Robo-AO, SAPHIRA
produced its first science publication~\citep{Han2017}. ESO has
also been evaluating SAPHIRA arrays~\citep{Finger2012}, and they are deployed to the Very
Large Telescope's GRAVITY 
instrument for wavefront sensing at the individual telescopes and fringe tracking at the combined 
focus~\citep{Mehrgan2016}. Lastly, First Light Imaging has developed a commercial system called
C-RED One that combines the SAPHIRA detector, camera and associated cooling, and readout
electronics~\citep{Greffe2016}.

In this paper, we report on the operation modes and intricacies of behavior of SAPHIRAs. 
Unless stated otherwise, the data presented below were collected at a temperature of 85 K
and at unity avalanche gain (i.e. a bias voltage of 2.5 V) with Mark 13 and 14 arrays on
ME1000 ROICs.

\section{Readout Modes}\label{sec:readoutmodes}
The readout modes available on SAPHIRA detectors differ according to when and the number of times that
pixels are read between resets. 
We have operated the SAPHIRA in three different readout
schemes: 1) sampling up-the-ramp mode, which utilizes an initial reset of all pixels
followed by multiple reads; 2) read-reset-read mode, which reads a row of pixels, resets it, and reads
it again before clocking to the next row; and 3) read-reset mode, whereby a row is read and
then reset before moving to the next row, and a reference image is subtracted in order to remove 
the pedestal voltage (sometimes called the fixed-pattern noise or bias). These readout modes are 
independent of subarray size. In practice, we use read-reset mode for AO-related observations because
it provides regular time sampling (unlike sample up-the-ramp mode) and better noise performance
and double the frame rate compared to read-reset-read mode.
In the following sections, we discuss each readout mode in turn. 

\subsection{Sampling Up-the-ramp Mode}\label{sec:utr}
In sampling up-the-ramp (SUTR) mode, the entire array is reset, and then it is non-destructively read a 
user-specified number of times~\citep{Chapman1990}. These reads can be processed in one of two ways. 
First, the user can fit a line to the ``ramp'' (flux vs.~time) behavior for each pixel. This removes
the pedestal voltage (fixed-pattern noise) and reduces the white read noise variance as $1/N$, where 
$N$ is the number of times the array is read per ramp. 
A mathematical analysis of SUTR (and the conceptually similar Fowler sampling)
is given by Garnett and Forrest (1993)~\citep{Garnett1993}. Additionally, if a pixel saturates partway 
through the ramp due to high flux or a cosmic ray 
strike, the reads in which the pixel is saturated can be excluded from the fit. 
When SUTR data are reduced by line-fitting, one ``science'' frame is produced for each ramp. This is
well-suited for low-background observations in which read noise dominates and the exposures are long, but
it is poorly suited for instances where fast analysis of the images is required (such as AO wavefront
sensing).

The second way to reduce SUTR data is to subtract subsequent reads. If being used for AO wavefront
sensing, this enables more frequent updates to the correction. By subtracting temporally adjacent frames,
one removes the pedestal voltage and $kTC$ (reset level) noise because it is present in both images,
leaving only the flux that accumulated in the pixel between those two exposures and the read noise.
However, this mode causes three challenges. First, it requires that the dynamic range
be rationed across the ramp (if there are $N$ reads per ramp, then no individual read can utilize more
than $1/N$ of the dynamic range, or else the final reads of the pixel will contain no new flux). 
Second, there is irregular time sampling because it wouldn't make sense 
to subtract the read immediately before a reset from the one immediately after the reset because
this would result in negative flux and $kTC$ noise.
This is problematic for applications that require regular temporal sampling.
Third, due to the intrinsic nonlinearity of infrared
arrays, the responsivity of the detector decreases as the pixels accumulate flux. This causes frame pairs 
toward the end of the ramp to incorrectly report less flux than an equally-illuminated pair at the 
beginning of the ramp. One can compensate for this last effect with a standard linearity correction, 
however.

Often, the frame read immediately after a reset contains excess noise due to settling effects. In 
practice, we typically exclude the first frame following each reset from analysis. This
exacerbates the irregularity of time sampling in SUTR mode and reduces its duty cycle, particularly 
when there  are few reads between resets. 

In both methods of reducing SUTR data, variations in the voltage of a pixel after one
reset compared to its voltage after another reset has no effect because it is in all reads of 
a given ramp. This is called $kTC$ noise because its variance is given by
\begin{equation}
\sigma_{kTC}^2 = k_{B} T C
\label{eqn:ktc}
\end{equation}
where $k_{B}$ is the Boltzmann constant, $T$ is temperature, and and $C$ is node capacitance. We 
typically operate the detector at $T=85$ K, and 
$C=28$ fF 
for SAPHIRA, so $\sigma_{kTC}\approx36$ e$^-$. 
Unlike read noise, $kTC$ noise is not affected by the pixel read rate. 
Because SUTR readouts are not affected by $kTC$ noise, they enable the lowest noise of the 
readout modes discussed here. SUTR readouts are best-suited for low-background, long integrations.

\subsection{Read-reset-read Mode}\label{sec:rrr}
In read-reset-read mode, a row is read, reset, and then read again before clocking to the next
one. The read immediately following the reset (which contains only the pedestal voltage) is then 
subtracted from the read preceding the next reset after scanning through the array
(which contains the pedestal voltage plus flux).
Therefore, there are two reads per science frame. This mode removes the effects of $kTC$
noise.

\begin{figure}[!ht]
\begin{centering}
\includegraphics[width=\textwidth]{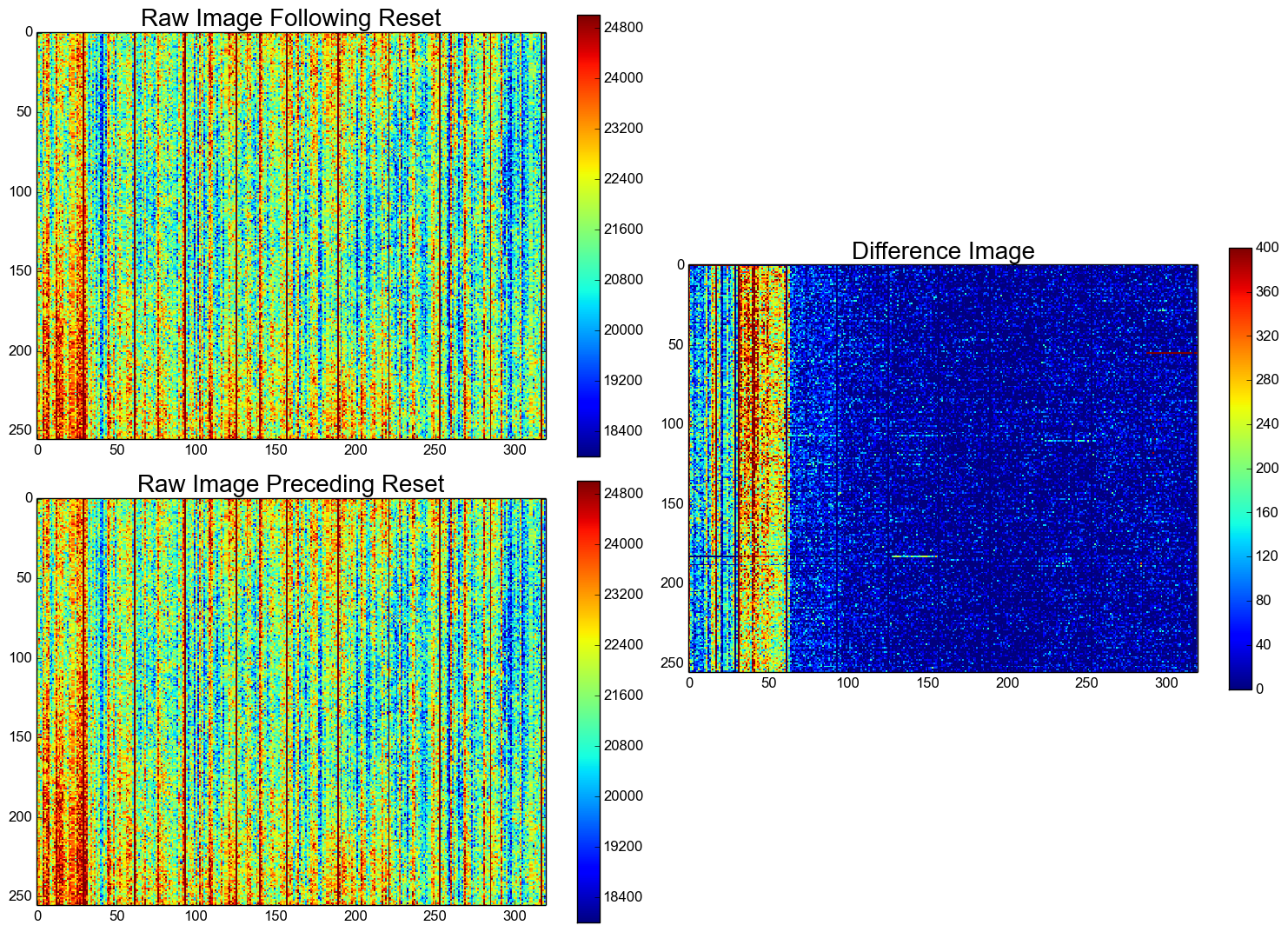}
\caption[The problem with read-reset-read mode]{On the left are consecutive unilluminated raw frames from SAPHIRA after (top) and before 
(bottom) the reset in read-reset-read mode. Approximately 2.7 ms has elapsed between the two reads. As 
is normal for images from infrared arrays, the pedestal noise dominates both images. However, when the
before image is subtracted from the after image, not all the noise  disappears. An entire
row is reset at a time, the detector clocks from left to right, and the column blocks read immediately
following the reset have spurious flux. Operating the SAPHIRA with a subarray that excludes these left
columns does not solve the problem; the column blocks with spurious flux simply move to the edge of the
subarray being read, since they are read immediately following the reset. All images have units of 
ADUs. We have noticed that this effect is reduced on other SAPHIRA arrays. Because of this post-reset 
setting problem, we do not typically operate SAPHIRA arrays in read-reset-read mode.}
\label{fig:rrr}
\end{centering}
\end{figure}

In practice, read-reset-read mode presents two problems. First, and more importantly, settling effects
immediately following the reset lead to spurious flux in the first column blocks read (a column block is a 32-pixel-wide row of pixels. These are read simultaneously.). This problem is
illustrated in Figure~\ref{fig:rrr}. The amplitude of this effect varies from one SAPHIRA array to 
another, and we theorize that it can be mitigated by inserting a delay between the reset and first read,
though we have not tested this because it would slow the frame rate. In our typical 1 MHz clocking
rates, there is only 850 ns between the end of the reset pulse and the sampling of the pixels in
the first column block.
Second, because there is read noise associated with both reads, these add in quadrature for the 
science frame. For the $\sim 100$ kHz clocking rates used for most IR detectors, the read noise is much
less than the $kTC$ noise. However, the read noise approximately scales with the square root of the 
sampling frequency, and for the 1-10 MHz sampling frequencies used for SAPHIRA wavefront sensing, 
these two
noise sources become comparable. Because of the spurious flux and replacement of $kTC$ noise with
similar-magnitude read noise, we do not commonly use read-reset-read mode.

\subsection{Read-reset Mode}\label{sec:rr}
If the read noise of one frame is comparable to its $kTC$ noise, then the correlated double sampling
of read-reset-read mode (which produces an image free of $kTC$ noise but containing read noise contributed
by both frames) is no better than subtracting a reference dark frame that has been 
produced from averaging together many frames (and therefore contains negligible $kTC$ and read noise)
from a single read (which does contain these noise sources).
The read noise depends on the sampling rate and specifics of the readout electronics, but 
during typical usage of SAPHIRA, it is of the same order as the 36 $e^-$ $kTC$ noise calculated
from Equation~\ref{eqn:ktc}. 

In read-reset mode, the pedestal voltage of the detector is removed by averaging 
together many unilluminated frames and then subtracting this reference frame from each individual frame.
During science operation, these dark frames can be collected before or after the data frames. However,
a new reference frame needs to be collected for each bias voltage (the bias voltage controls the avalanche 
gain), since the pedestal voltage pattern on the detector changes with bias voltage. Read-reset mode is 
our standard mode for AO observations because its noise
is similar or lower than that of read-reset-read mode, it produces a ``science'' frame for every read,
and the temporal sampling is regular. Additionally, 
read-reset mode does not exhibit the post-reset settling problem (the gradient of spurious flux) of 
read-reset-read mode. This is because read-reset-read mode has a read immediately following the reset,
and the pixels are still settling during this time; on the other hand, in read-reset mode, the pixels
have the entire integration time to settle.

\subsection{Comparison of the Readout Modes}\label{sec:comparison}
A comparison of the loop update rate and noise sources for each of the readout modes is
provided in Table~\ref{table:readoutmodes}. In this section, we compare these modes and their
associated noise sources.

\begin{sidewaystable}
\caption{Comparison of noise sources for different SAPHIRA read modes}
\vspace{1ex}
\centering
\begin{tabular}{ |c|c|c|c| } 
 \hline
	& \textbf{Maximum loop}  
    & \textbf{Noise variance}
	& \textbf{Comments} \\
    & \textbf{update rate}   
    & $\sigma_{total}^2=\sigma_{photon}^2+$	
    &		  
    \\ \hline 
      
	\textbf{Up the ramp}	
    & Every read, with 
    & 
    & Dynamic range must be split
    \\     
    (CDS pairs processed	
    & interruption 
    & $2\sigma_{RN}^2 $ 
    & between the reads in the ramp.
    \\
    independently)
    & following reset
    &	
    & 
    \\ \hline 
     
  	\textbf{Up the ramp} 
    & 
    & 
    & Cosmic rays and saturation \\
    (flux determined by
    & Every ramp
    & $\sigma_{RN}^2/N_{reads}$ 
    & can be mitigated by excluding \\
    fitting to ramp) 
    &				
    & 
    & those reads from fitting.
    \\ \hline 
     
	\textbf{Read-reset-read} 
    & Every two reads 
    & $2\sigma_{RN}^2 + \sigma_{spurious}^2$ 
    & The first few column blocks \\ 
    
    & 
    & 
    & may contain spurious flux.
    \\ \hline 
    
	\textbf{Read-reset} 
    & Every read 
    & $\sigma_{RN}^2 + \sigma_{settling}^2 + \sigma_{kTC}^2 $ 
    & Requires reference dark frame.
    \\ \hline 
\end{tabular}
\quote{A comparison of the different readout modes available for the SAPHIRA detector.
$\sigma_{RN}^2$ is the variance in read noise, $N_{reads}$ is the number of reads between resets
in a ramp, $\sigma_{kTC}^2$ is the kTC noise, $\sigma_{spurious}^2$ is the noise associated with 
post-reset settling in read-reset-read mode,
$\sigma_{settling}^2$ is noise due to thermal drifts and is negligible if the data and reference
frames are collected near in time to each other or both are collected after at least two hours of
detector operation. $\sigma_{RN}^2$ is assumed to be white. Noise due to radio frequency 
interference (Section~\ref{sec:rfi}) is a component of $\sigma_{RN}^2$.}
\label{table:readoutmodes}
\end{sidewaystable}

The SAPHIRA detector can be operated with global resets or line-by-line resets. The user can
configure which regions are reset; one can save time by not resetting portions of the array
not being read. We typically operate the detector with global resets
when sampling up-the-ramp and line-by-line resets in read-reset and read-reset-read mode. However,
line-by-line resets can be used in all three modes. ME1000 and ME1001 ROICs support both global 
and line-by-line resets, whereas ME911 ROICs only support global resets.

Due to the fact that an entire row is reset at once, but then 32 pixels are read in each increment along
the row, in read-reset mode, one side of the detector will see less flux. In other words, for pixels on
one side of the array, there is a greater interval between the read and reset than for pixels on the
opposite side of the array. This effect is worse for a short, fat subarray than for a tall, thin
one. For a 320$\times$256 pixel array, the difference in flux from one side to the other is 
$\sim$0.3\%. For a 320$\times$64
subarray, however, this effect is 1.3\%. For most astronomical applications, a standard 
flat field correction effectively compensates for this effect. This effect is not relevant to
sampling up-the-ramp readouts because one is calculating the slope of the flux between reads, and it
does not affect read-reset-read images because the read of a given pixel is delayed relative to the 
reset an equal amount in both frames.

\begin{figure}[!ht]
\begin{centering}
\includegraphics[width=\textwidth]{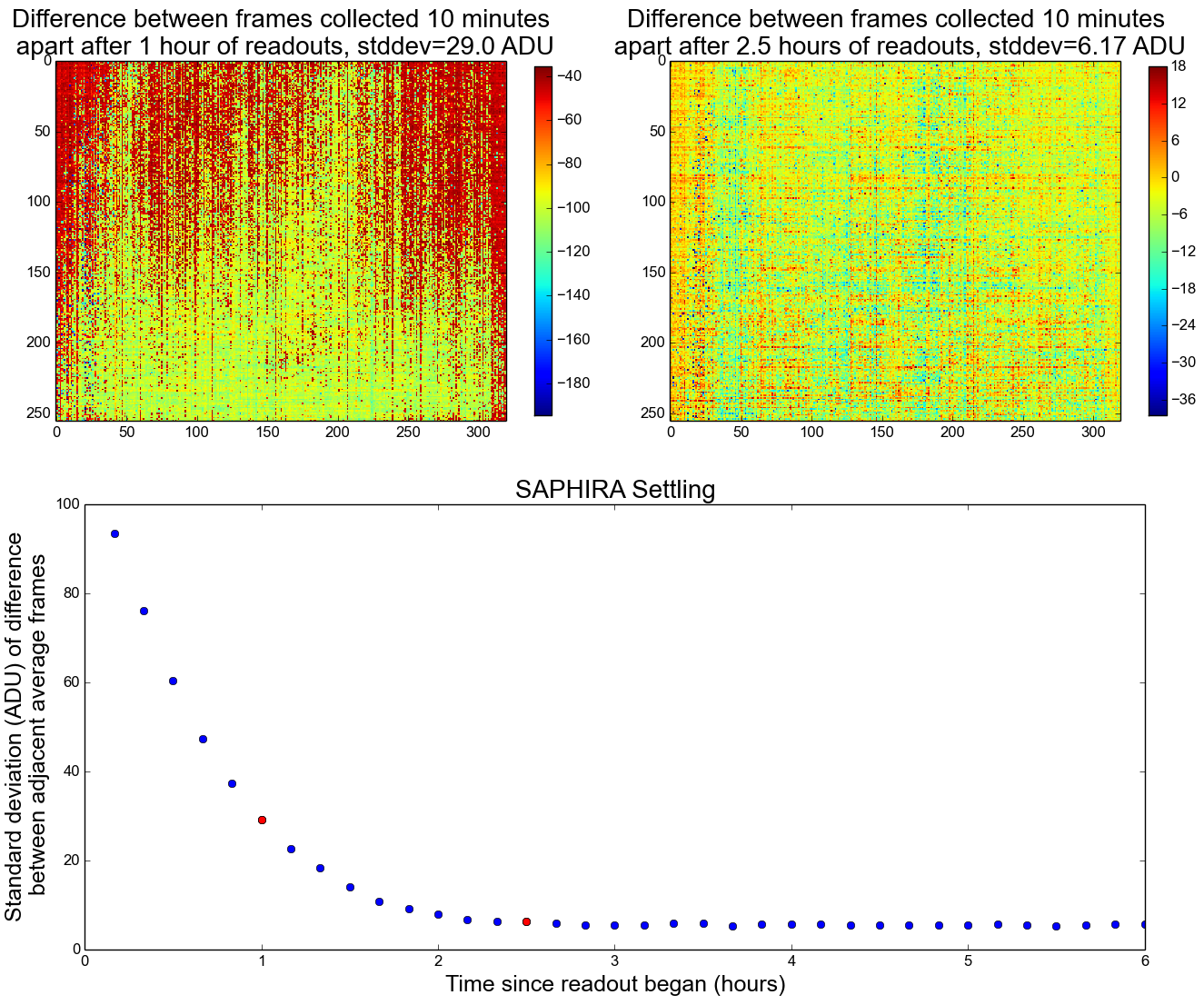}
\caption[SAPHIRA temporal settling]{While the SAPHIRA continuously streamed images, frames were saved every 10 minutes for
several hours. At each interval, 1000 frames were saved and averaged together in order to reduce
read noise. Each average frame was then subtracted from the average frames collected 10 minutes
earlier and later to see how the pedestal voltage of the detector changed. For
approximately the first two hours of image streaming, the pedestal voltage exhibited
slow drifts (as shown in the top left figure). However, after about 2-2.5 hours, the detector
had largely stabilized (top right). The bottom plot shows how the spatial standard deviation of 
these difference frames decreases with time. The two example frames shown at the top are identified
with red points in the bottom plot. For this reason, it is important to clock the detector for
at least two hours before observations that require good stability.}
\label{fig:settling}
\end{centering}
\end{figure}

In addition to the post-reset settling effects discussed in Sections~\ref{sec:utr} and~\ref{sec:rrr},
we have also observed a smaller amplitude but longer duration thermal settling in the array. When
an array begins clocking after sitting idle, the power dissipation in it changes significantly. This 
causes thermal variations in the array, and it can take hours for these to reach equilibrium. During
this time, pixels exhibit slow drifts. We performed a test in which we read out
an unilluminated array for several hours in read-reset mode. Every ten minutes, we saved 1000 frames 
and averaged them to produce an image of the detector's pedestal voltage with minimal read noise. 
We then subtracted
subsequently collected average frames and computed the standard deviation to see how much the
pedestal voltage drifted. The result is shown in Figure~\ref{fig:settling}. Again, an approximately two-hour 
settling timescale
was observed. For this reason, it is important to run an ``idle sequence'' before observations in which 
no data are saved but the  detector clocks as it would during science operations. The SAPHIRA detector
should be operated like this for at least two hours prior to observations where temporal stability
is required (such as before collecting the reference frames for subsequent read-reset mode 
observations). It is worth noting that clocking-caused thermal drifts are not unique to SAPHIRA arrays;
clocking the same way when idling and observing is key to obtaining good performance from CMOS 
detectors in general.

This same two-hour timescale for SAPHIRA thermal drifts was also observed in dark current 
measurements performed by~\citet{Atkinson2017}. While operating SAPHIRA arrays using
up-the-ramp sampling, they noted that it took approximately two hours of continuous clocking before the dark
current stabilized. They directly confirmed this thermal time constant during measurements to calibrate the 
detector temperature offset from the usual temperature monitoring point at the thermal control sensor/heater
assembly. This was done using a temperature sensor permanently mounted to the SAPHIRA leadless chip carrier
ceramic (this sensor
cannot be powered up during science observations because the glow levels are prohibitive). For the temperature 
offset calibration, the system was allowed to stabilize with the SAPHIRA powered down (except for the 
temperature sensor) to establish the offset between the two temperature sensors. The SAPHIRA was then powered
on and clocked normally. After it stabilized, which took approximately two hours, the on-chip sensor was used
to measure the temperature offset.

\section{Subarray Operation and Calculation of Frame Rate}\label{sec:subwindows}
There are 10 32-channel column blocks across the SAPHIRA, and it can be subarrayed to read
integer numbers of column blocks (i.e. the frame size can be adjusted by 32-pixel increments in the x
direction). On the other hand, the detector will read any number of pixels in the y direction. Multiple
subarrays can be supported at once. The 32 outputs on the SAPHIRA read out 32 adjacent pixels in a 
row at a time.
Greatly increased frame rates during subarray readouts are possible because the interleaved nature of the
outputs enables all of them to still be used.
To illustrate this, consider the case of a 128$\times$128 pixel subarray. A HAWAII-2RG has 2048 pixels per side
and 32 outputs (therefore 64 columns per output), or a HAWAII-4RG has 4096 pixels per side and 64 outputs 
(so again 64 columns per output). Either HAWAII array would utilize two outputs to read the 128$\times$128
subarray, whereas the SAPHIRA could read it using all 32 outputs. At identical pixel clocking rates, the 
SAPHIRA could read it 16 times faster.

The SAPHIRA frame rate scales roughly inversely with the number of pixels in the window. This rule is 
approximate because the array must be reset and there are extra clock cycles at the ends of rows and frames; 
the full equation for duration of one frame when operating in read-reset mode is given by 
Equation~\ref{eqn:timage-rr}.
\begin{equation}
t_{img,rr} = t_{fd} + t_{fc} + n_{r}(t_{rd}n_{cb} + t_{rst} + t_{prst} + t_{rc} + t_{ra})
\label{eqn:timage-rr}
\end{equation}
where $t_{img,rr}$ is the time to collect one image in read-reset mode and the other parameters and
their typical values are summarized in Table~\ref{table:timings}.
These clock pulses are all necessary for the detector to operate reasonably and
cannot be set to 0; the acceptable ranges are specified in the SAPHIRA manual. 
Read-reset-read mode is similar, but each pixel is read twice, and there is an 
extra $t_{rc}$ per row.
\begin{equation}
t_{img,rrr} = t_{fd} + t_{fc} + n_{r}(2t_{rd}n_{cb} + t_{rst} + t_{prst} + 2t_{rc} + t_{ra})
\label{eqn:timage-rrr}
\end{equation}
For sample up-the-ramp mode, the time to collect a frame (not including the time to reset, 
since that typically is not done every frame) is
\begin{equation}
t_{img,sutr} = t_{fd} + t_{fc} + n_{r}(t_{rd}n_{cb} + t_{rc})
\label{eqn:timage-utr}
\end{equation}
Naturally, the framerate $f$ is
\begin{equation}
f = t_{img}^{-1}
\label{eqn:framerate}
\end{equation}
The timing values for our current 1 MHz pixel rate are summarized in Table~\ref{table:timings}. These 
are subject to tuning but can be used as typical values. In line-by-line reset modes, these enable a 
1.69
kHz frame rate for a 
128$\times$128 pixel subarray or 339
Hz frame rate for 320$\times$256 pixel full frames.
\begin{sidewaystable}
\caption{SAPHIRA clocking timings}
\vspace{1ex}
\centering
\begin{tabular}{ |c|c|c|c| }  \hline
\textbf{Parameter} & \textbf{Definition} & \multicolumn{2}{|c|}{\textbf{Value}} \\ \hline
 &  & \multicolumn{2}{|c|}{Frame size} \\
 \cline{3-4}
 & & 128$\times$128 px & 320$\times$256 px \\ \hline
$t_{fd}$ & duration of frame demand clock pulse & 310 ns & 310 ns \\ \hline 
$t_{fc}$ & duration of frame completion clock pulse & 410 ns & 410 ns \\ \hline 
$n_{r}$ & number of rows being read & 128 & 256 \\ \hline 
$t_{rd}$ & time that an output spends on each pixel & 1 $\mu$s & 1 $\mu$s \\ \hline 
$n_{cb}$ & number of 32-pixel column blocks in the image & 4 & 10 \\ \hline 
$t_{rst}$ & time to reset a row & 100 ns & 1 $\mu$s \\ \hline 
$t_{prst}$ & post-reset delay & 400 ns & 400 ns \\
 (global resets)  & &  &  \\ \hline
$t_{prst}$ & post-reset delay & 120 ns & 120 ns \\ 
 (line-by-line resets)  & &  &  \\ \hline
$t_{rc}$ & duration of row completion clock pulse & 210 ns & 210 ns \\ \hline
$t_{ra}$ & duration of row advance clock pulse & 210 ns & 210 ns \\ \hline
\end{tabular}
\quote{A summary of the various times for a full-frame and 128$\times$128 subarray readout of the 
SAPHIRA. These times can be used in Equations~\ref{eqn:timage-rr}-\ref{eqn:framerate} to calculate the expected frame rate. Except for the time to reset the detector, 
the timings are independent of window size. Except for the post-reset dead time, the timings 
are independent of reset mode.}
\label{table:timings}
\end{sidewaystable}

It should be noted that in all readout modes described above, the detector is continuously clocking 
and reading
pixels. This is known as a rolling shutter; different parts of the frame are read at different
times. This can be problematic for observations for which the flux is rapidly changing and
synchronization is important. For example, in present pyramid wavefront sensors deployed on-sky, 
the beam is modulated in a circle around the tip of the pyramid in order to increase the linearity 
of the wavefront sensor's response. This causes each pupil image to be illuminated at a different
time. In SCExAO, the beam completes one full modulation per EMCCD image, and the EMCCD has a fast
frame transfer which functions as a global shutter, so the detector does not notice the pupils 
being illuminated at different times. On the other hand, a detector with a rolling shutter might
see gradients across the pupil images caused by the modulation, interpret it as an optical
aberration, and drive the deformable mirror to a suboptimal shape. The SAPHIRA in the Keck
Planet Imager solves this challenge 
by synchronizing the start of each frame's readout to the modulation so that the part of the frame 
being read does not include the pupil which is changing in illumination. The time to modulate is
slightly greater than the time to read one image, but the dead time between reads is minimized
for the reason described in the next paragraph.

The integration times of an image given by Equations~\ref{eqn:timage-rr}-\ref{eqn:timage-utr} 
depend entirely on the 
durations of the various clock pulses and number of pixels being read. A user could insert
an additional delay in order to increase the detector's integration time. However, unless storage space
is limited, this is suboptimal. Instead of inserting a delay to make each exposure longer, 
the user does better to clock the detector at its maximum rate, increase the avalanche gain
to use the entire well depth with each integration (assuming read-reset or read-reset-read 
mode), and then average images together. This reduces the read noise and therefore improves
the signal to noise ratio. This contrasts with a normal (non-avalanche multiplying) detector,
wherein reading and resetting the array when very little target flux has accumulated and 
then coadding the exposures results in the read noises adding in quadrature and therefore 
a reduced signal-to-noise ratio. (In both normal and avalanche-multiplying detectors, 
non-destructively reading up the ramp at maximum readout rate until the entire well depth
has been used results in optimal signal to noise, assuming that ROIC glow is negligible.
For SAPHIRA, one can trade off the ramp length and avalanche gain; the optimal 
signal-to-noise ratio can be obtained by
selecting the avalanche gain that produces the minimum dark current and then using a
ramp length that fully utilizes the detector's dynamic range.)

\section{Radio Frequency Noise and Cryogenic Preamplifiers}\label{sec:rfi}
Radio frequency interference (RFI) noise is the dominant noise source for SAPHIRA's deployments
to Subaru Telescope with the SCExAO instrument, but it is negligible in other environments. 
In the laboratory, the measured SAPHIRA CDS read noise~\citep[9 $e^-$ RMS,][]{Atkinson2014} at unity 
avalanche gain and a Generation III Leach Controller~\citep{Leach2000} at a 265 kHz clocking rate is very 
comparable to the read noise of conventional HAWAII series arrays~\citep{Hall2016B, Fox2012}.
The read noise was similar in telescope environments such as the NASA IRTF and Kitt Peak. However when 
mounted to the SCExAO instrument at the Subaru telescope, the noise is increased by a factor of 4.5, and 
we have been unable to reduce this through the normal procedures. The SAPHIRA is ground isolated within 
the cryostat with the analog and digital grounds brought out separately to the controller and then to a 
``star'' ground point. The internal radiation shield and outer vacuum vessel are tied separately to this
same star ground to provide concentric Faraday cages. The entire camera system is isolated from the 
SCExAO bench except through a tie from the star ground to the primary SCExAO ground. 
SCExAO is characterized by an extremely challenging RFI environment and many of the other instruments
experience higher levels of noise than they did during laboratory testing elsewhere.

\begin{figure}
\begin{centering}
\includegraphics[width=0.93\textwidth]{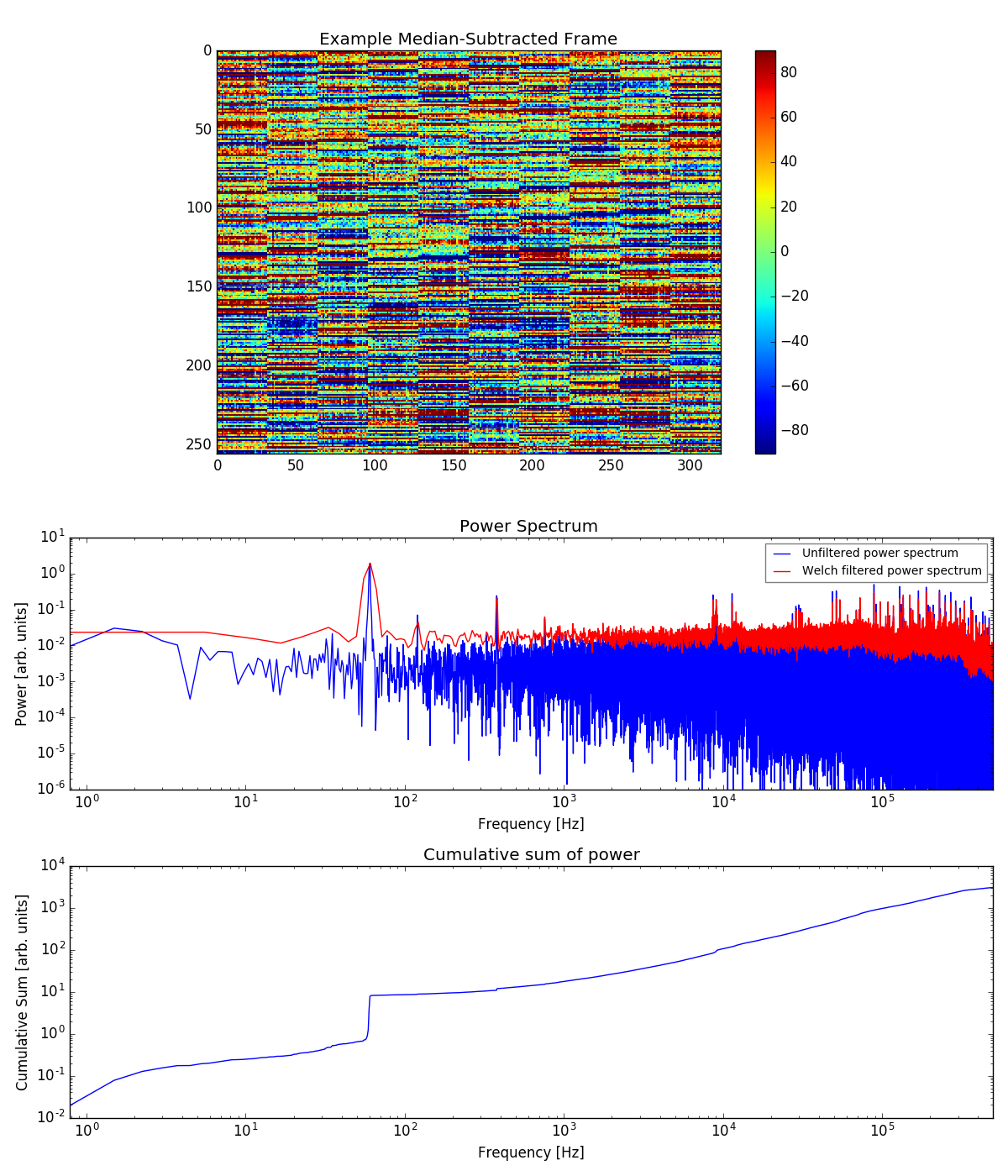}
\caption[Image and power spectrum of radio frequency noise]{Shown at top is a sample single frame (with the pedestal noise subtracted) which exhibits
the radio frequency interference of the SCExAO environment. In the middle is a power spectrum 
calculated from the corresponding full image 
cube. Apart from a major spike at 60 Hz (likely corresponding to ground contamination from
other electronics), the noise is largely white. Plotted over the power spectrum is a 
Welch-filtered version. The Welch filter~\citep{welch67} reduces noise in a power spectrum, but this 
comes at the
expense of spectral resolution. As a result, the lowest frequency bins of the Welch-filtered
data are much broader. At the bottom is the cumulative sum of the unfiltered power spectrum.}
\label{fig:rfi1}
\end{centering}
\end{figure}

RFI shows up as 32-pixel-wide blocks that oscillate in value (Figure~\ref{fig:rfi1}). A 
power spectrum can reveal some of the sources of RFI.
In our current SCExAO deployment, the noise is 
white except for a 60 Hz spike. In previous deployments, it was dominated by a few discrete 
frequencies (and therefore could be mostly filtered out from the data). The RFI changes with time, 
grounding setups, and the arrangement of wires on the outside
of the camera. We found that RFI was decreased by increasing the shielding of cables, using twisted-pair
cables, eliminating ground loops, and using a ``clean'' (i.e. with minimal other electronics connected)
ground point. RFI behaves as read noise in the noise comparisons of Section~\ref{sec:comparison}.

We are presently testing cryogenic preampliers that were developed at Australia National University. These
sit near the detector inside the camera and amplify its electrical signals prior to the wires to the readout
electronics. We anticipate that these will greatly reduce the RFI noise. Additionally, our maximum pixel rate
until now has been limited to about 1 MHz per output due the ROIC output drivers causing settling issues with
our $\sim 1$ m cable lengths. We expect that the preampliers will enable faster readout rates.

\section{Avalanche Gain, Charge Gain, and the Excess Noise Factor}~\label{sec:gains}
%
%
In a conventional source-follower HgCdTe array such as the Teledyne HxRG family~\citep{Beletic2008} 
or the Raytheon Virgo~\citep{Starr2016}, there are two gains are of interest: the voltage gain (which has 
units of $\mu$V/ADU) and the charge gain (which has units of $e^-$/ADU). The charge
gain is related to the voltage gain through $Q = C V$ where $C$ is the integrating node capacitance 
(the sum of the photodiode and ROIC capacitances) and has a typical value of 20 to 40 fF. The
voltage gain is a property only of the ROIC. In the previously mentioned infrared detector arrays, the
bias voltage is low enough that there is minimal change in the photodiode capacitance as the pixel 
accumulates photons, and this effect can be compensated for with a linearity correction.
However, the
situation is somewhat more complicated for SAPHIRA. The user adjusts the bias voltage across the 
detector over a 20 V range in order to set
the avalanche gain. At bias voltages too low to produce any avalanche gain ($\lesssim1$ V), operation
is similar to the conventional source-follower arrays. However, as the bias voltage
of SAPHIRA increases, avalanche multiplications begin to occur and the diode capacitance, and therefore
the charge gain, decreases.
Both the avalanche multiplication and the decreasing charge gain should be taken into account when 
converting between collected photons, electrons, and ADUs for data from SAPHIRA detectors.

During SAPHIRA operation, the input node of each pixel is reset to the selected bias voltage relative to
the detector common
voltage. After the reset is lifted, photoelectrons discharge the integrating node, and the pixel 
saturates when the node voltage equals common. The node voltage is read out by a ROIC with 
a typical gain of 0.8 to 0.9 (for the single source follower design of the ME1000/ME1001)
and is 
further amplified by the controller preamplifier chain before conversion to ADU. The voltage gain of the 
system (in units of $\mu$V/ADU) can be calibrated by varying the reset voltage in known increments and
measuring changes in the corresponding digitized signal.
The charge gain is normally determined through the signal vs.~variance 
method, although~\citet{Finger2005} directly measured a $V$ vs $Q$ relation for an H2RG
detector by connecting it to a calibrated large external capacitor and measuring voltage changes in each
while pixels were exposed and then reset.

\begin{figure}[!ht]
\begin{centering}
\includegraphics[width=0.9\textwidth]{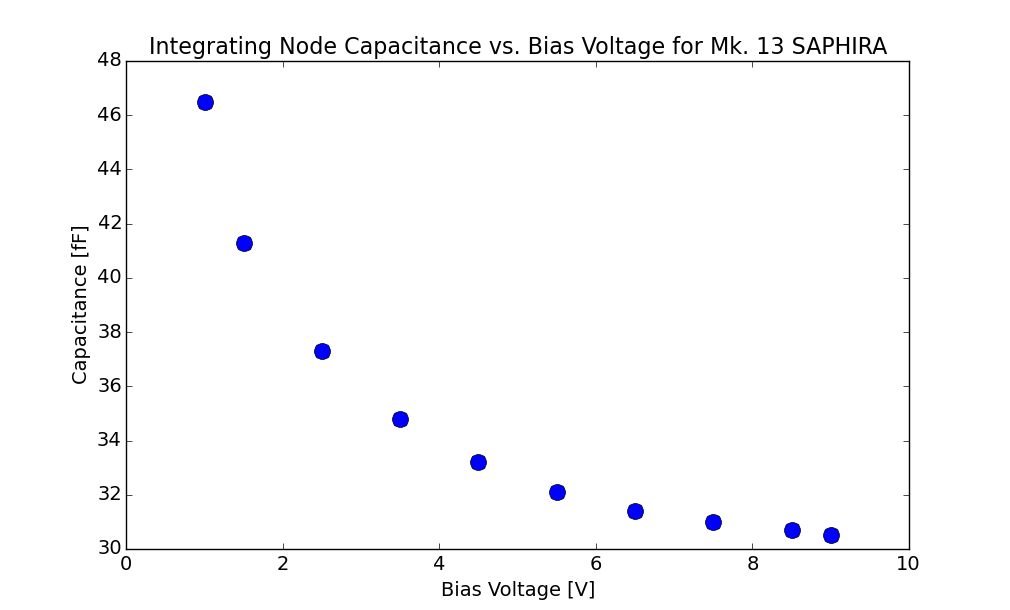}
\caption[SAPHIRA integrating node capacitance plotted against bias voltage]
{The measured integrating node capacitance vs.~bias voltage for a Mk.~13 SAPHIRA detector. The
data are from Dr. Ian Baker (private communication). 19.9 fF of the integrating node capacitance is
due to the silicon readout, and the remainder is the diode. It is important to take into account the 
increasing capacitance at low bias voltages when measuring avalanche gains.}
\label{fig:capacitance}
\end{centering}
\end{figure}

The photodiode capacitance of SAPHIRA increases as the node is discharged, 
and this causes a flux vs.~time plot to depart from linearity as the pixel accumulates flux.
Therefore, for photometry applications, it is important to calibrate the linearity of
the detector's response for each bias voltage used.
Figure~\ref{fig:capacitance} depicts the photodiode capacitance of a Mark 13 SAPHIRA from 1 V (where
the avalanche 
gain is negligible but the capacitance varies significantly with bias voltage) to 9 V 
(where the diode is fully depleted so the capacitance is insensitive to bias, but the avalanche gain is 
$\sim5$). 
Avalanche gain becomes appreciable at a bias voltage of 2-3 V, and by biasing the detector by up to 20 
V, avalanche gains of $\sim 600$ can be obtained. In all cases, the unit cell source
follower still limits dynamic range to a few hundred mV. 
The avalanche gain is usually measured by varying the bias voltage under constant detector illumination 
and defining that the flat portion of the response curve (which occurs at 1-2.5 V bias) corresponds to
an avalanche gain of 1.
From Figure~\ref{fig:capacitance}, it is evident that for avalanche bias voltages $\lesssim 10$ V, the 
change in capacitance with bias must be included in this avalanche gain calculation.

Read noise and RFI noise are independent of avalanche gain, so one can improve the signal to noise
ratio of observations by amplifying the signal of photons with a higher avalanche gain. However, as
avalanche gain increases, the dynamic range (in photons) of the
detector is reduced because the well depth in electrons remains constant, but more electrons are produced
per photon. At sufficiently high bias voltages, tunneling current (a type of dark 
current) increases dramatically~\citep{Atkinson2017}, can dominate over the read 
noise, and thereby can degrade the signal to noise ratio of observations~\citep{Finger2016}.

The excess noise factor $F$ characterizes noise that occurs during the multiplication process of
avalanche photodiodes.
$F$ is the ratio of the signal to noise of the photon induced charge to that of the avalanche-multiplied
charge~\citep{McIntyre1966}; noise-free amplification produces $F=1$.
The advantage of HgCdTe over other semiconductors is that only electrons participate in the avalanche
process, and this has the potential for noise-free amplification. \citet{Finger2016} measured
an excess
noise factor of $F=1$ at 60 K at all gains and a maximum of $F=1.3$ at an avalanche gain of 421 at 90 K.

\section{Conclusions}
For most AO-related observations, we operate SAPHIRA in read-reset mode, wherein each row is read 
before being reset, and the pedestal noise of the detector is removed by subtracting
a reference dark frame. This mode enables the maximum usable frame rate and regular time sampling. 
When a minimum of noise is desired and long periods between detector resets are acceptable, the SAPHIRA
can be operated in up-the-ramp mode, wherein there are many reads between resets. The user then
fits a line to the flux of each pixel over time, which reduces read noise. If a pixel saturates, one
can estimate the correct flux by fitting only to the reads that occurred before the saturation. 
Read-reset-read mode, wherein a row was read, reset, and read again, and this repeats for
the next line, failed to deliver the hoped-for low noise due to a post-reset settling
effect that caused spurious flux in the columns read in the first few microseconds after
resets. For this reason, we do not commonly use read-reset-read mode. 

Because the pixel rate is determined by the readout electronics (and therefore relatively constant) 
and the outputs are designed to read contiguous 32-pixel blocks,
higher frame rates can be achieved by reading $32x \times y$ pixel subarrays, where $1 \leq x \leq 10$,
and $1 \leq y \leq 256$. For example, at a 1 MHz clocking rate in read-reset mode, one can read the 
full 320$\times$256 pixel frame at a rate of about 339 Hz or a 128$\times$128 subarray at about 
1690 Hz. The detector itself is capable of 10 MHz pixel rates, and ESO operates their SAPHIRA
arrays at typically 5 MHz rates.


SAPHIRA detectors enable the potential for new observing modes due to their high frame rate,
low noise, and infrared sensitivity. We hope that this paper can assist users in
utilizing the detectors to their maximum potential. 

\section*{ACKNOWLEDGMENTS}       
The authors acknowledge support from NSF award AST 1106391, NASA Roses APRA award NNX 13AC14G, and the 
JSPS (Grant-in-Aid for Research \#23340051 and \#26220704). Sean Goebel acknowledges funding support
from Subaru Telescope and the Japanese Astrobiology Center.

{\bibliographystyle{apj}}
{\bibliography{ch2bib}}
\chapter{Commissioning of cryogenic preamplifiers for SAPHIRA detectors}\label{chapter:preamps}

Note:~This chapter originally was submitted to the ``High Energy, Optical,
and Infrared Detectors for Astronomy VIII" 2018 SPIE proceedings
with co-authors Donald N.B. Hall, Shane M. Jacobson, Annino Vaccarella, 
Rob Sharp, Michael Ellis, and Izabella Pastrana. It will be published
as~\citet{Goebel2018spie}.

\section*{Abstract}
SAPHIRA detectors, which are HgCdTe linear avalanche photodiode arrays manufactured by
Leonardo, enable high frame rate, high sensitivity, low noise, and low dark current 
imaging at near-infrared wavelengths.
During all University of Hawaii Institute for Astronomy lab testing and observatory
deployments of SAPHIRA detectors, there was approximately one meter of cables between the
arrays and the readout controllers. The output drivers of the detectors struggled to
stably send signals over this length to the readout controllers. As a result, voltage
oscillations caused excess noise that prevented us from clocking much faster than 1 MHz.
Additionally, during some deployments, such as at the SCExAO instrument at Subaru 
Telescope, radio-frequency interference from the telescope environment produced noise 
many times greater than what we experienced in the lab.
In order to address these problems, collaborators at the Australia National University 
developed a cryogenic preamplifier system that holds the detector and buffers the signals
from its outputs. During lab testing at 1 MHz clocking speeds,
the preamplifiers reduced the read noise by 45\% relative to data collected using the
previous JK Henriksen detector mount. Additionally, the preamplifiers enabled us to increase
the clocking frequency to 2 MHz, effectively doubling the frame rate to 760 Hz for a full
(320$\times$256 pixel) frame or 3.3 kHz for a 128$\times$128 pixel subarray. 
Finally, the preamplifiers reduced the noise observed
in the SCExAO environment by 65\% (to essentially the same value observed in the lab)
and eliminated the 32-pixel raised bars characteristic
of radio-frequency interference that we previous observed there. 

\section{Introduction}\label{sec:intro} 
The SAPHIRA detector is a 320$\times$256@24 $\mu$m pixel mercury cadmium telluride
(HgCdTe) array produced by Leonardo (formerly Selex) with $0.8 - 2.5 \; \mu$m 
sensitivity\citep{Baker2016}. 
SAPHIRA is a linear avalanche photodiode (L-APD),
meaning that it has a multiplication region over which the avalanche gain multiplies
the photon signal. At $\sim$2 V bias,
SAPHIRA operates as a traditional HgCdTe array with no avalanche multiplication.
The bias voltage can be increased to $\sim$20 V, where the photon signal is
multiplied by a factor of several hundred~\citep{Atkinson2016}. The read noise is unaffected
by the avalanche gain, so at moderate to high bias voltages, the gain-corrected
read noise is sub-electron~\citep{Finger2016}. Equivalently, the signal of an individual
photon exceeds the read noise; this is called photon counting~\citep{Atkinson2018}.
Unlike Geiger-mode APDs~\citep{Renker2006}, SAPHIRA does not need to be reset after each 
photon arrival. SAPHIRA has 32 outputs and is optimized for high frame rates. The outputs are
interleaved (the 32 outputs read 32 adjacent columns), so all are still used when 
the detector is operated in subwindow mode.
Therefore, the frame rate increases approximately as the inverse of the number
of pixels read~\citep{Goebel2018jatis}. 

The low noise, high sensitivity, and high frame rate makes
SAPHIRA well-suited for wavefront sensing for adaptive optics (AO). Presently
most wavefront sensing is done at optical wavelengths with CCDs, and the science
instruments behind AO systems most commonly observe at near-infrared (NIR)
wavelengths. 
Wavefront sensing at NIR wavelengths is of particular interest because it
would reduce chromatic aberrations
caused by wavelength differences between the science instrument and wavefront
sensor, and it would enable observations of targets with low optical emission
but high NIR emission such as M dwarfs~\citep{Wizinowich2016}. 
SAPHIRA also 
features a very low dark current (\citet{Atkinson2017} measured an upper limit
of the dark current at 62.5 K of 0.0014 e$^{-}$ s$^{-1}$ pix$^{-1}$), which 
offers promise for low-background imaging, but this application is presently limited by
the relatively small format of the detector. However, a 1K$\times$1K@15 $\mu$m pixel
SAPHIRA is currently in development~\citep{Hall2016}.

Much of this paper
focuses on the SAPHIRA deployment to the Subaru Coronagraphic Extreme Adaptive Optics 
(SCExAO) instrument at Subaru Telescope. SCExAO, as the name implies, provides
extreme adaptive optics corrections using a 2000-element deformable mirror
(typically operated at 2 kHz) and is optimized for coronagraphic observations
of extrasolar planets and debris disks~\citep{Jovanovic2015}. SCExAO constantly undergoes changes
and serves as a testbed for new technologies, such as SAPHIRA, the MKIDs
energy-resolving superconducting detector~\citep{Szypryt2017}, phase-induced amplitude 
apodization (PIAA) coronagraphs~\citep{Martinache2012}, and a photonic spectrograph
fed by a single-mode fiber~\citep{Jovanovic2017}. At SCExAO, SAPHIRA 
has been used for speckle lifetime measurements~\citep{Goebel2016} and as a
demonstrator for near-infrared pyramid wavefront sensing. 

SAPHIRA detectors have been deployed to a number of other facilities. In part because 
of the capabilities SAPHIRA demonstrated at SCExAO, it is in the 
process of being deployed in the Keck Planet Imager~\citep{Mawet2016}, where it will
be used behind pyramid optics for wavefront sensing. 
SAPHIRA arrays have also been used for
lucky imaging at the NASA Infrared Telescope Facility~\citep{Atkinson2014} and tip/tilt tracking
at the Robo-AO instrument at Kitt Peak~\citep{Salama2016,Baranec2015,Jensen-Clem2018}, where
it produced its first science publication~\citep{Han2017}. Several SAPHIRA detectors are 
deployed at the Very Large Telescope's GRAVITY instrument for wavefront
sensing and fringe tracking~\citep{Finger2016}. Additionally, First Light Imaging has marketed
a commercial camera with readout electronics based around the SAPHIRA detector
called the C-RED One~\citep{Greffe2016}.

\section{SAPHIRA without Preamplifiers}
In all SAPHIRA deployments except at the GRAVITY instrument, the signals from the
detector's read out integrated circuit (ROIC) were not amplified prior to being 
digitized by the readout electronics.
Institute for Astronomy SAPHIRA arrays were operated using Gen III Leach 
controllers~\citep{Leach2000} clocking
at 265 kHz or Pizza Box controllers developed there which clocked at 1 MHz. 
For these setups, there was 
about 1 m of cabling between the camera and the electronics. A 1 MHz clocking frequency
enabled a full-frame (320$\times$256 pixel) readout rate of 380 frames per second or 
a 128$\times$128 pixel subarray read rate of 1.68 kHz~\citep{Goebel2018jatis}. 

The detector itself is designed for up to 10 MHz clocking, but we encountered problems 
when we attempted to clock detector significantly faster than 1 MHz.
If we attempted to do this, ``ringing in the cables''
became a problem.\footnote{The C-RED One is able to clock faster than 1 MHz without
a preamplifier because the cables between the detector and readout electronics are 
shorter and lower impedance than ours.} Essentially, the cables were high impedance and reflections
occurred at the ends, causing the voltage seen by the readout electronics to oscillate. 
This effect manifested itself in images as high noise, and the noise
became catastrophically limiting if we tried to clock the detector much faster than 1 MHz.

\begin{figure}
   \begin{center}
   \includegraphics[width=0.85\textwidth]{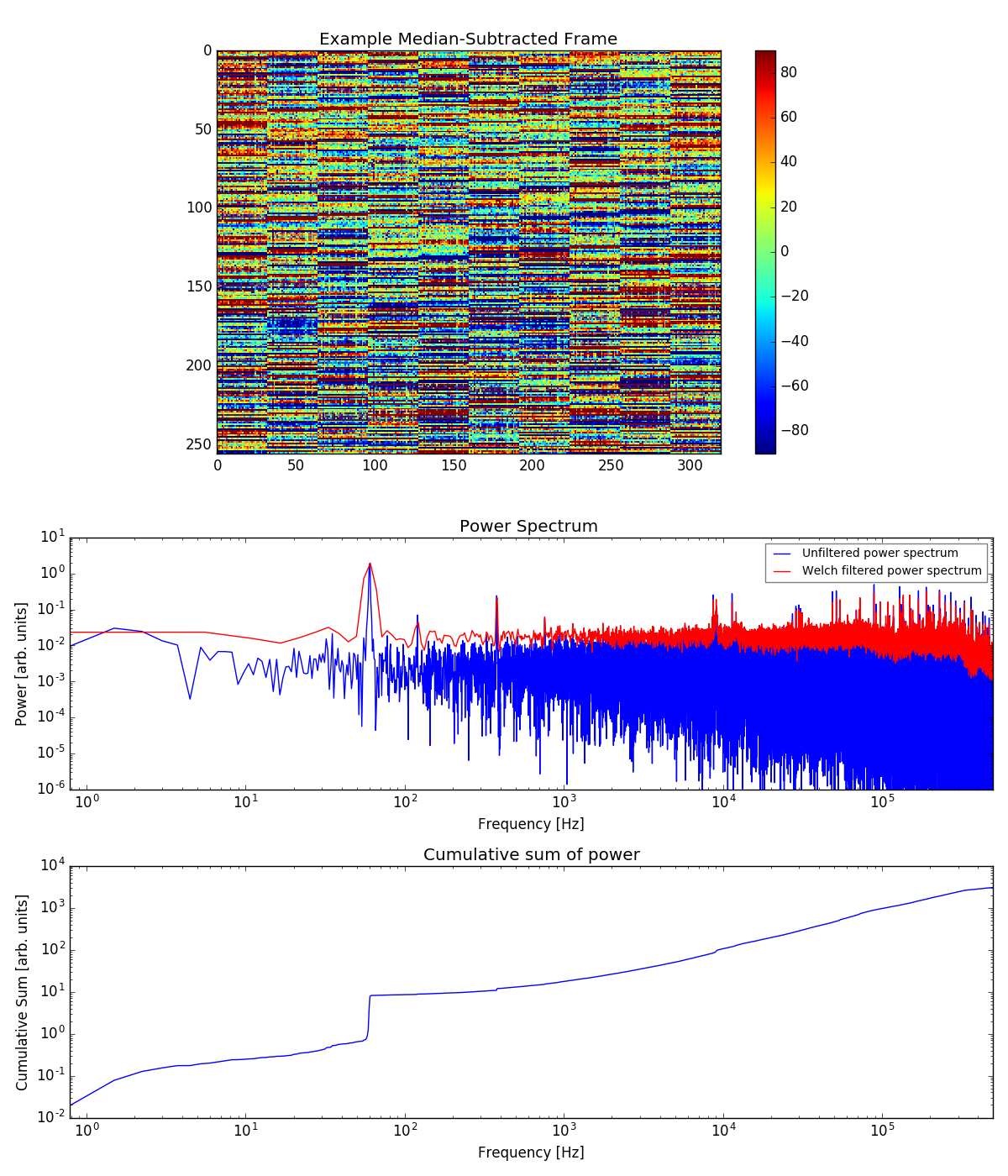}
   \end{center}
   \caption[Image and power spectrum of radio frequency noise.] 
   {The top panel shows an unilluminated SAPHIRA image which exhibits radio frequency noise.
   This was collected at SCExAO using a Pizza Box controller. Many images were averaged
   together to produce a map of the detector's fixed-pattern noise (pedestal voltage) and then
   subtracted from this frame in order to remove the structure that otherwise would dominate.
   The second panel shows the power spectrum of the noise produced by carefully taking into account
   how the readout temporally sampled the array in order to string together many
   images and sample over large temporal baselines. The red line is the power spectrum after Welch
   filtering~\citep{welch67}; this reduces noise but also decreases the temporal resolution.
   Finally, the bottom panel shows the cumulative sum of the power spectrum; it is dominated
   by a single spike at 60 Hz.
   This figure reproduced from~\citet{Goebel2018jatis}.
\label{fig:rfi2}}
\end{figure}

Second, we encountered extremely high noise  when we deployed
the SAPHIRA camera to observatories. This was the worst at the SCExAO instrument at
Subaru Telescope, where we encountered read noises in excess of 300 e$^-$.
For comparison, with the same setup, we measured to 9 e$^-$ in the lab in Hilo~\citep{Atkinson2014}.
At SCExAO and other observatories, rapidly changing voltage offsets were introduced
during the digitization of pixel
values. Because 32 adjacent pixels are read at the same time, this appears as
horizontal 32-pixel bars in images and is illustrated in the top panel of
Figure~\ref{fig:rfi2}. 

The SAPHIRA camera is electrically isolated from the SCExAO bench. The camera
is supposed to act as a Faraday shield for the detector, though it is imperfect
because the detector is located close to the entrance window and thus not surrounded
by metal on all sides.
Early SCExAO deployments used a Gen III Leach controller, which had seven
cards plugged into a black plane board and therefore was susceptible to excess noise
caused by imperfect grounding. We were able to reduce but not eliminate the noise by 
eliminating ground loops, covering the open side of the Leach controller in foil, and 
wrapping foil around the unshielded cables from it to the camera. However, even after 
this, the noise was still a factor of $\sim 10$ higher than lab values. Much of the 
noise power appeared at discrete frequencies, so we were able to partially mitigate 
it after observations by digitally Fourier filtering the images.

\begin{figure}
   \begin{center}
   \includegraphics[width=0.65\textwidth]{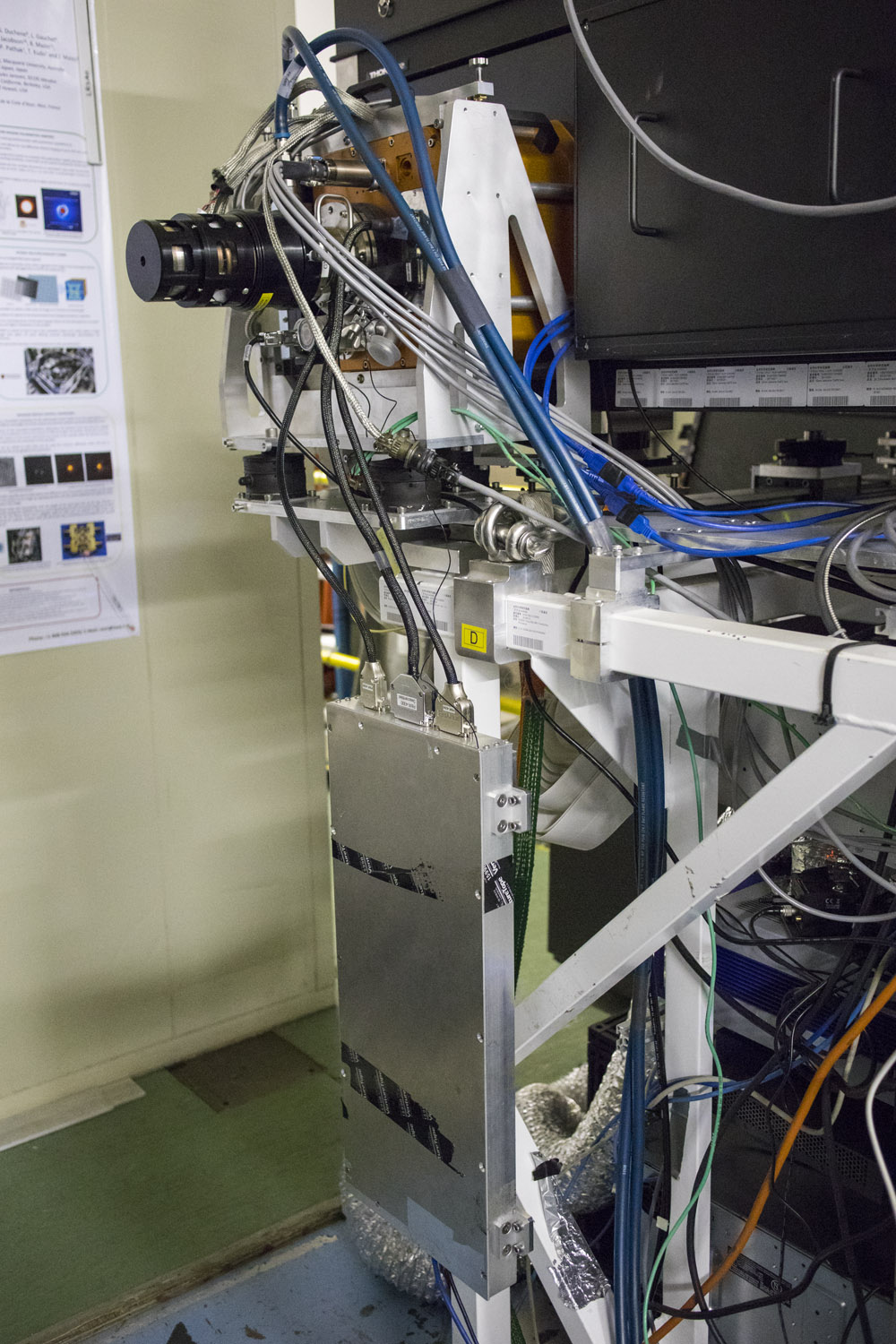}
   \end{center}
   \caption[Photograph of SAPHIRA camera deployed to SCExAO]
   {Shown here is the SAPHIRA camera (orange) at the SCExAO instrument. The
Pizza Box readout electronics is the rectangular box below it. The camera
and Pizza Box are electrically isolated from the frame.
\label{fig:cameraatscexao} }
\end{figure}

We further reduced the noise when we switched to a Pizza Box controller (which
has a single ground plane), began using cables with a grounded shield around them,
and grounded the camera and readout to Subaru's ``clean'' instrument-only ground
instead of the standard building ground. However, the noise was at best still a
factor of $\sim 3$ higher than that of the lab. Figure~\ref{fig:cameraatscexao} 
shows the typical setup of the SAPHIRA system on the SCExAO instrument. By taking 
into account the amount of time the readout spent on each pixel, ends of rows, and 
ends of frames, we
stitched together frames and produced power spectra of the noise with frequency
coverage from 1 Hz to 500 kHz (Figure~\ref{fig:rfi2} middle and bottom panels).
After performing the grounding and shielding steps described above, the noise was
relatively white except for a spike at 60 Hz.

SAPHIRA is not the only instrument at Subaru to experience excess noise; the CHARIS
integral field spectrograph and HiCIAO imager also had this problem, and others likely
did as well. Subaru has a large number of instruments in relatively tight proximity
to each other, there are high magnetic fields in the facility, and the ground beneath
the building is volcanic and inherently a poor conductor. Additionally, when we
tested SAPHIRA at Keck Observatory, the noise was double that of Hilo.

\section{The ANU Preamplifier}
In order to reduce the radio frequency noise that appeared during telescope
deployments and avoid the ringing in the cables that limited our maximum clocking
frequency, we concluded that we needed to
amplify the detector's signal prior to the cables to the readout electronics.
To achieve this, we collaborated with Australia National University (ANU), which 
developed a cryogenic preamplifier for SAPHIRA. We tested it at IFA Hilo at cryogenic 
temperatures and then deployed it to SCExAO at Subaru Telescope. The ANU preamplifier
replaced the JK Henriksen detector mount that we previously used.

\begin{figure} [ht]
   \begin{center}
   \includegraphics[width=0.8\textwidth]{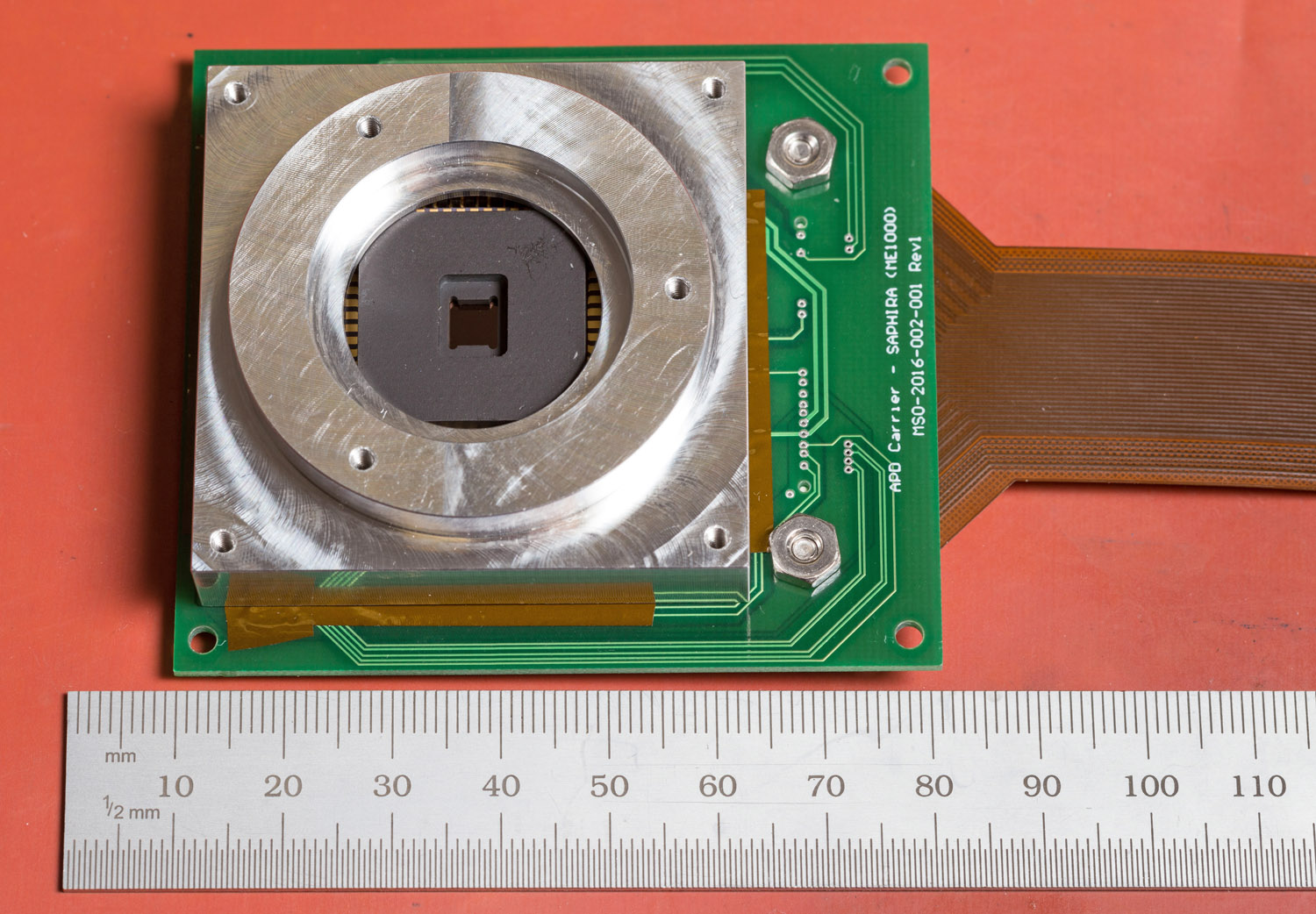}
   \end{center}
   \caption[Photograph of the detector carrier board of the ANU preamplifier]
   {Pictured here is the APD carrier board. A ruler provides scale.
   A shaped mask covers the edges of the detector in order to provide image structure
   during illuminated tests and reduce ROIC glow during dark tests. The flex cable 
   exiting the right side of the image connects to the
   similarly-sized board that holds the actual preamplifying circuitry. 
   \label{fig:preamp}}
\end{figure}

The ANU preamplifier consists of two boards connected by a flex cable: the APD carrier,
which holds the detector and replaces the JK Henriksen mount, and a second board
which actually carries the preamplifiers. The APD carrier board is shown in 
Figure~\ref{fig:preamp} and measures approximately $63 \times 71$ mm.
The preamplifier board measures approximately $64 \times 70$ mm. These two
boards are collectively referred to as the ANU preamplifier throughout this paper.
The preamplier requires a $3-5.5$ V power supply and draws $\sim$0.14 A at 4.00 V.
The boards use Hirose connectors, but in future revisions, these may be 
replaced with AirBorn connectors for improved connective reliability. 
We have tested it successfully at temperatures as low as 58 K, but we have not made
a concerted effort to find its minimum operating temperature.

The ANU preamplifier passes through the clocking pulses to the detector 
without affecting them. It provides filtering to the bias supplies on the underside of
the carrier board as close as possible to the detector (except for Bias\_ILIMIT, 
IMPIX1\_OR, and IMPIX2\_OR, which are disconnected). The filtering is placed on the
underside of the board to mitigate the risk of glow from warm components being seen
by the detector. The preamplifier buffers the output signals from the ROIC to clean them and 
enable longer cable runs to the readout circuitry. Previously, the output amplifiers of the
ROIC struggled to drive signals over our meter-long cables, and the preamplifier eases 
that. The ANU preamplifier 
was designed to have a gain of unity so that users would not need to adjust the dynamic 
range of the readout's analog-to-digital converters.

\section{Lab Testing}\label{sec:lab}
\begin{figure} [ht]
   \begin{center}
   \includegraphics[width=\textwidth]{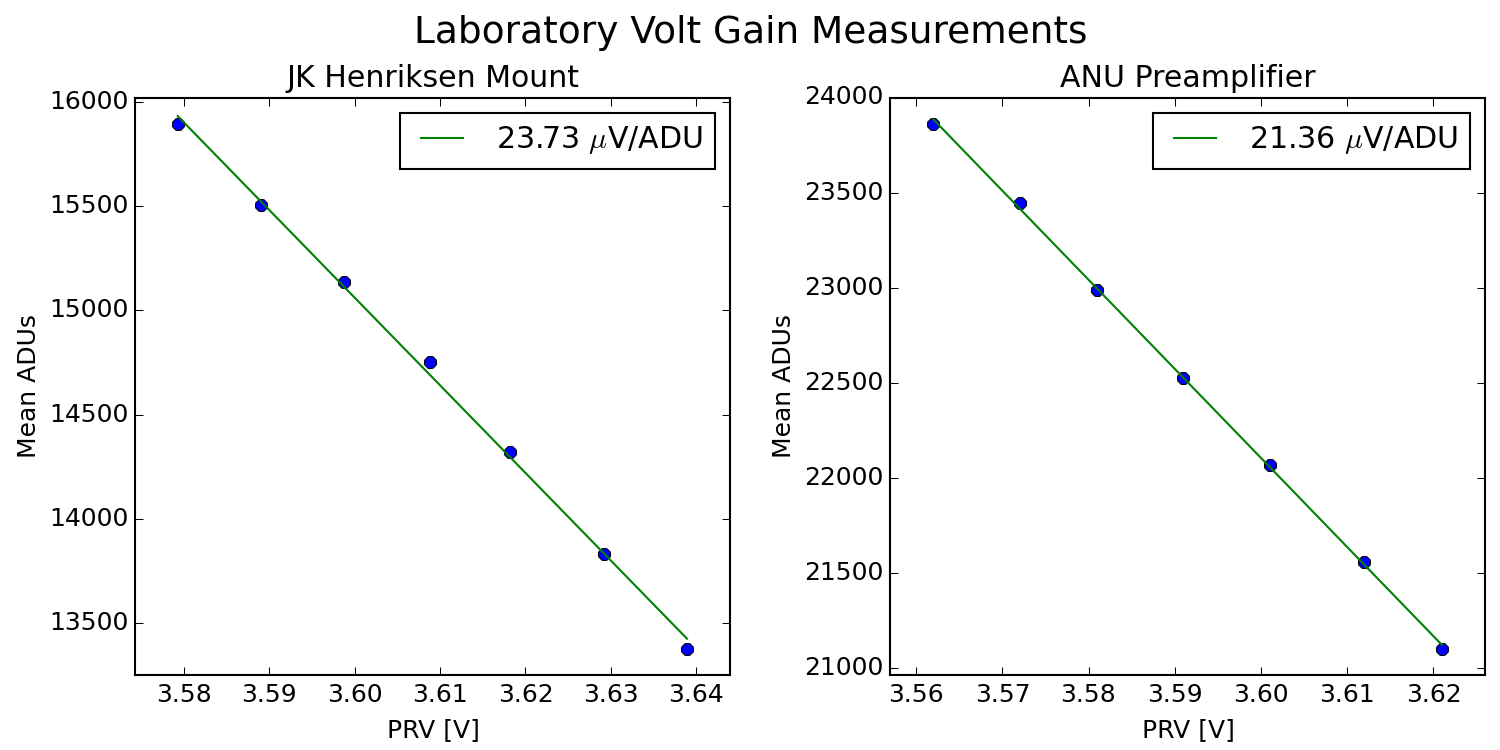}
   \end{center}
   \caption[Laboratory volt gain measurements before and after installation of the ANU preamps]
   {Shown here are our volt gain measurements for the lab characterization
   of the ANU preamplifier. We varied the pixel reset voltage (PRV) in 10 mV increments
   and measured the average ADU value of the unilluminated array. Several thousand frames
   were averaged for each data point.
   \label{fig:vglab}}
\end{figure}

After performing preliminary functionality tests with the cryogenic preamplifier in the 
lab at IFA Hilo, we looked at its impact on read noise. We collected frames using a Rev.~4
Pizza Box controller in low gain mode and clocking at 1 MHz, and an unilluminated
Mk.~15 SAPHIRA array on a ME1001 ROIC at 60 K. The detector had a 1.5 V bias (i.e.~no
avalanche gain) and was operated in sample up the ramp readout mode. We subtracted adjacent
frames (i.e.~$F_{N+1} - F_{N}, F_{N+3}-F_{N+2}, \cdots$, where 
$F_N$ is the $N$th frame in the ramp) in order to remove the pedestal noise. We then
dropped the first 70 frames after each detector reset and computed the standard deviation
of the remaining correlated double sample (CDS) pairs with no masking of bad pixels. 
We observed 4.593 analog-to-digital units (ADU) of noise when using the ANU preamplifier
and 7.459 ADU when using the traditional JK Henriksen detector mount. To convert these
numbers to physical units and 
compensate for any gain in the preamplifier, we measured the volt gain in order to find
the conversion from ADU to volts. We varied the pixel reset voltage (PRV) in 10 mV
increments and measured the average ADU (DC offset) of the unilluminated detector. 
The result is plotted in Figure~\ref{fig:vglab}. We measured a gain of 23.73 $\mu$V/ADU using 
the JK Henriksen mount and 21.36 $\mu$V/ADU using the preamplifier. Therefore the 
cryogenic preamplifier attenuated the signal from the detector by a factor of 
$21.36 / 23.73 = 0.900$ at a temperature of 60 K. Putting all this together,
the noise was 177 $\mu$V with the JK Henriksen mount and 98.1 $\mu$V with the 
preamplifier. In other words, in this experiment the preamplifier reduced the read noise 
by 45\%. We did not have a charge gain measurement for this detector, and it has a 
significantly different architecture than detectors that we have measured charge gains 
for, so we cannot accurately convert this read noise to units of electrons.

\begin{figure} [ht]
   \begin{center}
   \includegraphics[width=0.9\textwidth]{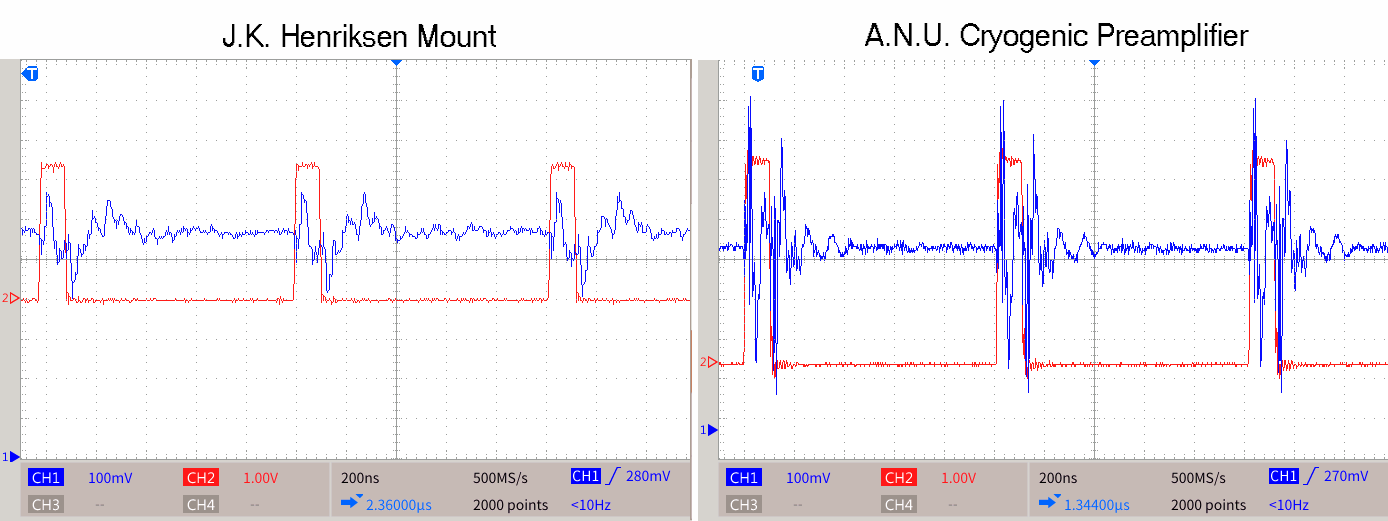}
   \end{center}
   \caption[Oscilloscope plot of pixel settling during clocking]
   {Plotted above are the pixel clock pulse (red) and the pixel output voltage (blue) with
   and without the cryogenic preamplifier. In both setups, the clocking frequency is 1 MHz 
   (i.e.~the interval between clock pulses is 1 $\mu$s). The voltage of a pixel oscillates after the
   readout clocks onto it, and we insert a delay between clocking onto a pixel and sampling it
   in order to permit its voltage to settle. A faster settling
   enables a higher clocking speed at a given noise level. The preamplified detector has a higher
   initial amplitude of oscillations, but these oscillations decay faster, and the voltage is more
   stable after $\sim$300 ns. Additionally, in the near future we will add additional filtering in
   order to reduce the voltage oscillation further. The vertical and horizontal scalings of both 
   plots are the same;
   the clocking voltage has a smaller amplitude in the left plot because it was sampled at
   a different location on the Pizza Box. 
   \label{fig:pixelsettling}}
\end{figure}

Second, we looked at whether the cryogenic preamplifiers enabled higher clocking frequencies
and thereby faster frame rates. The Pizza Box was originally intended for 2 MHz clocking
speeds, but we encountered very high levels of noise at this speed, so in all SAPHIRA Pizza Box 
deployments, the detector was clocked at 1 MHz. At 2 MHz sampling frequencies, the voltage from 
each pixel had inadequate time to settle, and this manifested itself in images as high read noise.
Figure~\ref{fig:pixelsettling} shows an oscilloscope trace of the voltage from a pixel with and
without the preamplifiers. The preamplifiers cause a high voltage oscillation when the readout
initially clocks onto a pixel, but this decays in $\sim$300 ns, and the voltage of the pixel is
more stable than it was at that same point in time without the preamplifier. This is why the
read noise was reduced in the 1 MHz clocking speeds described above. We successfully clocked
the detector at 2 MHz with the preamplifier; however, we encountered computer input/output
limitations due to the increased data rate and therefore could not reliably make a noise 
measurement. Once the necessary fixes are implemented, we anticipate operation at 2 MHz during
future AO deployments.

\section{SCExAO Deployment}\label{sec:scexao}
\begin{figure} [ht]
   \begin{center}
   \includegraphics[width=\textwidth]{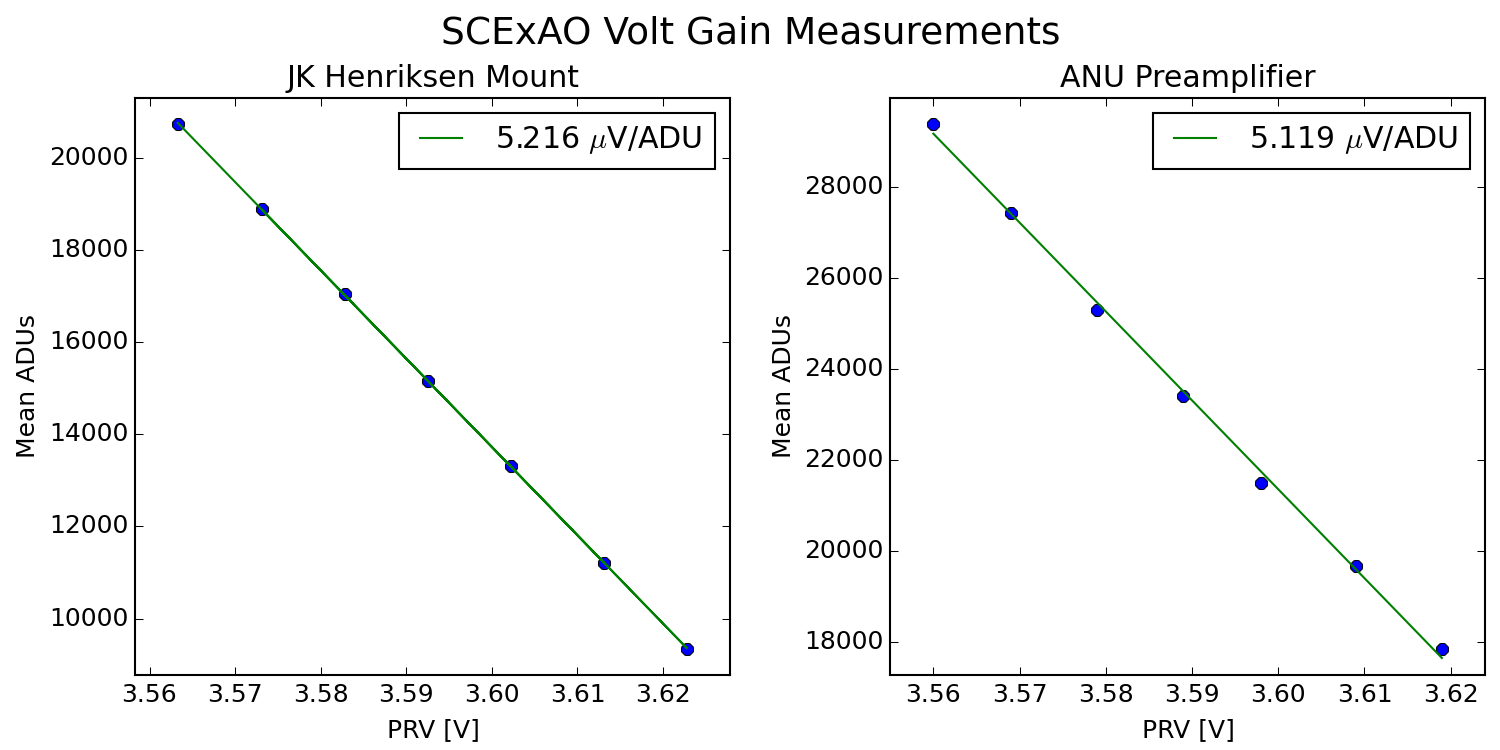}
   \end{center}
   \caption[SCExAO volt gain measurements before and after installation of the ANU preamps]
   {Shown here are our volt gain measurements for the SCExAO deployment
   of the ANU preamplifier. We varied the pixel reset voltage (PRV) in 10 mV increments
   and measured the average ADU value of the unilluminated array. Several hundred frames
   were averaged for each data point. The gains are a factor of $\sim$4 different from that
   of Figure~\ref{fig:vglab} because the Pizza Box operated in low gain mode in the lab
   and high gain mode for this test.
   \label{fig:vgscexao}}
\end{figure}

Next, we modified our GL Scientific Sterling-cooled cryostat to accommodate the cryogenic
preamplifier and deployed the camera system to SCExAO. Because of the high noise we 
experienced in that environment, we were particularly interested in the preamplifier's 
performance there. This system used a Mk.~14 SAPHIRA detector on a ME1000 ROIC. Following
the procedure described in Section~\ref{sec:lab}, in order to convert from ADUs to volts, we 
performed a volt gain measurement before and after the installation of the preamplifier; this
is shown in Figure~\ref{fig:vgscexao}.
In this case, the ratio of the volt gain with the JK Henriksen mount to the gain with the 
preamplifier is 0.98. This contrasts with the 0.90 ratio observed in our laboratory testing. 
This suggests that the preamplifier has a temperature dependent gain, since the laboratory 
testing was at 65 K and the SCExAO testing was at 85 K.

To measure the noise, we operated the detector in read-reset mode at 1 MHz
clocking speed and collected large cubes of frames in order to allow the detector to settle.
For a discussion of various readout modes and the noise sources of each, 
see~\citet{Goebel2018jatis}. Once the detector was stable, we selected 
500 subsequent frames and then subtracted the time average of these from each frame in order
to remove its pedestal voltage. We then computed the standard deviation of each frame and
took the median of these standard deviations. Using this method, we saw a noise of 126 ADU
with the JK Henriksen mount and 45 ADU with the preamplifiers. Taking into account the
volt gains derived in Figure~\ref{fig:vgscexao}, this is 657 $\mu$V and 230 $\mu$V, respectively,
corresponding to a 65\% reduction in noise! These numbers are not directly comparable
to those reported in Section~\ref{sec:lab} because in that case the detector was operated in
sample up the ramp readout mode. However, data collected in the lab with the
same readout mode, Pizza Box settings, and reduction method as reported in this section
had 211 $\mu$V standard deviation of noise. The difference between the two can be accounted for
by the reduction in kTC noise due to the lower operating temperature in Hilo. 
Pastrana et al.~(in preparation) measured
an avalanche-corrected charge gain for this laboratory Mk.~13 SAPHIRA at the same (2.8 V) bias 
voltage of 1.2 e$^-$/ADU. If we assume that the Mk.~14 detector deployed to SCExAO has the 
same charge gain (a reasonable but not exact assumption), this corresponds to a
noise standard deviation of 55 e$^-$. However, because the detector was operated in
read-reset mode instead of correlated double sample mode, this noise number includes 
both kTC noise (which is most of it) 
and also the read noise (a comparatively small fraction). As stated before, the
avalanche gain multiplies the photon signal but has no impact on the read and kTC noise, so an
avalanche gain above 55 would enable the equivalent of sub-electron noise.

\begin{figure} [ht]
   \begin{center}
   \includegraphics[width=\textwidth]{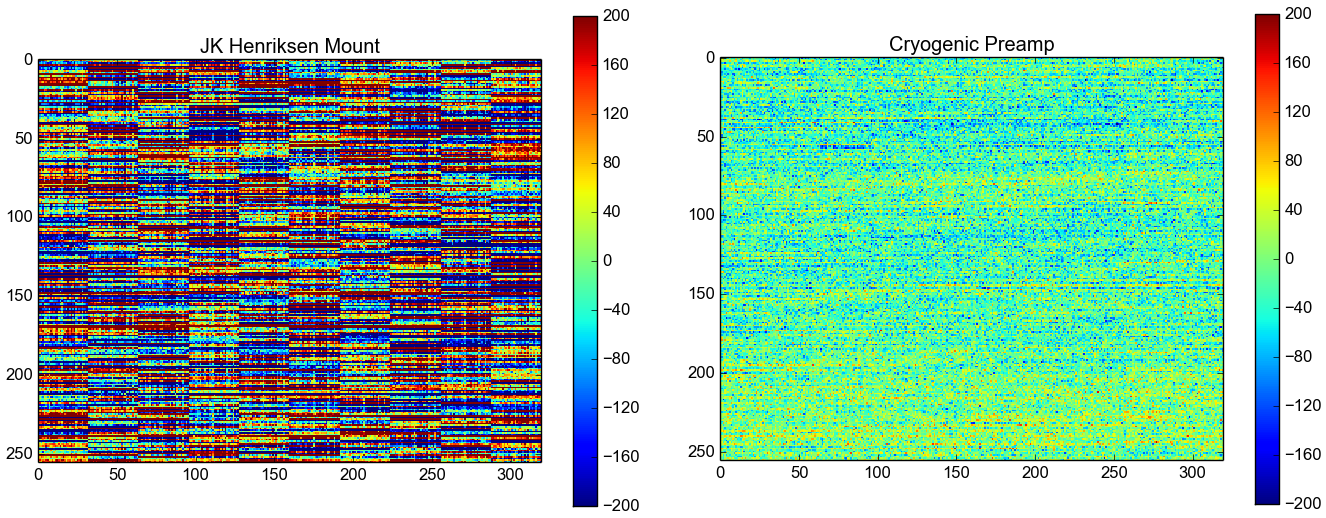}
   \end{center}
   \caption[Images showing the RFI noise before installation of the ANU preamplifiers and
   lack thereof afterward]
   {Illustrated are single images collected at SCExAO using SAPHIRA in read-reset
   mode with the JK Henriksen detector mount (left) and the ANU preamplifier (right).
   The images have identical processing and scaling in units of ADUs. In both cases, a
   mean frame was subtracted in order to remove the pedestal voltage pattern from the detector. 
   \label{fig:beforeafter}}
\end{figure}

The noise in the images at SCExAO with the preamplifiers appeared white and did not exhibit
the 32-pixel-wide raised bars characteristic of
the previous radio frequency interference (Figure~\ref{fig:rfi2}). Additionally, we tested
various grounding setups at SCExAO and wrapped the flex cable from the camera to the 
Pizza Box in foil (it was previously not shielded), and there was very little change in 
this noise. This contrasts sharply with the previous situation at SCExAO, where every
minor change in grounding made a large impact on the noise. A comparison of image quality
before and after installation of the preamplifier is shown in Figure~\ref{fig:beforeafter}.

\section{Conclusions}
We tested cryogenic preamplifiers that were developed at ANU for SAPHIRA detectors.
In lab testing, they reduced read noise at 1 MHz clocking by 45\% and enabled us to 
increase the clocking
frequency of SAPHIRA from 1 to 2 MHz. A 2 MHz frequency enables a full frame 
(320$\times$256 pixels) to be read at a rate of approximately 760 Hz or a 
128$\times$128 pixel subarray to be read at 3.3 kHz. When deployed to SCExAO, where
we struggled for years to minimize excess noise caused by radio-frequency 
interference, the cryogenic preamplifiers reduced noise by 65\%, thereby enabling
essentially the same noise performance we measured in the lab. This improvement
greatly enhances the sensitivity of SAPHIRA arrays and enables higher frame rate 
and/or higher signal-to-noise observations during future deployments.


\section*{ACKNOWLEDGMENTS}       
The authors acknowledge support from NSF award AST 1106391, NASA Roses APRA award NNX 13AC14G, and the 
JSPS (Grant-in-Aid for Research \#23340051 and \#26220704). Sean Goebel acknowledges funding support
from Subaru Telescope and the Japanese Astrobiology Center. 

\bibliographystyle{apj}
\bibliography{ch3bib}

\chapter{Measurements of Speckle Lifetimes in Near-Infrared Extreme Adaptive Optics Images for Optimizing Focal Plane Wavefront Control}\label{chapter:specklelives}

Note: this chapter has been submitted to PASP and is currently under review. Co-authors include Olivier Guyon, Donald N.B. Hall, Nemanja Jovanovic, Julien Lozi, and Frantz Martinache. A formal reference and DOI for this paper is not yet available.

\section*{Abstract}
Although extreme adaptive optics (ExAO) systems can greatly reduce the effects of atmospheric 
turbulence and deliver diffraction-limited images, our ability to observe faint objects such
as extrasolar planets or debris disks at small angular separations is greatly limited by the 
presence of a speckle halo caused by imperfect wavefront corrections. These speckles
change with a variety of timescales, from milliseconds to many hours, and various techniques
have been developed to mitigate them during observations and during
data reduction. Detection limits improve with increased speckle reduction, so an
understanding of how speckles evolve (particularly at near-infrared wavelengths, which is
where most adaptive optics science instruments operate) is of distinct interest.
We used a SAPHIRA detector behind Subaru Telescope's SCExAO instrument to
collect $H$-band images of the ExAO-corrected PSF at a frame rate of 1.68 kHz. We
analyzed these images using two techniques to measure the timescales over which the
speckles evolved. In the first technique, we analyzed the images in a manner applicable to 
predicting performance of real-time speckle nulling loops. We repeated this analysis
using data from several nights to account for varying weather and AO conditions. 
In our second analysis, which follows the techniques employed by Milli et al.~(2016) 
but using data with three orders of magnitude better temporal resolution, we 
identified a new regime of speckle behavior that occurs at timescales of milliseconds and
is clearly due to atmospheric (not instrumental) effects. We also observed an exponential
decay in the Pearson's correlation coefficients (which we employed to quantify the change
in speckles) on timescales of seconds and a linear decay on timescales of minutes, which
is in agreement with the behavior observed by Milli et al.
For both of our analyses, we also collected similar datasets using SCExAO's internal light 
source to separate atmospheric effects from instrumental effects. 

\section{Introduction}
Extreme adaptive optics (ExAO) images consist of a point spread function (PSF)
and a surrounding halo of speckles. The PSF's size is determined by the
diffraction limit of the telescope and the wavelength of observations, 
and the speckle halo has an angular size
corresponding to that of the natural atmospheric seeing. The speckle halo
is composed of speckles with intensities up to a few $10^{-3}$ times the intensity
of the PSF core. Speckle causes include diffraction within the optical train
and imperfect AO corrections due to time lag between wavefront sensing and correction,
non-common path aberrations introduced between the wavefront sensor and science camera,
and imperfect wavefront measurements due to noise. These speckles evolve on a variety 
of timescales, from hours (diffraction from the telescope spiders, for example) to 
millisecond (the deformable mirror's imperfect corrections for atmospheric turbulence)~\citep{Macintosh2005}.

These speckles reduce the ability to detect faint features near a star such as extrasolar
planets or debris disks, and their effect is orders of magnitude higher than photon
noise. \citet{Racine1999} showed that the limiting brightness for detection is
proportional to $(1-S)/S$ where $S$ is the Strehl ratio, which emphasizes the need for
ExAO instruments producing high-Strehl corrections. The Strehl ratio is the comparison
of the peak
intensity of the measured PSF to the intensity of a PSF diffracted by the telescope
pupil but otherwise unaberrated~\citep{Strehl1902}.
The Strehl ratio will never be 100\% for ground-based astronomical 
observations, so there will always be speckles, and
several techniques have been developed to differentiate between
them and actual structure in the science target. These include angular differential 
imaging~\citep{marois06} (which takes advantage of the rotation of the sky relative to an 
alt/az telescope)
and spectral differential imaging~\citep{smith87}
(which utilizes the fact that the speckle halo expands with increasing
wavelengths, but features in the target do not). However, because these techniques are
implemented after the data have been collected, they do not remove the shot noise caused by 
the speckles. Additionally, because they are typically applied to long exposures, they have
no effect on quickly-changing speckles. These techniques have been applied to obtain
contrasts on the order of $10^{-6}$ at several tenths of an arcsecond from the PSF 
core.
However, Earth-like planets in the habitable zones of M dwarfs have contrasts
on the order of $10^{-7}$ to $10^{-8}$~\citep{Guyon2012}, and an Earth-like planet around a Sun-like star 
has a contrast of $10^{-10}$, so 
better speckle reduction strategies need to be implemented if such planets are to be detected.

Various techniques have been proposed for real-time speckle nulling 
loops~\citep[e.g.][]{Borde2006,Giveon2007,Sauvage2012,Guyon2004,Baudoz2006,Serabyn2011}. 
These provide two advantages over post facto speckle mitigation
techniques such as those discussed above. First, they are able to reduce some of the dynamic 
speckles in addition to the static ones. However, the temporal bandwidths of the speckle 
nulling loops that have been implemented on-sky have been relatively slow~\citep{Martinache2014},
and these loops did
nothing for speckles with lifetimes shorter than several seconds. A loop's temporal 
bandwidth determines to what extent the short-lived speckles can be reduced. 
The second advantage of real-time speckle nulling loops is that, because they utilize 
interference principles to null the speckles, they remove the speckles' shot noise as well.

The lifetimes of speckles determine a given technique's ability to mitigate them, so
several attempts have been made to define and measure speckle lifetimes. 
\citet{Vernin1991} was interested in speckle interferometry (a high-resolution technique
that predates adaptive optics) and performed early speckle lifetime measurements.
\citet{Macintosh2005} simulated speckle lifetimes in coronagraphic and noncoronagraphic
data, and~\citet{Stangalini2017} measured speckle lifetimes in AO-corrected visible-wavelength
images using techniques more commonly applied to solar photospheric characterization.
\citet{Milli2016} quantified the temporal changes in speckles by calculating Pearson's
correlation coefficients~\citep{Pearson1895} for frames separated by varying amounts
of time. This analysis was most similar to ours; 
we describe their methods and results in Section~\ref{sec:milli}.
In this paper, we describe an experiment which utilizes high frame rate SAPHIRA speckle 
observations to identify the timescales over which speckles change and predict the 
contrast improvement that could be enabled by a real-time speckle nulling loop.

\section{Experimental Setup and Observations}
The Subaru Coronagraphic Extreme Adaptive Optics instrument~\citep{Jovanovic2015scexao} (SCExAO) is
an instrument at Subaru Telescope atop Maunakea in Hawaii. It utilizes a pyramid wavefront sensor
and 2000-element deformable mirror (DM) updating at typically 2 kHz. AO188~\citep{minowa10}, Subaru's
facility AO system,
provides preliminary corrections, and then SCExAO provides high-order corrections in order to 
achieve Strehls of 80-90\% at $H$ band~\citep[e.g.][]{Kuhn2018}. SCExAO is designed to observe high-contrast objects such as
extrasolar planets, brown dwarfs, and debris disks~\citep{Jovanovic2013,Jovanovic2016}. Numerous different modules can utilize SCExAO's
corrected beam. Visible light is sent to the VAMPIRES~\citep{norris15} and FIRST~\citep{huby12} 
aperture-masking interferometers. Infrared light can be directed to the CHARIS integral field 
spectrograph~\citep{groff17}, First Light Imaging C-RED 2 internal science camera~\citep{Gach2018}, or the SAPHIRA
camera.

SAPHIRA~\citep{Baker2016} is a 320$\times$256@24 $\mu$m pixel HgCdTe linear avalanche photodiode detector 
manufactured by Leonardo. It has low dark current~\citep{Atkinson2017} and noise and an adjustable
gain, and it is optimized for high frame rates (up to 10 MHz per channel clocking frequencies). 
At bias voltages of $\sim 18$ V, SAPHIRA detectors have a multiplication gain of several 
hundred~\citep{atkinson16}. Because this gain multiplies the signal but not the read noise,
SAPHIRA has the potential for photon-counting performance~\citep{Atkinson2018,finger16}. 
SAPHIRA detectors can be read using 1, 4, 8, 16, or 32 outputs. Regardless of whether it is
a full frame or subarray being read, the outputs read adjacent pixels. Therefore, unlike HAWAII
detectors, all outputs are used even when reading a subarray, and therefore the frame rate scales
approximately inversely with the number of pixels being read~\citep{Goebel2018}. We operated the
SAPHIRA using a controller developed  at the University of Hawaii Institute for Astronomy
affectionately called the ``Pizza Box'' due to its shape.

During SCExAO engineering nights on UTC May 31, August 13, August 15, and September 11, 2017, we recorded SAPHIRA
images of unresolved stars. These observations are summarized in Table~\ref{table:obs}.
No coronagraph was used.
The detector was subwindowed to $128\times128$ pixels (approximately a
1 arcsecond square) in order to achieve a frame rate of 1.68 kHz. During the first three nights, 
the PSF core was saturated by a
factor of $\sim10$ in order to obtain a better signal to noise ratio (SNR) for the speckles;
this data were used in the analysis of Section~\ref{sec:sl1}.
On the night of September 11, most of the dynamic range of the detector was used, but the PSF
core was not saturated; this data were used for the analysis of Section~\ref{sec:sl2}. On all
nights, the 
SAPHIRA detector was operated in ``read-reset'' mode, wherein each line was read once and 
then reset. This enabled a maximum effective frame rate, optimal duty cycle, and full use of 
the dynamic range per image~\citep{Goebel2018}. We later collected dark/bias frames and subtracted
these in order to remove the detector's pedestal pixel-by-pixel offset pattern. In our previous speckle
lifetime measurements~\citep{goebel16}, we operated the detector in up-the-ramp mode (multiple reads 
followed by a reset) and then subtracted adjacent reads. This resulted in irregular time 
sampling and much poorer SNR due to having to split the flux over multiple reads. Additionally,
in the previous observations, we used the PSF core to align the images in post-processing, 
so we did not saturate it during observations. We have since developed other techniques as
described below for tip/tilt correction in post-processing. For these reasons, our more recent
observations are much higher quality than the older ones.

\begin{deluxetable}{lcccc}
\tablecaption
{A summary of observations.
\label{table:obs}}
\tablecolumns{5}
\tablewidth{0pt}
\tablehead{
\colhead{UTC Date} & \colhead{Target} & \colhead{Seeing at 0.5 $\micron$} & \colhead{Wind Speed} & \colhead{Strehl Ratio} \\
\colhead{} & \colhead{} & \colhead{(arcsec)} & \colhead{(m/s)} & \colhead{}
}
\startdata
     May 31, 2017 & Vega & 0.9 & 0 & 92\% measured  \\
     August 13, 2017 & Altair & 0.35 & 0 & 70\% estimated \\
     August 15, 2017 & $\beta$ And & Unreported & 5 & 90\% measured \\
     September 11, 2017 & 63 Cet & 0.35 & 1 & 90\% estimated \\ 
\enddata
\tablecomments{The Strehl ratios
were calculated from individual sub-millsecond exposures at $H$ band, so they do not include the
blurring effects of tip and tilt. The seeing measurements were produced by the Canada France 
Hawaii Telescope (CFHT) MASS/DIMM monitor, and the wind speeds are those recorded by the CFHT
weather station. This is located approximately 750 m from Subaru Telescope.}
\end{deluxetable}

The May and August nights produced the data used in the analysis presented in 
Section~\ref{sec:sl1}. On those nights, datasets were collected with SCExAO+AO188 (extreme 
adaptive optics) corrections, AO188 corrections only, and with no adaptive optics. In each 
regime, approximately one minute of data was collected with each of a 10 nm bandpass filter 
centered at 1550 nm, a 50 nm bandpass filter centered at 1550 nm, and a full $H$ filter 
(about 260 nm bandpass). The narrower bandpass filters enabled better resolving of the speckles 
because they reduced the speckles' chromatic elongation. The three bandpasses in each of the
three AO regimes produced nine datasets per night. In order to align images and provide a flux
calibration, we applied an astrometric grid~\citep{Jovanovic2015speckles} during observations.
In short, sine waves in orthogonal
directions with frequencies optimized to the DM actuator pitch were applied to
the DM. This generated an artificial speckle in each corner of the detector. By alternating 
the phase of the sine waves between 0 and $\pi$ on timescales shorter than individual SAPHIRA
exposures, the astrometric speckles became effectively incoherent with the speckle 
halo~\footnote{The speckle phase was modulated with a frequency four times SAPHIRA's
average frame rate. It was not perfectly synchronized to the exposures, so some very
small residual coherence remained. However, this effect did not matter because the
region used for speckle lifetime calculations did not include the astrometric speckles.}. This
avoided coherent interference between the electric fields of the artificial and natural speckles
that could distort the position and lifetime measurements. These astrometric speckles were 
$5.1\times 10^{-3}$ 
times the brightness of the PSF core for the May 31 observations and 
$2.6\times10^{-3}$ 
the brightness of the PSF core during the August observations. We used these astrometric
speckles to 1) align the images in order to remove tip/tilt, and 2) photometrically calibrate
the speckle brightnesses.

The data from September 11 were used for the analysis presented in Section~\ref{sec:sl2}.
We were interested in longer-term speckle evolution, so we collected seven minutes of 
ExAO-corrected PSF data. We again used the 50 nm 
bandwidth filter centered at 1550 nm; this bandwidth provided a good balance between
throughput and chromatic speckle elongation. We did not use the astrometric
speckle grid on this night; we aligned the images during processing by cross-correlating
the speckle patterns from one image to the next.

SCExAO has an internal broadband light source and optics which simulate the telescope
pupil~\citep{Jovanovic2015scexao}. This produces a nearly-static PSF whose temporal evolution 
is solely due to instrumental effects.
We observed this light source using the same camera settings as were used on-sky
and then reduced the data using the same techniques as were employed for the on-sky 
observations. This enabled us to separate instrumental effects from actual atmospheric
behavior. However, because this light source is internal to SCExAO, we were not able to
probe the behavior of the telescope, image rotator, atmospheric dispersion corrector,
or AO188 apart from the on-sky observations.

\section{Speckle Lifetime Measurements for Real-Time Speckle Nulling}
\subsection{The Benefits and Techniques of Real-Time Speckle Nulling}
Unlike speckles subtracted away during post-observation image processing, speckles 
interferometrically destroyed during observations do not leave behind photon noise. Because
a speckle is fundamentally scattered starlight, it is possible to apply equal-amplitude but
out-of-phase starlight on top of it and thereby destructively interfere it away. On the
other hand, light from a substellar companion is not coherent with the starlight, so it cannot
be destructively nulled away.

\begin{figure}[!ht]
 \begin{centering}
 \includegraphics[width=0.8\textwidth]{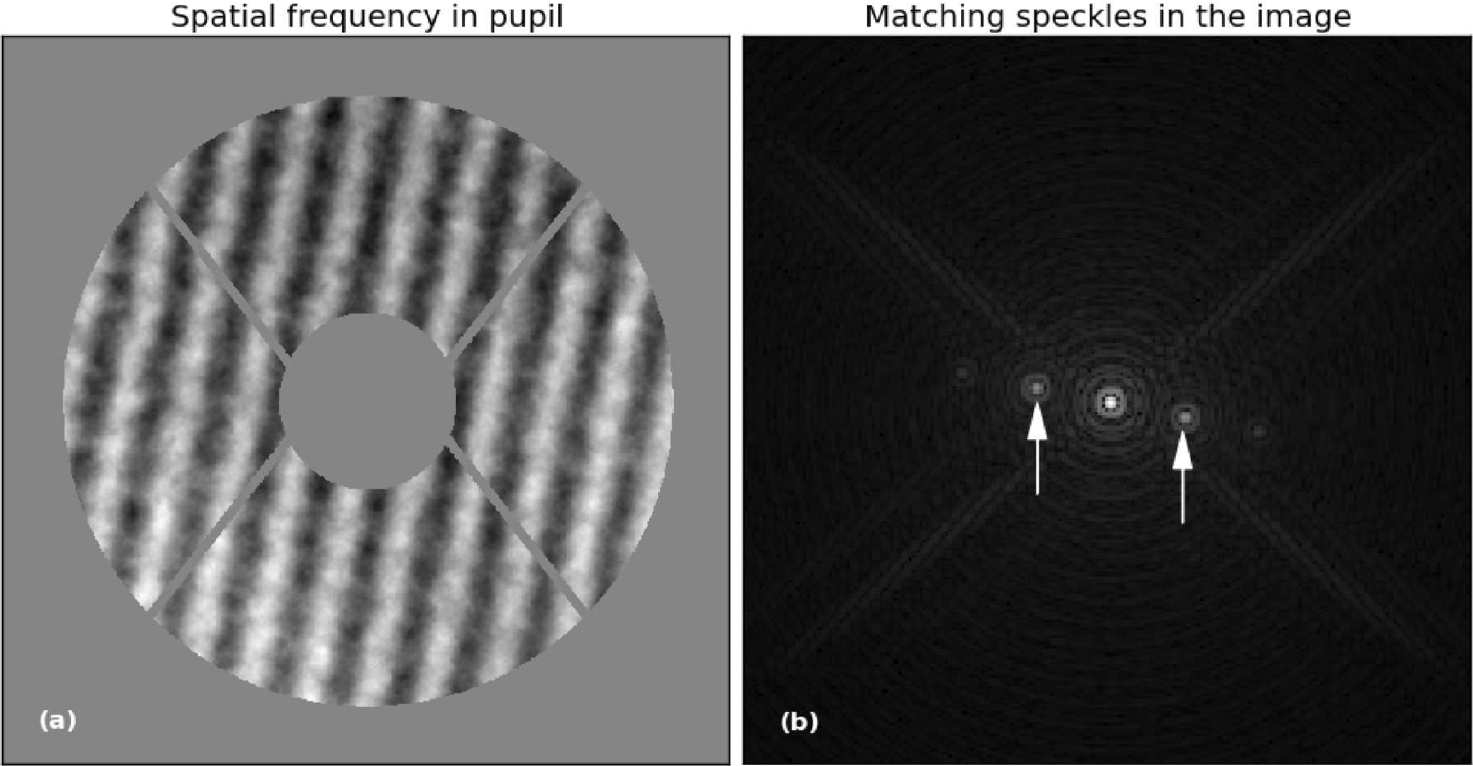}
 \caption
 [Sine waves on the deformable mirror produce speckles at the focal plane]
 {On the left is a map of deformable mirror displacement, and on the right are the resulting
 speckles produced in the focal plane. By varying the orientation, frequency, amplitude, and phase
 of the sine wave on the DM, one varies the location, spacing, brightness, and phase of the
 artificial speckles, respectively. Multiple sine waves can be applied in order to produce multiple
 speckles. Figure reproduced from~\citet{Martinache2014}.}
 \label{fig:pupilfocalplanes}
 \end{centering}
 \end{figure}

A number of techniques have been proposed to differentiate between speckles caused by scattered
starlight and substellar companions and then mitigate the speckles. The pioneering work in
real-time speckle nulling algorithms and conceptually most straightforward method was proposed 
by~\citet{Borde2006}. The focal plane, which is where the science detector is located, 
shows the Fourier transform of the pupil plane, which is where the deformable mirror is
located. A sine wave on the DM generates delta functions (PSFs) on the science detector. This is
shown in Figure~\ref{fig:pupilfocalplanes}. The amplitude, phase, orientation, and frequency of 
the sine wave on the DM determines the brightness, phase, location, and spacing of the artificial
speckles on the science detector. Therefore, one can generate an artificial speckle in the same
location as an existing speckle, scan through phase space to find the opposite complex amplitude, 
and thereby
null it away. The upside of this technique is that it is robust and does not require a detailed
system model. The primary downside is that several phase measurements need to be made before a
speckle can be nulled, so it is not as fast as other algorithms. \citet{Martinache2014} 
implemented this speckle nulling algorithm on SCExAO. In practice, it worked well on the SCExAO
internal light source for quasi-static speckles, but because the camera used operated
at $\sim 170$ Hz and was noisy, and the loop required many phase measurements per iteration, it 
was not fast enough to affect the dynamic speckles when tested on-sky.

A more complex but potentially faster speckle nulling method known as ``electric field conjugation"
was proposed by~\citet{Giveon2007}. This method places small displacements on the
deformable mirror to derive the electric field and thereby phase of speckles. It has produced
the deepest contrasts (on the order of $10^{-9}$) in highly stable high contrast imaging 
laboratories, but it
requires a detailed system model and high signal to noise because the actuator pokes are quite
small. The theory of this technique forms the basis of several other proposed speckle nulling
ideas.
Other techniques for real-time speckle discrimination include 
coronagraphic focal plane wavefront estimator for exoplanet imaging
(COFFEE)~\citep{Sauvage2012},
the self-coherent camera~\citep{Baudoz2006},
the phase-shifting interferometer~\citep{Serabyn2011,Bottom2017}, 
synchronous interferometric speckle subtraction~\citep{Guyon2004},
and linear dark field control~\citep{miller2017}. 
However, apart from the implementation by~\citet{Martinache2014} of the technique 
of~\citet{Borde2006}, none of these methods have been used for speckle nulling
on-sky. This is primarily because they 
are optomechanically complex, require a higher
degree of PSF stability than can be achieved on-sky, require very high Strehl ratios, or
require deformable mirror perturbations that would be detrimental to science observations.
\citet{Jovanovic2018} provides an informative review of these techniques.

Regardless of the particular algorithm, our speckle lifetime measurements are critical to
quantify closed loop speckle nulling performance. 
As loop bandwidth increases, shorter-lived speckles can be
destroyed, and therefore the focal-plane contrast and detection limits improve. An 
understanding of the rates at which speckles change is important because it influences
the feasibility of implementing the loops described above and gives the bandwidth necessary
to reach a given performance level.

\subsection{Our Observations and their Interpretation}\label{sec:sl1}
\begin{figure}
 \begin{centering}
 \includegraphics[width=\textwidth]{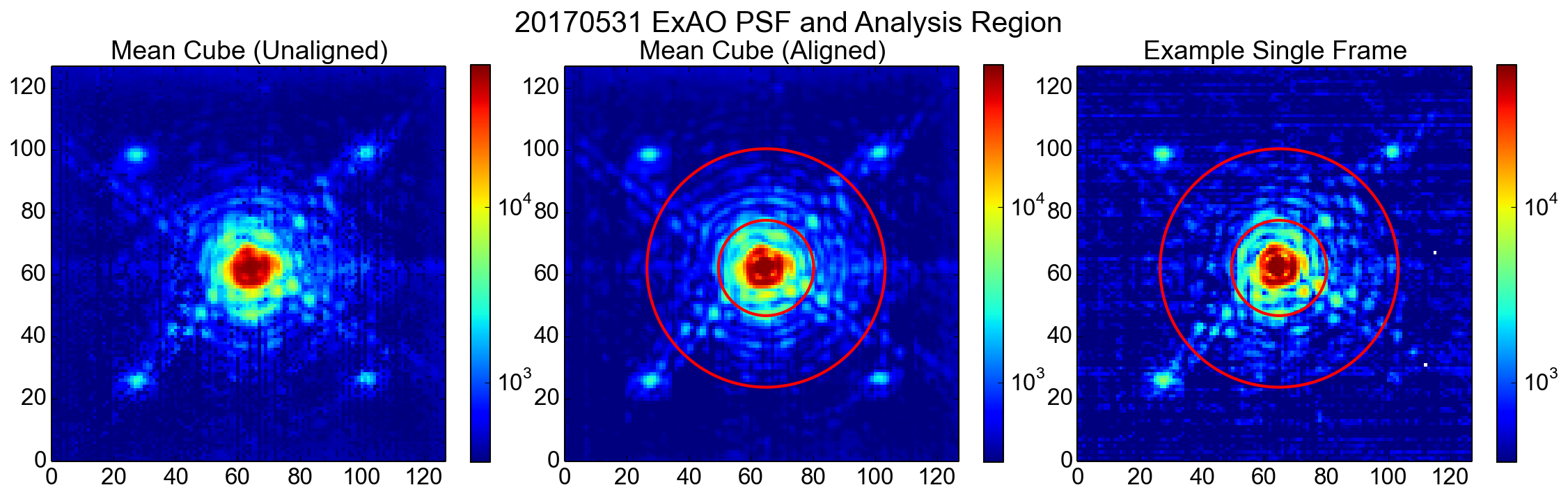}
 \caption
 [Example PSF images before and after alignment]
 {Shown are three sample logarithmically scaled ExAO-corrected PSFs from the May 31,
 2017 dataset. The four bright speckles in the corners are the artificial astrometric speckles
 we created for image registration and photometry (the PSF core couldn't be used because it was saturated). Overplotted
 on the second and third images are circles indicating $r=4 \lambda / D$ and 
 $r=10 \lambda / D$ (the region within which we analyzed speckle lifetimes). The left image
 shows the result of coadding 10,000 frames (6 s of data), and the middle one shows
 the same coaddition after the images were aligned. Due to the alignment, the middle image exhibits
 many more airy rings and better-defined speckles than the left one. The 
 right image is a sample individual frame. One astrometric speckle appears to be missing due to a combination of a rolling shutter effect and the astrometric speckle modulation.}
 \label{fig:alignment}
 \end{centering}
\end{figure}

Our goal is to measure how speckles evolve as a function of time and brightness.
We quantified this using a new technique described below.
The single largest detriment to image quality at SCExAO is tip/tilt caused by vibrations from
the telescope and instrument~\citep{Lozi2016}, which causes translation of the speckles
instead of evolution, so it needed to be mitigated. Therefore, we
began by aligning all the images. The results of the image alignment are shown in 
Figure~\ref{fig:alignment}. We located the artificial astrometric speckles and aligned images using
those, and as a check also cross-correlated each image against the next to detect shifts. Both
techniques registered images to $\sim 0.01$ pixel accuracy and produced similar results. 

\begin{figure}
 \begin{centering}
 \includegraphics[width=0.72\textwidth]{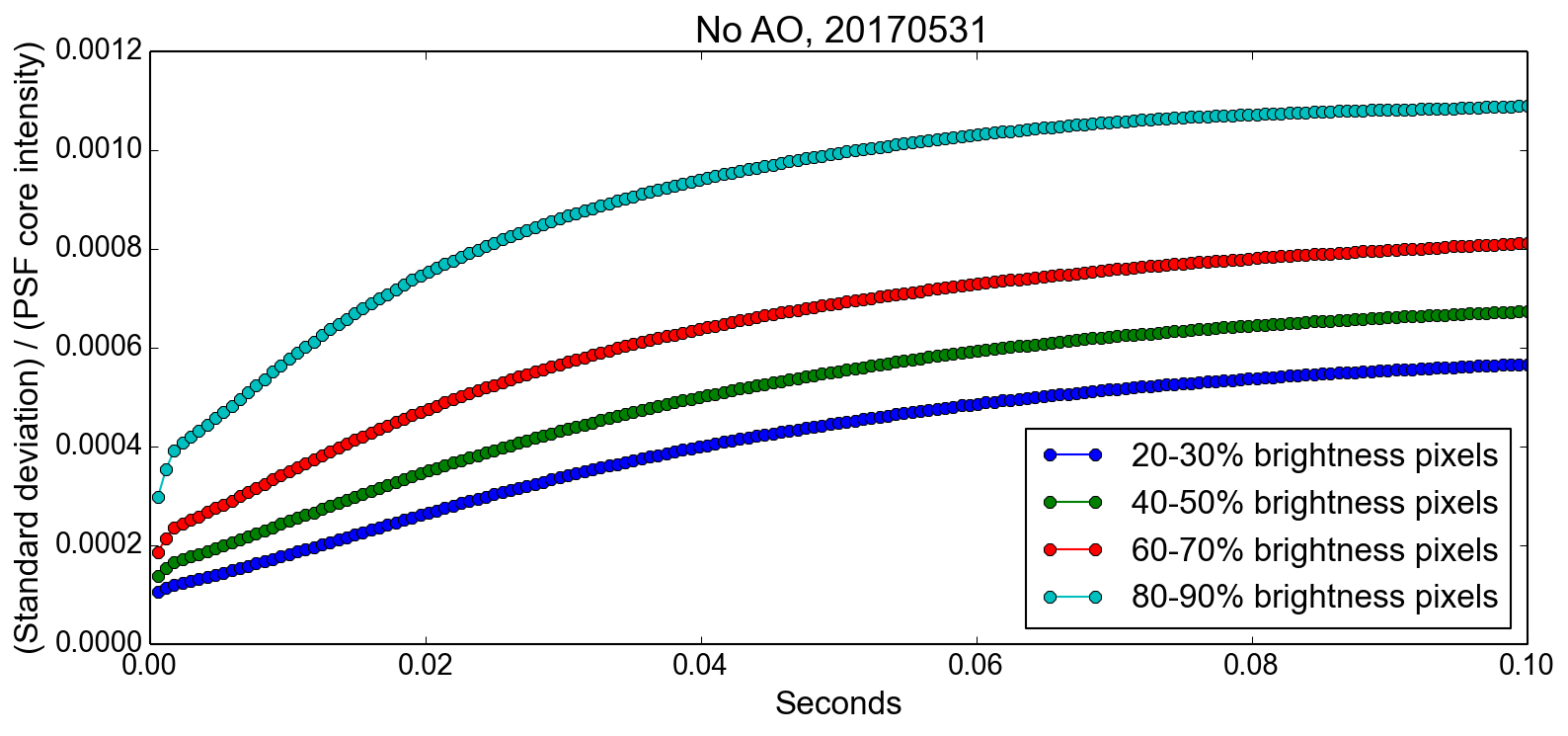}
 \includegraphics[width=0.72\textwidth]{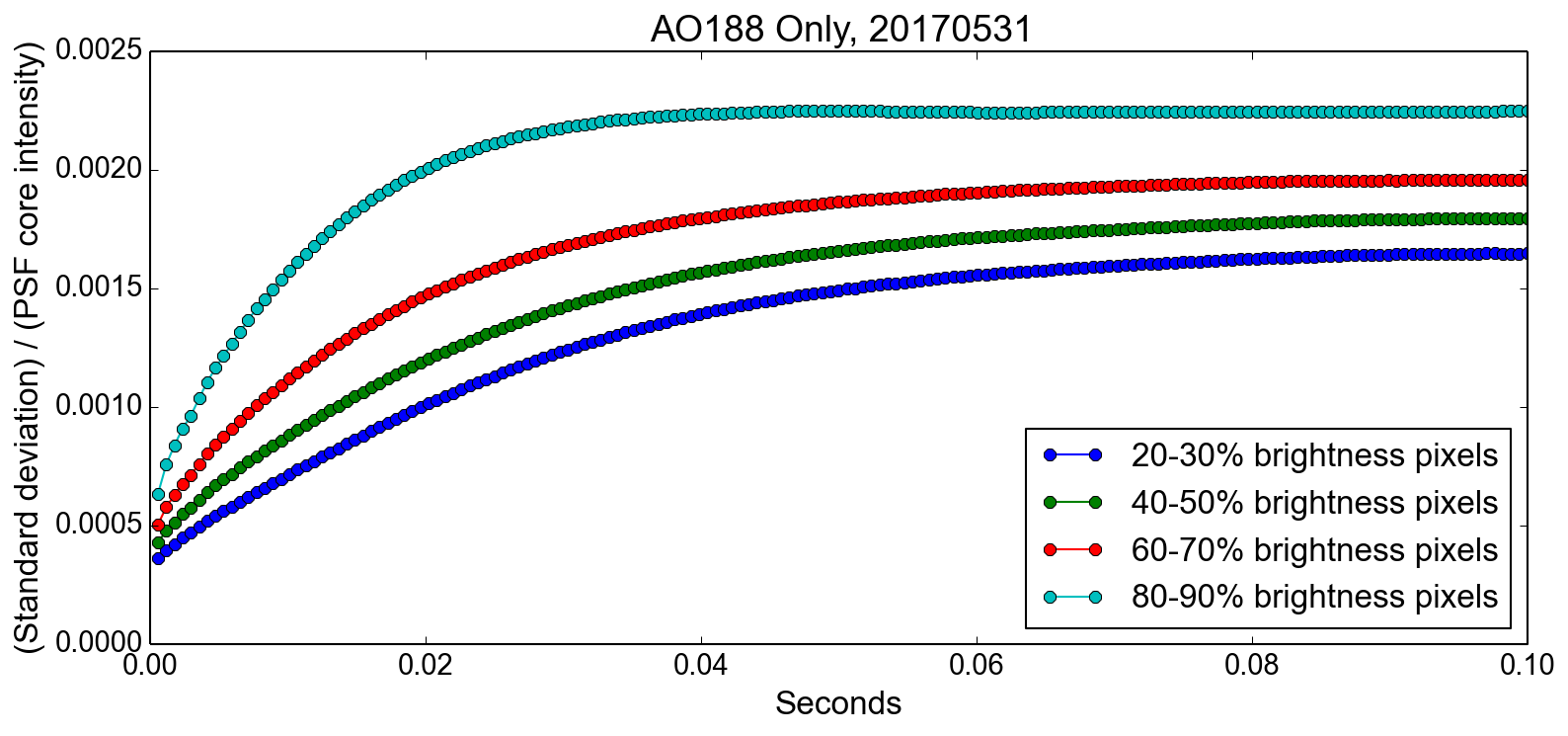}
 \includegraphics[width=0.72\textwidth]{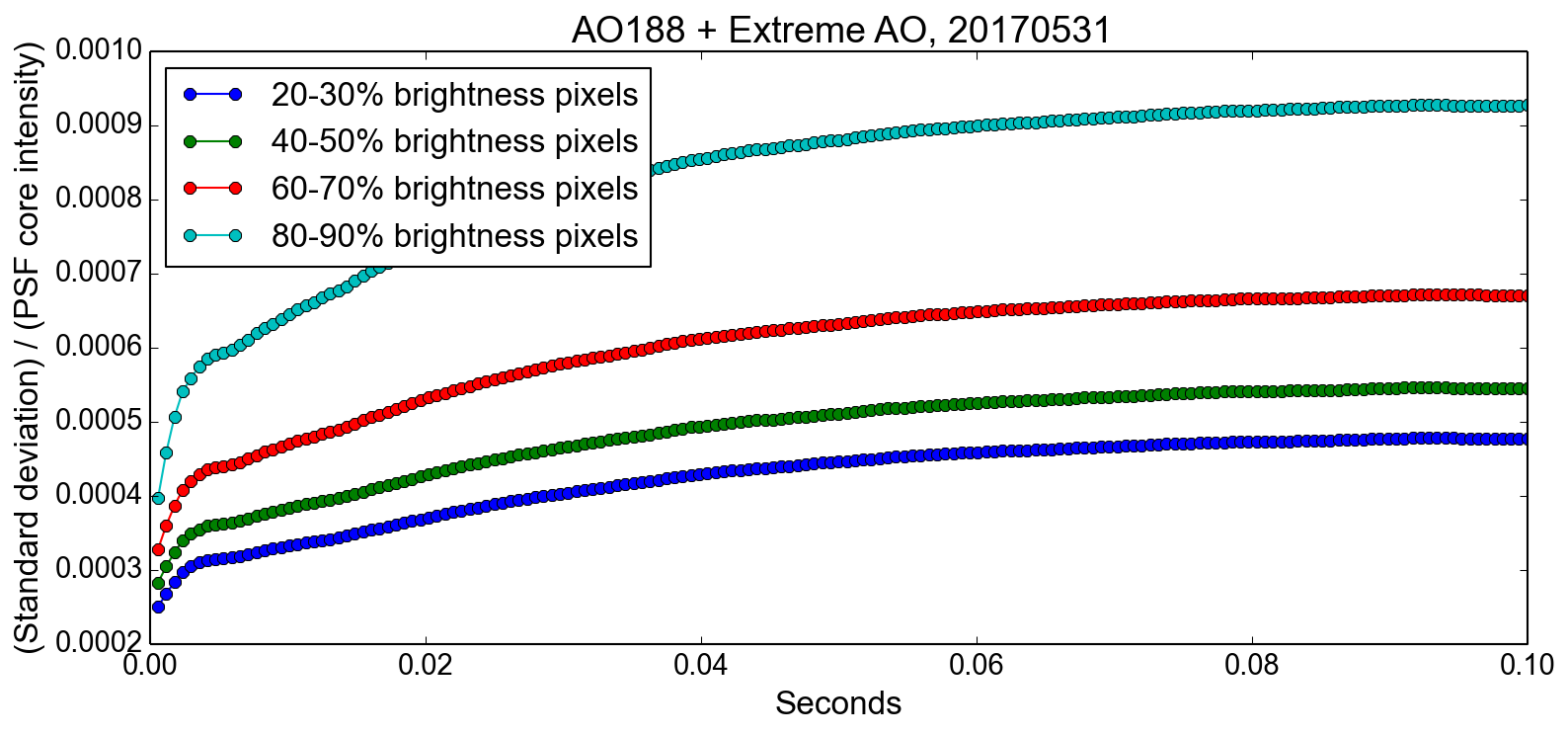}
 \caption
 [Speckle lifetime measurements on May 31, 2017 in the regimes of Extreme AO, AO188 only, and no AO]
 {Shown here are the speckle evolution plots for four different brightness bins and
 the three different AO regimes (no AO, AO188 only, and AO188+ExAO) on the night of
 May 31, 2017. AO188 corrections reduced the standard deviation in speckle brightness relative
 to having no AO, and ExAO again reduced it relative to AO188. We divided the calculated
 standard deviations by the PSF core brightness in order to express them in units of contrast.
 It should be noted that the non-AO images were nowhere near diffraction-limited, so 
 dividing the standard deviations by the peak brightness of the image resulted in
 misleadingly optimistic contrasts. The plots approach asymptotic
 values at large temporal separations because the speckles have entirely changed. 
As predicted by Equation~\ref{eqn:didt}, the brighter speckles change more rapidly than the dimmer ones.}
 \label{fig:lifetimes1}
 \end{centering}
 \end{figure}

\begin{figure}[!ht]
 \begin{centering}
 \includegraphics[width=0.75\textwidth]{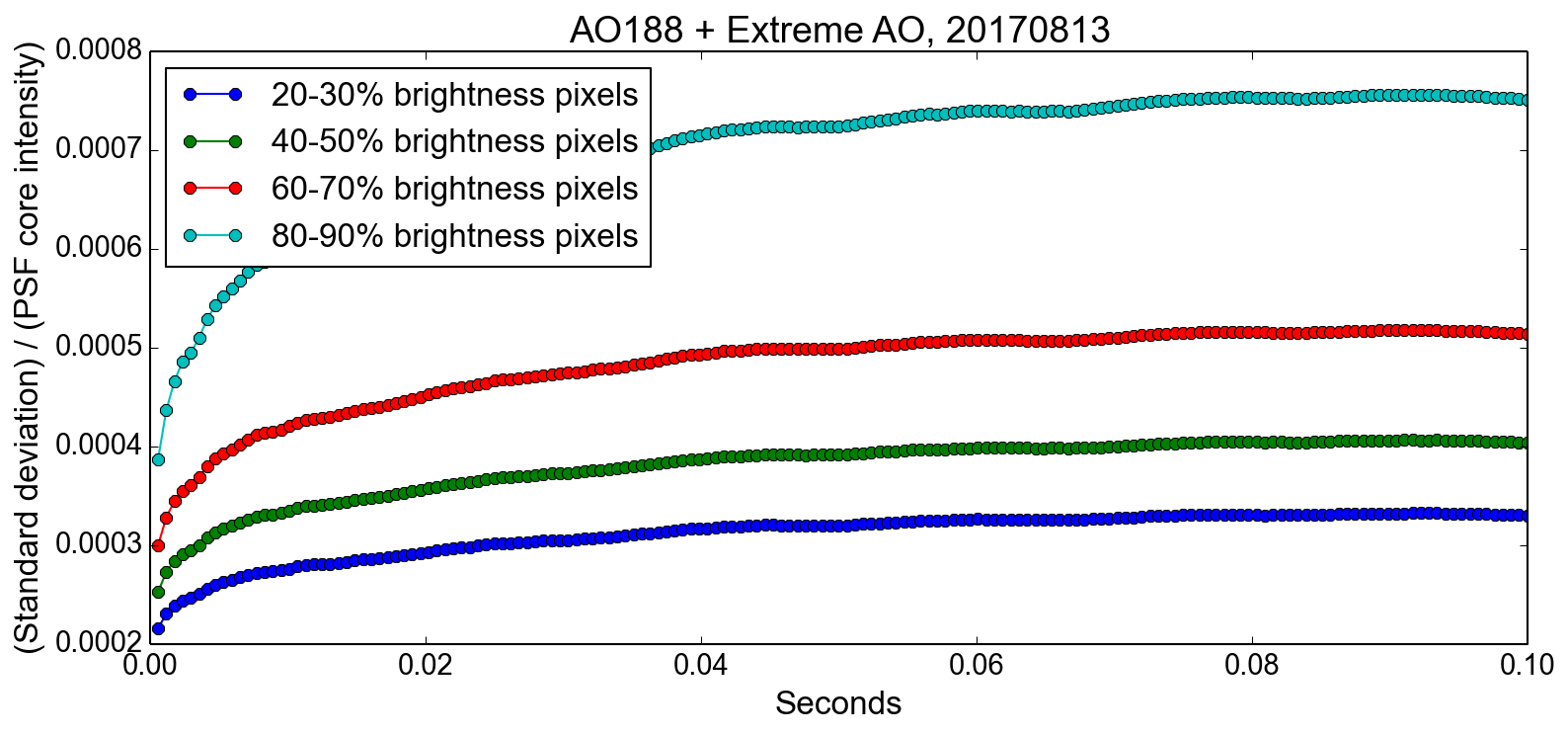}
 \includegraphics[width=0.75\textwidth]{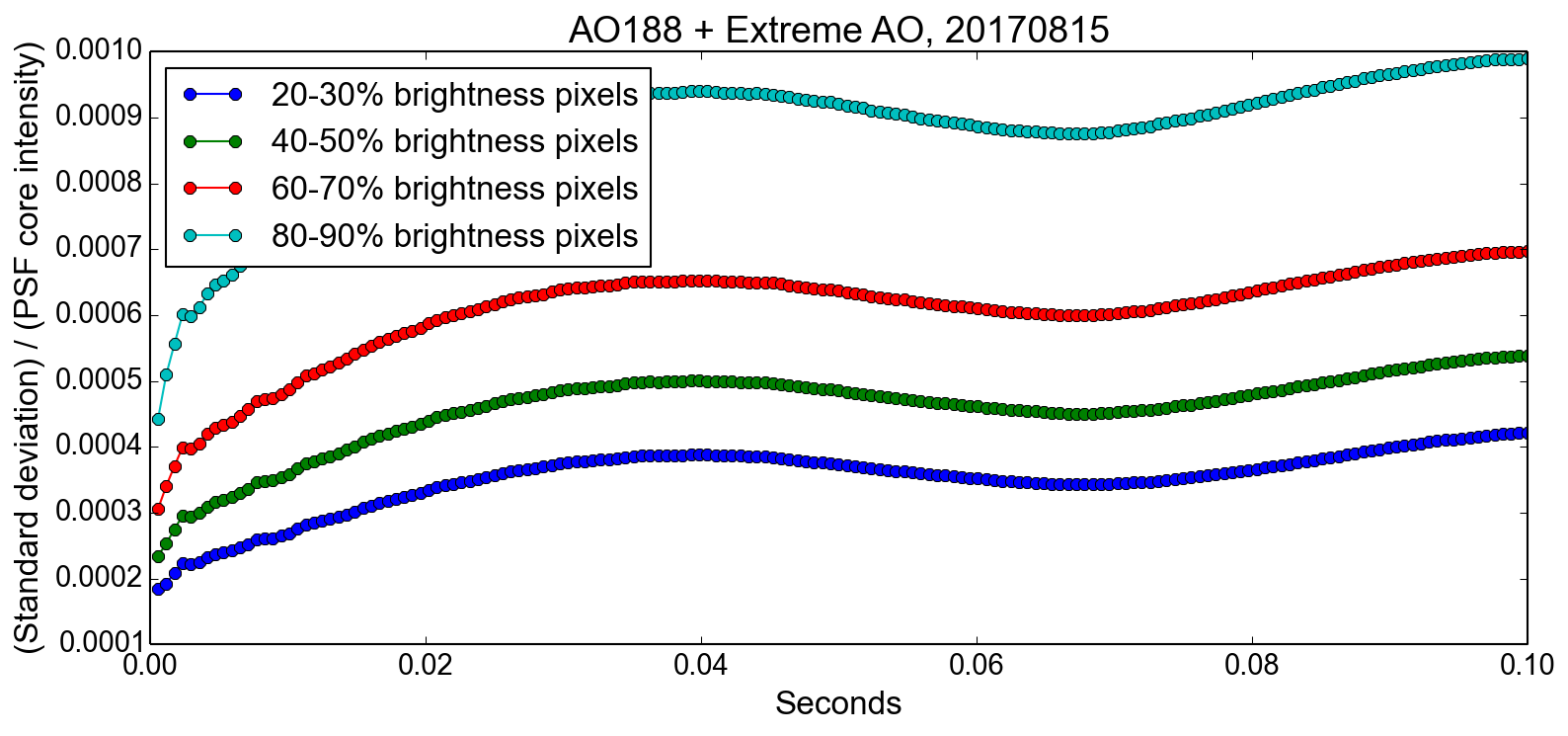}
 \caption
  [Speckle lifetime measurements on August 13 and 15, 2017 in Extreme AO images]
 {These two plots are equivalent to the bottom plot in Figure~\ref{fig:lifetimes1},
  but were produced using the extreme AO data collected on August 13 (top) and 15 (bottom), 
  2017. The seeing and AO tuning were significantly different in each of the three nights.
  The minor oscillation of the August 15 plot is likely due to a vibration that was not 
  completely removed by our image alignment. We divided the calculated standard deviations 
  by the PSF core brightness in order to express them in units of contrast.}
 \label{fig:lifetimes2}
 \end{centering}
\end{figure}

Second,
we selected all pixels within a radius of $4 \lambda / D < r < 10 \lambda / D$ of the PSF core.
Given that the observed intensity $I$ of a speckle is the square of the modulus of complex 
amplitude $\mathbf{A}$,
\begin{equation}
I = |\mathbf{A}|^2 \; ,
\label{eqn:ia}
\end{equation}
it follows that
\begin{equation}
\frac{dI}{dt} = 2\mathbf{A} \frac{d \mathbf{A}}{dt} \;.
\label{eqn:didt}
\end{equation}
$\mathbf{A}$ is a random variable with a characteristic evolutionary timescale (temporal 
derivative). This is due to linearity between $\mathbf{A}$ and turbulence in the pupil, which to 
first degree, over short timescales, has a linear temporal evolution (fixed derivative).
Therefore, $d\mathbf{A}/dt$ is static over short timescales, and brighter (larger $I$) 
speckles will change more quickly than dimmer ones. Because of this,
we sorted the pixels into four brightness bins (we selected the 20-30 percentile brightness pixels,
40-50 percentile, etc.). For each selection of pixels, we subtracted those pixels from the same
pixels time $t$ later. This formed a difference image. Finally, we calculated the standard 
deviation of these difference images in order to measure how the pixels changed in brightness
over that time interval. If a pixel did not change in brightness, this quantity would be 0. We
divided the standard deviation by the peak brightness of the PSF core so that it had units of contrast.
We repeated this process for each image compared to $1-260$ frames after it, and then
averaged all the measurements at a given temporal separation to improve the signal to noise.
We plotted this quantity against time for each brightness bin
in Figures~\ref{fig:lifetimes1} (extreme AO, AO188 only, and no AO on the night of May 31, 2017)
and~\ref{fig:lifetimes2} (extreme AO on August 13 and 15, 2017).

\begin{figure}
 \begin{centering}
  \includegraphics[width=0.75\textwidth]{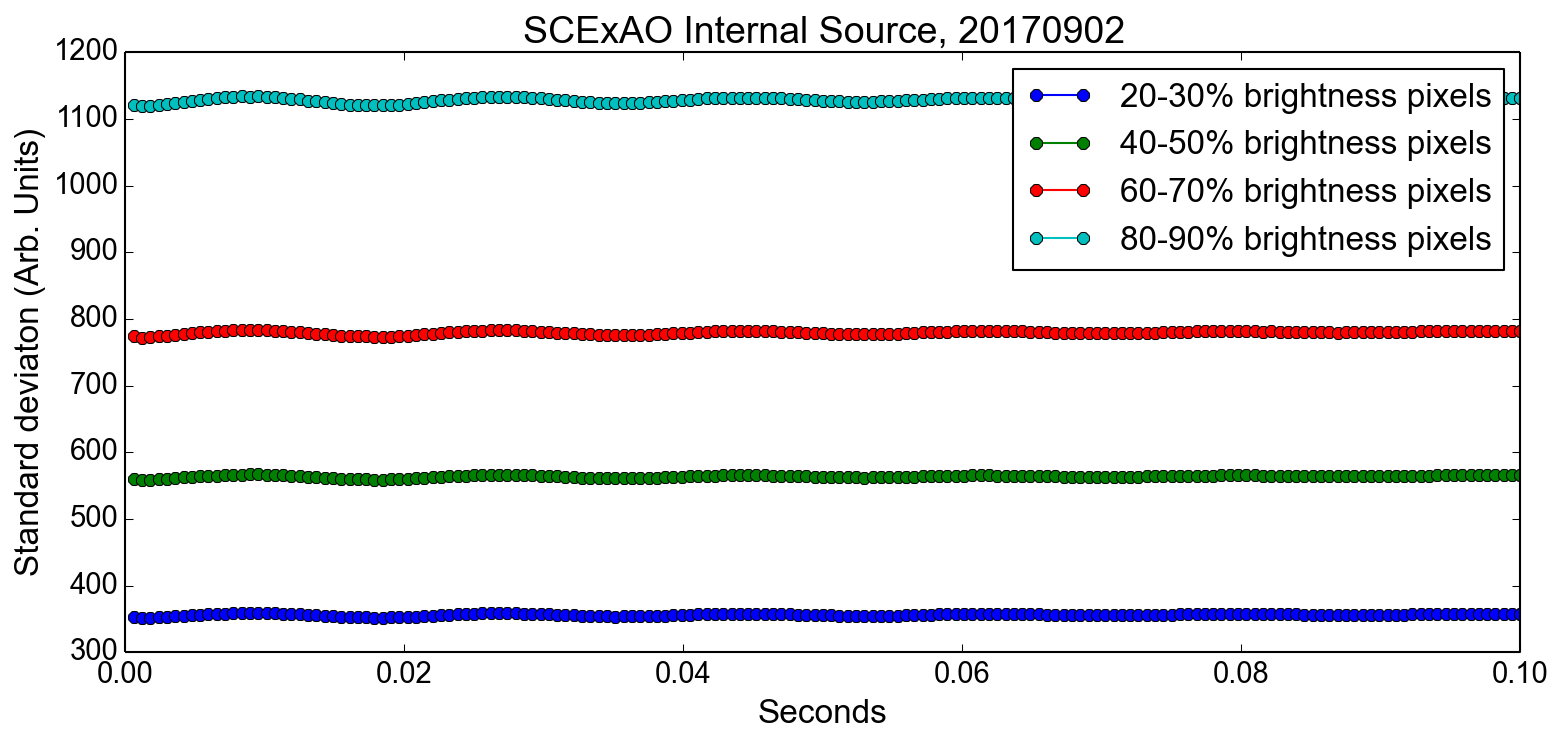}
  \includegraphics[width=0.75\textwidth]{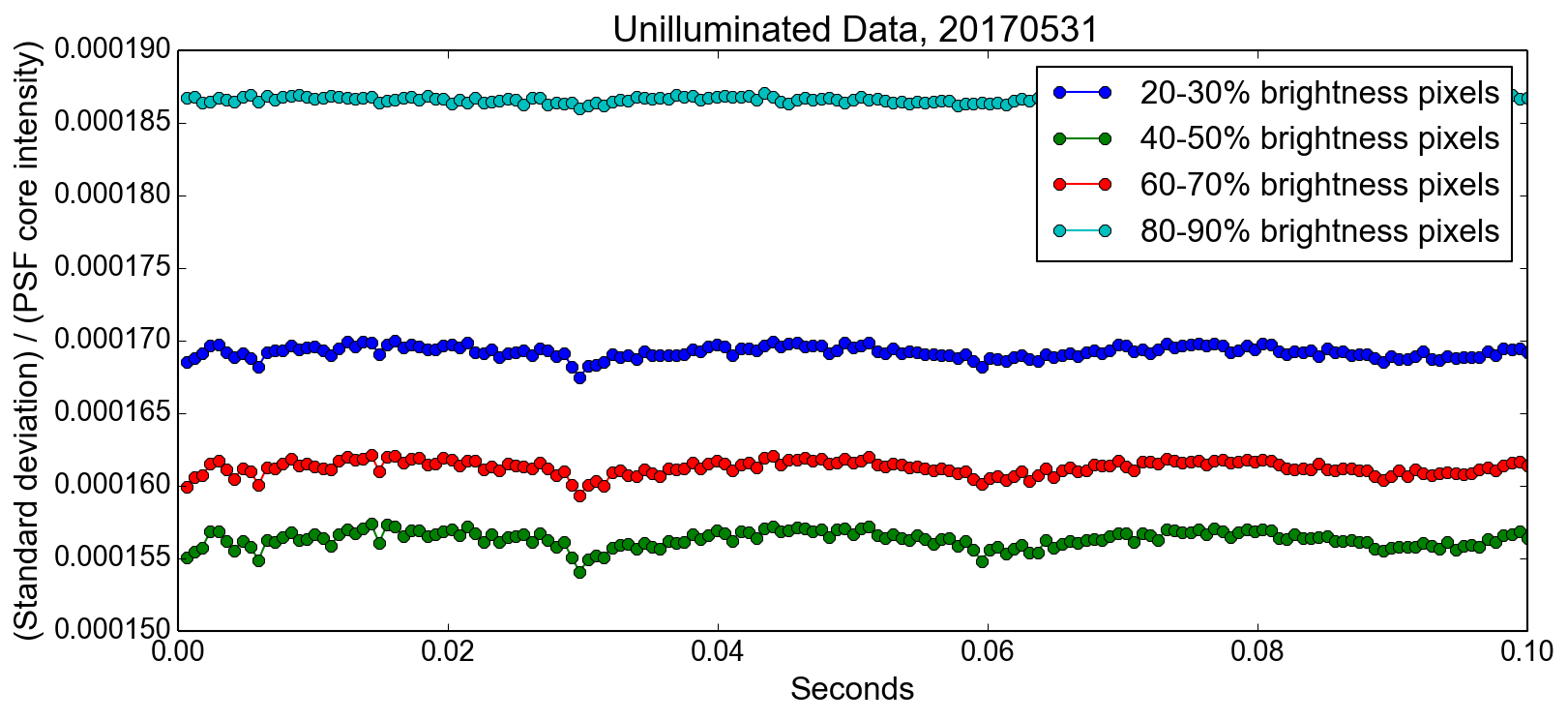}
 \caption
 [Speckle lifetime measurements on internal source data and unilluminated images]
 {In order to understand whether the behavior illustrated in 
 	Figures~\ref{fig:lifetimes1} and~\ref{fig:lifetimes2} was due to intrinsic atmospheric
    speckle evolution, we computed the standard deviations using images of the SCExAO
    internal light source and also dark frames collected immediately following the May 31
    on-sky observations. The internal PSF data are reassuringly flat with time, indicating that
    the speckle evolution of the previous plots is not due to instrumental effects. A
    photometric calibration was not available for this data, so it is expressed in
    arbitrary brightness units.
    In the lower plot, the noise data are also flat with time. This has been scaled by the
    PSF brightness from the ExAO images collected a short time earlier in order to express
    it in units of contrast.
    The order of the relative brightnesses in this second plot is different from the other plots.
    This is most
    likely due to the behavior of pixels on the tails of the responsivity distribution;
    well-behaved pixels have the lowest standard deviations when unilluminated, and ``bad''
    pixels are more prone to wandering and therefore higher standard deviations.
 }
 \label{fig:lifetimes3}
 \end{centering}
\end{figure}

To understand whether the behavior of the standard deviations illustrated in 
Figures~\ref{fig:lifetimes1} and Figures~\ref{fig:lifetimes2} was in fact due to speckle behavior
or merely instrumental effects, we observed the SCExAO internal light source. There were no
moving components in this setup. As is illustrated in Figure~\ref{fig:lifetimes3}, indeed that
speckle behavior is entirely static over the timescales examined. Second, we ran unilluminated
frames through the above analysis pipeline in order to see how noise affected the plot's behavior.
Again, the standard deviations were static in time. This means that the behavior in 
Figures~\ref{fig:lifetimes1} and Figures~\ref{fig:lifetimes2} was indeed caused by the temporal
evolution of the on-sky speckles.

In all cases, the brighter speckles change in brightness more rapidly than the dimmer ones.
This confirms the behavior we expected from Equation~\ref{eqn:didt}. The implication of this
is that once a dark hole is present in an image, it is easier for a speckle nulling loop to
keep it dark. Because a dimmer speckle changes more slowly, more speckle nulling loop 
iterations can be carried out before it changes. Therefore, the performance of a real-time
speckle nulling loop will be highly affected by the Strehl ratio and/or presence of a 
coronagraph that creates dark holes in the image.

On sufficiently short timescales, the complex amplitude can be approximated as behaving 
linearly with time, i.e.
\begin{equation}
\mathbf{A}(t) \approx a t + a_0 + b t i + b_0 i
\end{equation}
where $a$ and $b$ are the rates at which the real and imaginary parts of $\textbf{A}$ change,
respectively, and $a_0$ and $b_0$ are their initial values. Inserting this into 
Equation~\ref{eqn:ia} and computing the measured
intensity as a function of time $I(t)$ produces
\begin{equation}
I(t) = (a^2 + b^2) t^2 + (2 a a_0 + 2 b b_0)t + (a_0^2 + b_0^2) \;.
\label{eqn:expectedintensity}
\end{equation}
Therefore, the measured intensity of a speckle is a quadratic function of time, and its 
behavior depends on the complex amplitude's initial value and temporal rate of change.
The standard deviation of pixel intensity, which is what we plotted in 
Figures~\ref{fig:lifetimes1}-\ref{fig:lifetimes3}, behaves the same way. For example,
if a speckle is entirely static, then $a=b=0$, and $I(t) = a_0^2 + b_0^2$, and
this produces a flat line in the plot with a y-intercept which depends on the 
brightness the speckle. This exact case is illustrated in the top panel of 
Figure~\ref{fig:lifetimes3}. On the other hand, the other simple case occurs when
the speckles are changing (i.e.~$a \neq 0$ and $b \neq 0$) but start from zero
intensity ($a_0 = b_0 = 0$).
In this case, the plot would exhibit an upward parabola centered at the origin. We
approach this limit (but do not reach it) in the on-sky observations by selecting 
increasingly dim speckles (i.e. $a_0 \to 0$ and $b_0 \to 0$) for analysis.

Depending on the temporal sampling of the data, the initial brightness of the speckles, and
the rates at which they change, this upward parabolic behavior may not be obvious in real-world
measurements. The first few data points in the May 31
ExAO plot appear linear, the first few points in the August 13 plot appear downwardly parabolic,
and only in the August 15 plot are the first few points plausibly upwardly parabolic. It is
possible that our data are not fast enough to recover this predicted effect, and it would become
apparent in data with even faster temporal sampling. Additionally, the behavior described above
assumes that the complex amplitude at a pixel changes linearly with time, and it certainly 
departs from this after a few milliseconds. Indeed, at sufficiently long timescales, the speckles
have entirely changed from their initial state, causing the measured standard deviations to
plateau. This is why our plots approach asymptotic standard deviations at increasing time.

If the behavior described by Equation~\ref{eqn:expectedintensity} was evident in
the images, one could predict the performance of real-time speckle nulling loops from it. Since
it is unclear whether our data reveal the predicted behavior, we will not do that, but we will
explain how it could be done. If our analysis described above could be
carried out at even higher temporal sampling or in an extremely high Strehl case, and the 
upward parabolic behavior revealed itself, one could fit a second-order polynomial to it. Solving
this quadratic equation would provide the contrast that could be achieved at a given speckle 
nulling loop bandwidth. The y-intercept in this fit would
correspond to the read and photon noise (i.e.~the noise that would be present if it was 
possible to image repeatedly with zero time between the images and with zero speckle evolution). 
A real-time speckle nulling loop would zero the linear and constant terms with each update 
by forcing $a_0 = b_0 = 0$. In this case, $I(t) \propto t^2$. Therefore, 
the achievable contrast would increase with the square of the speckle nulling loop bandwidth
(where bandwidth is $1/t$).
This result emphasizes the major impact that speckle nulling loops will have on detection limits.

\section{A different approach: extending the analysis of Milli et al.}
\subsection{Milli et al.'s Technique and Results}\label{sec:milli}
\citet{Milli2016} employed a different technique for measuring the timescales over which
speckles evolve. Their observations were comparable to ours, so we analyzed our comparatively
higher frame rate data using their methods. They collected $H$-band images at a frame
rate of 1.6 Hz of coronagraphic
PSF images from the SPHERE ExAO instrument. They then quantified the change
in speckles over the 52 minutes of their on-sky observations by computing Pearson's
correlation coefficient~\citep{Pearson1895} for pairs of images.
Pearson's correlation coefficient $\rho(t_i, t_j)$ between two frames collected at times $t_i$ 
and $t_j$ can be defined as follows:
\begin{equation}
\rho(t_i, t_j) = \frac{\sum\limits_{x \in S}[I(x, t_i)-\overline{I_{t_i}}][I(x,t_j) - \overline{I_{t_j}}]}{\sigma_{t_i}\sigma_{t_j}N_{\rm pix}}
\label{eqn:corrcoeffs}
\end{equation}
where the sum occurs over all pixels in the region $S$, $I$ is the instantaneous intensity
at a given pixel with position $x$, $\overline{I}$ is the instantaneous average intensity over the 
region, $\sigma$ is the spatial standard deviation over the region,
and $N_{\rm pix}$ is the number of pixels in the region. The normalization is chosen so
that $\rho=1$ when $t_i = t_j$ (equivalently, if the speckles did not change at all
between images at $t_i$ and $t_j$, 
$\rho=1$). \citet{Milli2016} provided a derivation in their Appendix C demonstrating that 
this quantity is directly related to the contrast obtained after subtracting one image of 
the pair from the other. The signal to noise of $\rho$ for a single pair of frames is 
typically quite poor,
so they improved it by averaging the values of $\rho$ computed for many frame pairs 
separated by equal amounts of time.

After completing their on-sky observations, \citet{Milli2016} observed the 
instrument's internal light source and repeated the analysis in order to separate 
atmospheric effects from instrumental effects. In both cases, they observed two primary 
temporal regimes of speckle decorrelation.
Over the first few seconds of their observations, the speckles decorrelated rapidly
with an exponential decay. They fitted the first 30 seconds of their measurements with
the equation
\begin{equation}
\rho(t) = \Lambda e^{-t/\tau} + \rho_{0}
\label{eqn:expdec}
\end{equation}
and found best-fitting values of $\tau = 3.5$ s, $\Lambda=0.059$, and $\rho_0=0.713$
for their on-sky observations. They initially posited that these speckles were primarily 
residuals of uncorrected atmospheric turbulence, but were surprised when then data
collected using the internal light source exhibited the same behavior (in that case,
they found $\tau=6.3$ s). They performed additional tests but were not able to identify
the fundamental cause of this behavior, and in the end concluded that it was ``real 
[and] related to an internal effect in the instrument independent of the atmospheric
conditions or the telescope."

For images separated by greater than a minute, they observed a second regime of
speckle change. They observed a linear decorrelation with time, and this was again 
present in both the on-sky data and also their internal lamp data. They suggested that 
this change in speckles must be caused by mechanical changes in the instrument (in 
particular the motion of the image derotator and thermal expansion effects). When
they disabled the image rotator, this linear decorrelation of speckles
became flat with time, i.e.~the speckles were not measurably changing.
In
summary, they observed two regimes of speckle decorrelation, and these had
different timescales for the on-sky and off-sky observations, but both were 
fundamentally caused by instrumental and not atmospheric effects.

\subsection{Analysis of Our Data Using this Technique}\label{sec:sl2}
Our data also consisted of $H$-band images of an ExAO-corrected PSF, but it had three
orders of magnitude higher temporal resolution, enabling us to probe far shorter 
timescales. Additionally, since SCExAO is an entirely different instrument from SPHERE,
we hoped our data might shed light on~\citet{Milli2016}'s instrumental effects. 

Our data had some important
differences, however. First, the SPHERE data were collected
behind a Lyot coronagraph, whereas our data were noncoronagraphic. Second, their data
were full $H$-band, whereas our images used a 50 nm passband filter centered at 1550 nm, which
reduced the chromatic elongation of speckles and thereby enabled them to be better resolved.
Finally, while their data were collected over 52 minutes, our data were 7 minutes
in length, so we could not examine as long as timescales. We also recorded two 20 minute
datasets of the SCExAO internal light source in order to differentiate between
instrumental and atmospheric effects.

\begin{figure}[!ht]
 \begin{centering}
 \includegraphics[width=0.9\textwidth]{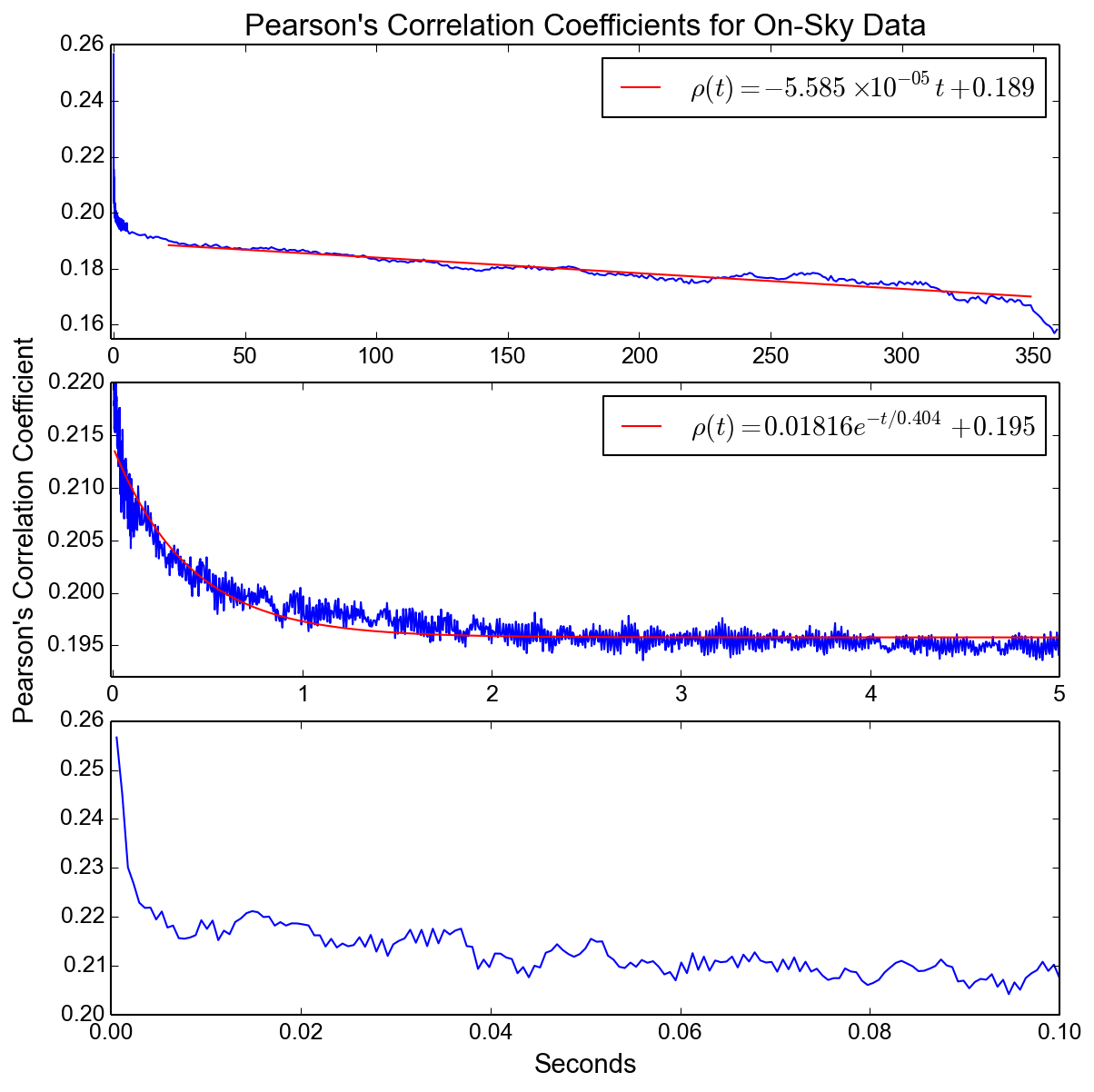}
 \caption
 [Pearson's correlation coefficients for on-sky data.]
 {Plotted above are the Pearson's correlation coefficients calculated from seven minutes
 of SCExAO on-sky data. The same data are shown for three different timescales. Linear
 and exponential decays have been fitted to the data points between 20 and 350 s (top plot) and
  0.1 and 5 s (middle plot), respectively. The coefficients of the fits are labeled. Unlike the plots of
 Section~\ref{sec:sl1}, less change in the speckles results in a higher value on the Y axis.}
 \label{fig:corrcoeffs_onsky}
 \end{centering}
\end{figure}

\begin{figure}[!ht]
 \begin{centering}
 \includegraphics[width=\textwidth]{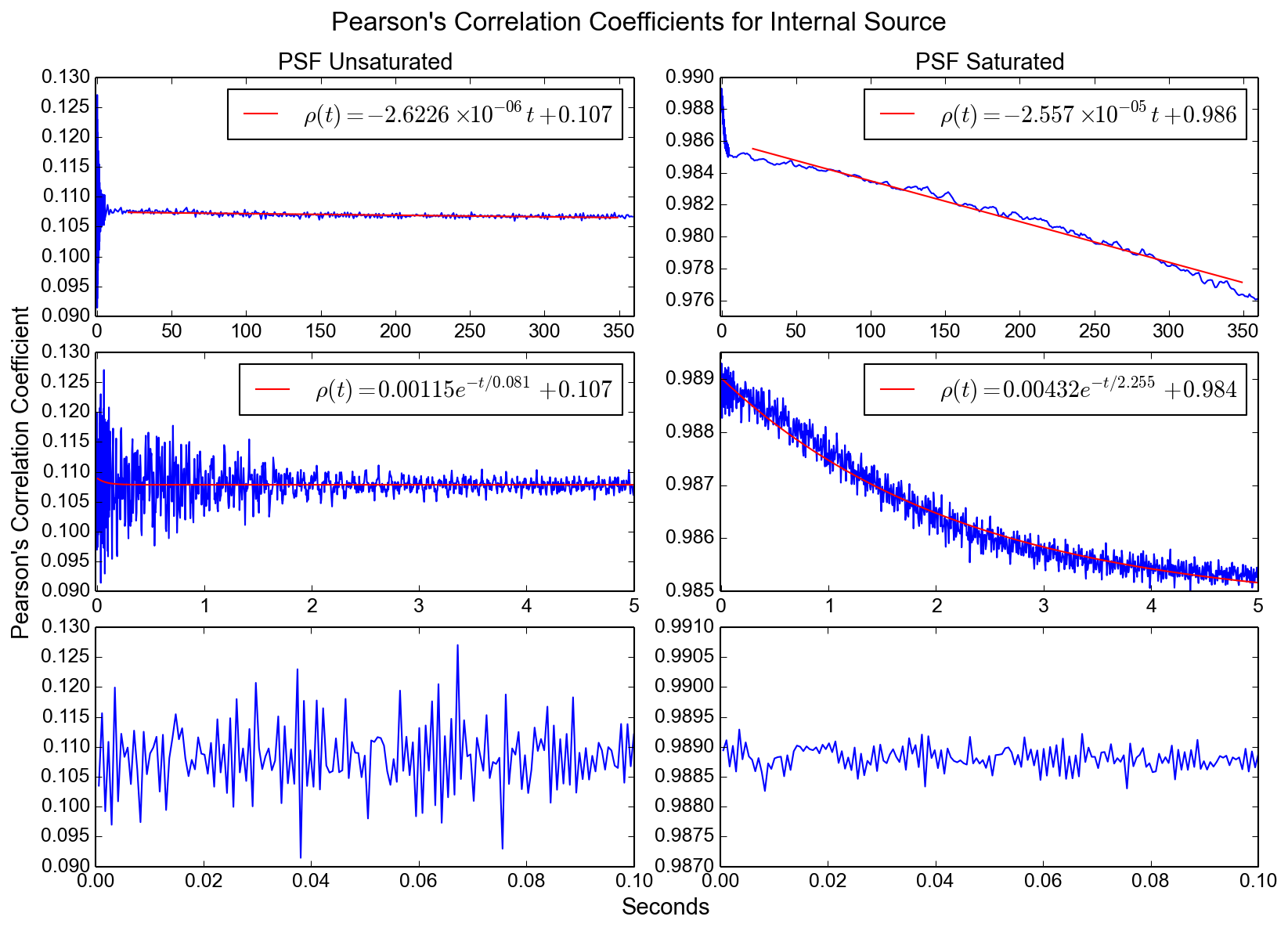}
 \caption
 [Pearson's correlation coefficients for internal source data.]
 {Plotted above are the Pearson's correlation coefficients calculated from two twenty-minute
 observations of the SCExAO internal light source. In the left plots, the PSF core was not saturated,
 mirroring our on-sky observations. However, because the PSF was nearly unaberrated, very few speckles
 were detectable, and no speckle evolution was measured. The dispersion in the correlation 
 coefficients in these left plots is primarily caused by vibrations that were not perfectly removed by
 the image alignment process. On the right, the PSF has been saturated by a factor of 100, thereby
 improving the SNR of the speckles. In this case, timescales of speckle evolution are revealed.
 In each dataset, the correlation coefficients are shown for three different timescales. Linear
 and exponential decays have been fitted to the data points between 20 and 350 s (top plots) and
  0.1 and 5 s (middle plots). The coefficients of the fits are labeled. Note that the high
 correlation between frames separated by $\lesssim 2$ ms of Figure~\ref{fig:corrcoeffs_onsky} is not present
 in this data, indicating that it is an atmospheric and not instrumental effect.}
 \label{fig:corrcoeffs_internal}
 \end{centering}
\end{figure}

We then calculated Pearson's correlation coefficients according to 
Equation~\ref{eqn:corrcoeffs} using the annular region centered on the PSF and defined by
radial separations $3 \lambda / D < r < 10 \lambda/D$. Comparing each of the 630 thousand 
frames for the on-sky data and
1.99 million frames for the internal source data with each of the other frames in that dataset
would have been computationally prohibitive. Therefore, we compared each frame against
$1, 2, 3, \mathellipsis, 200, 208, 216, \mathellipsis, 8400, 10080, 11760, \mathellipsis, 604800$ frames
later. This enabled us to have maximum signal to noise at each calculated temporal separation,
but we sacrificed temporal resolution at higher separations.
In other words, we wanted the best time resolution at the shortest possible intervals, but
settled for one Pearson's correlation coefficient data point every second at a time intervals
greater than six seconds. 
This simplification reduced the computational load by 99.9\%. Even after this, we calculated
852 million correlation coefficients for the on-sky data and 
6.04 billion for the internal source data. The data points at a given temporal separation
were then averaged to improve signal-to-noise. Because there are fewer frame pairs separated
by large amounts of time, signal-to-noise decreased at larger separations. The on-sky dataset
was seven minutes long, and we calculated correlation coefficients for data points up to six
minutes apart in order to still have reasonable signal to noise at the larger separations.
The computation of the Pearson correlation coefficients for each plot took several CPU-days.
Figures~\ref{fig:corrcoeffs_onsky} and~\ref{fig:corrcoeffs_internal} show the results for the
on-sky and internal light source data, respectively. 

We observed three timescales of speckle evolution in the on-sky data. First, and most 
interestingly, our high
frame rate data revealed an effect that~\citet{Milli2016} did not observe. Frames separated by 
$\lesssim 2$ ms are highly correlated in the on-sky data. 
After this time, the exponential decay behavior described in the next paragraph dominates.
The minimal change in data points separated $\lesssim 2$ ms is not present in the
internal source data, implying that this is an actual atmospheric effect. This timescale
is similar to the SCExAO loop bandwidth (2 kHz update rate and a $\sim 15 \%$ gain), so it is most
likely explained by the changes to the deformable mirror shape. Over timescales of $>2$ ms,
atmospheric speckles become de-correlated because they are corrected by the ExAO loop.
Temporal correlations that naturally exist at $>2$ ms in the atmospheric speckles are 
erased or greatly attenuated by the ExAO correction;
so there is a reduction in speckles with timescales of $>2$ ms.

We recorded two 20-minute datasets using the SCExAO internal source. The first internal 
source dataset had an unsaturated PSF core, mirroring our September 11, 2017 on-sky
observations. However, because this was very nearly 100\% Strehl
(the slight degradation from 100\% Strehl was due to imperfections in the optical elements of
the instrument), very few speckles rose above the read noise of the detector. As a result,
the calculation of the Pearson's correlation coefficients revealed no real speckle evolution
(Figure~\ref{fig:corrcoeffs_internal} left plots). This mirrors the findings 
of~\citet{Milli2016}, who saw no speckle evolution when they froze the operation of
SPHERE's optical elements.
In order to improve signal-to-noise in the
speckles, we then saturated the PSF core by a factor of 100 in the second internal source
dataset (Figure~\ref{fig:corrcoeffs_internal} right side). In this case, we observed
two timescales of decorrelation as described in the following paragraphs. Although the
PSF core of the on-sky data was unsaturated, we could measure speckle evolution there
because the Strehl was lower ($\sim$90\% instead of $\sim$100\%) and therefore the 
speckles were brighter.

We observed a second timescale on the order of minutes over which the speckles linearly
decorrelated in both the on-sky data and the saturated internal source data. 
We fitted a first-order polynomial to
data points between 20 and 350 seconds and found that the speckles decorrelated by 56 parts 
per million (PPM) on-sky and 26 PPM on the saturated internal source over this period. For
comparison,~\citet{Milli2016} measured 25 PPM and 9 PPM for these two timescales, respectively,
but their fit was performed over a different range of times.
Because this decorrelation is present in both the on-sky and internal source observations,
it is fundamentally caused by an instrumental effect. Unlike~\citet{Milli2016}, we did not use the image derotator; we had no moving components between the
light source and our detector. This decorrelation must therefore be due to thermal expansions
in the mechanics of SCExAO. This data depart from this linear decorrelation between 300 and
360 seconds in the on-sky data, but we think this is due to small number statistics (there are
fewer frame pairs with 6 minute separations than, for example, 1 minute separations) than any 
intrinsic speckle behavior. The internal source data did not exhibit this effect.

The third timescale we observed was an exponential decay in the Pearson's correlation
coefficients over the first few seconds for both the saturated internal source and the 
on-sky data. \citet{Milli2016}
also observed this behavior, but in our case it occurred much more quickly. We fitted
the points between 0.01 and 5 seconds with Equation~\ref{eqn:expdec} and found best-fitting
parameters of $\tau = 0.40$ s for the on-sky data and $\tau = 2.3$ s for the saturated 
internal source. Our measured values for $\Lambda$ and $\rho_0$
are provided on the plots, but these have fewer physical implications because they are 
influenced by the flux level on the detector, as demonstrated by 
Figure~\ref{fig:corrcoeffs_internal}.
In contrast,~\citet{Milli2016} found exponential decay terms of $\tau = 3.5$ s and $\tau = 6.3$ s for
their on-sky and internal light source data, respectively, when fitted to timescales between 1 and 30 
seconds\footnote{We performed the fits over a different range of times 
than~\citet{Milli2016} because in the case of the linear decay, we simply didn't have long
enough time coverage, and in the case of the exponential decay, the fit would have been poor
if we fitted to the same 1-30 s range as them.}. We observed a much more rapid decay than them, but the exponential decay behavior was present
in both observations. \citet{Milli2016} were unable to explain the cause of this effect.

Whereas the analysis presented in Section~\ref{sec:sl1} had applications to real-time speckle nulling,
the results presented here are primarily applicable to speckle reduction during post processing. If the
exposure times of individual images in an observing sequence are shorter than the timescales on 
which speckles evolve, then post-processing techniques such as KLIP~\citep{Soummer2012} and 
LOCI~\citep{Lafreniere2007} will be
better able to subtract the speckles. For this reason, it is advantageous to use short exposures while
conducting observations for which the speckle halo is relevant (this must be balanced against
the read noise of the detector, emphasizing the need for low-noise detectors such as SAPHIRA or
electron-multiplying CCDs). Although it is preferable to have the speckles as dim as possible, since
these have minimal photon noise, it is also ideal for the speckles that still exist to evolve as slowly
as possible, because this enables them to be mitigated during data reduction. This emphasizes
the need to have the instrument maximally mechanically and thermally stable. Indeed, SPHERE has
more thorough environmental/thermal stabilization than SCExAO, which likely explains why the 
speckle decorrelations measured in this section occur more rapidly than those of~\citet{Milli2016}.

\section{Conclusions}
We collected $H$-band images of the speckle halo surrounding the SCExAO PSF at a framerate of 1.68 kHz.
We then analyzed the data in order to see how speckles evolved as a function of time and brightness.
We performed two analyses: we used a new technique to quantify speckle lifetimes with implications 
to real-time speckle-nulling loops, and we performed a second analysis
with more general applications following the standard techniques in the literature. Our data used
in the second analysis had three orders of magnitude higher temporal resolution than that used for
previously published studies, and we identified a new timescale of speckle decorrelation. We
observed that speckles were relatively static for timescales of $\lesssim 2$ ms in on-sky data, and
we suggest that this timescale corresponds to the bandwidth of the ExAO loop. For timescales
on the order of seconds and minutes, we observe similar behavior to that identified by~\citet{Milli2016}, namely an exponential decay over the first several seconds, followed by a linear decay
over minutes. Mirroring the results of~\citet{Milli2016}, these two trends were present in both the on-sky
and internal source observations. They noted that the linear decorrelation over minutes
went away when the image rotator was stopped, but we observed this effect even without
an image rotator.

An understanding of the timescales over which speckles evolve can inform the development and
deployment of speckle mitigation techniques. A higher bandwidth speckle nulling loop can
destroy a greater fraction of speckles and thereby enable the detection of higher-contrast 
substellar companions. For about 300 nearby M dwarfs, the angular separation of an Earth-like
planet in the habitable zone at maximum orbital elongation would be at least 1 $\lambda/D$ in
the NIR for a 30-m-class telescope~\citep{Guyon2012}, and the contrasts of these targets are
in the range of $10^{-7} - 10^{-8}$. Although current techniques struggle to produce 
processed contrasts better than  $10^{-6}$ at much larger angular separations, ELTs and the implementation of improved speckle reduction techniques, both during observations and during
data processing, will enable observation of these targets.

\section*{ACKNOWLEDGMENTS}       
The authors acknowledge support from NSF award AST 1106391, NASA Roses APRA award NNX 13AC14G, and the JSPS (Grant-in-Aid for Research \#23340051, \#26220704, and \#$23103002$). This work was supported by the Astrobiology Center (ABC) of the National Institutes of Natural Sciences, Japan and the director's contingency fund at Subaru Telescope. FM acknowledges ERC award CoG 683029. The authors wish to recognize the very significant cultural role and reverence that the summit of Maunakea has always had within the indigenous Hawaiian community. We are most fortunate to have the opportunity to conduct observations from this mountain. 

\bibliographystyle{apj}
\bibliography{ch4bib}

\chapter{SCExAO/CHARIS Near-IR High-Contrast Imaging and Integral Field Spectroscopy of the HIP 79977 Debris Disk}\label{chapter:hip79977}

Note: this chapter has been submitted to AJ with co-authors Thayne Currie, Olivier Guyon, 
Timothy D. Brandt, Tyler D. Groff, Nemanja Jovanovic, N. Jeremy Kasdin, Julien Lozi, 
Klaus Hodapp, Frantz Martinache, Carol Grady, Masa Hayashi, Jungmi Kwon, Michael W. McElwain,
Yi Yang, and Motohide Tamura. A formal reference and DOI for this paper is not yet available.

\section*{Abstract}
We present new, near-infrared (1.1--2.4 $\micron$) high-contrast imaging of the bright debris disk surrounding HIP 79977 with the Subaru Coronagraphic Extreme Adaptive Optics system (SCExAO) coupled with the CHARIS integral field spectrograph.  SCExAO/CHARIS resolves the disk down to smaller angular separations of (0\farcs{}11; $r \sim 14$ au) and at a higher significance than previously achieved at the same wavelengths. The disk exhibits a marginally significant east-west brightness asymmetry in $H$ band that requires confirmation.   
Geometrical modeling suggests a nearly edge-on disk viewed at a position angle of $\sim$ 114.6$\arcdeg$ east of north. 
The disk is best-fit by scattered-light models assuming
strongly forward-scattering grains ($g$ $\sim$ 0.5--0.65) confined to a torus with a peak density at $r_{0}$ $\sim$ 53--75 au. We find that a shallow outer density power law of $\alpha_{out}=$-1-- -3
and flare index of $\beta = 1$ are preferred. Other disk parameters (e.g.~inner density power law and vertical scale height) are more poorly constrained. 
The disk has a grey to slightly blue intrinsic color and its profile is broadly consistent with predictions from birth ring models applied to other debris disks. 
While HIP 79977's disk appears to be more strongly forward-scattering than most resolved disks surrounding 5--30 Myr-old stars, this difference may be due to observational biases favoring forward-scattering models for inclined disks vs. lower inclination, ostensibly neutral-scattering disks like HR 4796A's.
Deeper, higher signal-to-noise SCExAO/CHARIS data 
can better constrain the disk's dust composition.

\section{Introduction} \label{sec:ch5intro}
Debris disks around young stars are signposts of massive planets \citep[e.g.][]{Marois2008a,Lagrange2010} 
and critical reference points for understanding the structure, chemistry, and evolution of the Kuiper belt \citep{Wyatt2008}. 
Debris disks may be made visible by recently-formed icy Pluto-sized objects stirring and causing collisions between surrounding boulder-sized icy planetesimals.   The luminosity
distribution of debris disks over a range of ages then traces the evolution of debris produced by icy planet formation.
\citep{Currie2008, Kenyon2008}).   Similarly, massive jovian planets  
may create gaps in some of these debris disks and sculpt the distribution of their icy planetesimals \citep{Mustill2009}.   

Resolved imaging of debris disks in scattered light has revealed dust sculpted in morphologies ranging from diffuse structures or 
extended torii to sharp rings;
disks exhibited scattering properties ranging from neutral to strongly forward 
scattering \citep[e.g.][]{SmithTerrile1984,Schneider1999,Schneider2005,Schneider2009,Kalas2005,Kalas2006,Kalas2007hd15745,Soummer2014, Currie2015a, Currie2017a}.  Furthermore, multi-wavelength imaging and spectroscopy of debris disks in scattered light provide further insights into the nature of debris disk properties.   The differing grain properties of debris disks can result in a spread in intrinsic disk colors from blue ~\citep[e.g. HD 15115,][]{Kalas2007}, where dust is reflecting light more efficiently at shorter wavelengths compared to what it receives from the star, to red~\citep[e.g. $\beta$ Pic,][]{Golimowski2006}. Detailed photometric color characterization provides insights into grain properties, and low-resolution spectroscopy (even as low as R $\sim$ 10) probes the presence of ices and organics \citep[e.g.][]{Debes2008,Rodigas2014,Currie2015a}.  

Extreme adaptive optics (ExAO) systems coupled with integral field spectrographs improve the ability to detect and characterize debris disks, especially at small angles.   For example, resolved imaging and spectroscopy of the HD 115600 debris disk with the \textit{Gemini Planet Imager}, the first object discovered with ExAO, revealed a sharp ring at $r$ $\lesssim$ 0\farcs{}5, modeling for which suggested neutral-scattering and possibly icy dust and a pericenter offset caused by a hidden jovian planet \citep{Currie2015a}.    \citet{Milli2017} resolved the well-known HR 4796A disk at far smaller angular separations than done previously.  They showed that a seemingly neutral-scattering dust ring has a strong forward-scattering peak at small angles, inconsistent with a single Henyey-Greenstein-like scattering function.  Resolved imaging and spectroscopy over a longer wavelength baseline enables better constraints on the properties of other debris disks \citep[e.g.][]{Rodigas2015, Milli2017}.

HIP 79977 is another young star whose debris disk can better understood using multi-wavelength imaging and spectroscopy with ExAO.  This is an F2/3V star (1.5 $M_{\sun}$) located $131.5 \pm 0.9$ pc away~\citep{Gaia2018} in the $\sim$ 10 Myr-old
Upper Scorpius association \citep{Pecaut2012}. Its infrared excess was detected by the $IRAS$ satellite, and the Spitzer Multiband Imaging Photometer associated it with a debris disk~\citep{Chen2011}. \citet{Thalmann2013} used Subaru's facility (conventional) AO188 adaptive optics system and the HiCIAO instrument at H band and produced the first resolved images of its debris disk. They revealed that it was viewed nearly edge-on 
($i=84^{+2}_{-3}\arcdeg$)
and had tangential linear polarization varying from $\sim$ 10$\%$ at 0\farcs{}5 to $\sim$ 45$\%$ at
1\farcs{}5. \citet{Engler2017} performed the first ExAO characterization of HIP 79977, observing it at visible 
wavelengths ($\lambda_c = 735$ nm, $\Delta \lambda = 290$ nm) using the SPHERE-ZIMPOL polarimeter. They
measured a polarized flux contrast ratio for the disk of 
$(F_{\rm pol})_{\rm disk}/F_{\star} = (5.5 \pm 0.9) \times 10^{-4}$ in that band and an increase in the thickness
of the disk at larger radii, which they explained by the blow-out of small grains by stellar winds.

These previous studies showed tension in some derived debris disk properties (e.g. the disk radius) and allowed a wide range of parameter space for others (e.g. the disk scattering properties).  No substellar companions were decisively detected in either publication.  However, \citet{Thalmann2013} did find a marginally significant point-like residual emission in their reduced image after subtracting a model of the debris disk's emission.

In this paper, we present the first near-IR resolved ExAO images of the HIP 79977 debris disk, using the Subaru Coronagraphic Extreme Adaptive Optics systems coupled with the CHARIS integral field spectrosgraph.   SCExAO/CHARIS data probe  inner working angles (0\farcs{}15--0\farcs{}2) comparable to those from SPHERE polarimetry reported in~\citet{Engler2017} and significantly smaller than that presented in~\citet{Thalmann2013}. Additionally, we present the first near-IR color analysis of the disk. 

The paper is 
organized as follows: in Section~\ref{sec:obs}, we describe the observations and the pipeline through
which the data was reduced and then PSF-subtracted; in Section~\ref{sec:geometry}, we describe the
basic morphology of the disk; then, in Section~\ref{sec:methodology} we discuss the process through which
we generated synthetic disks and propagated them through the same pipeline as the actual data in order 
to understand how the PSF subtraction attenuated the disk features; we provide the results of this
forward modeling in Section~\ref{sec:results}; finally, we describe the $J$-, $H$-, and $K_p$-band colors of the 
disk.

\section{SCExAO/CHARIS Data}\label{sec:obs}
\subsection{Observations and Data Reduction}
We targeted HIP 79977 on UT 14 August 2017 (Program ID S17B-093, PI T. Currie)
with Subaru Telescope's SCExAO~\citep{Jovanovic2015} instrument coupled
to the CHARIS integral field spectrograph, which operated in low-resolution ($R\sim 20$), 
broadband (1.13--2.39 $\micron$) mode \citep{Peters2012,Groff2013}.
SCExAO/CHARIS data were obtained using the Lyot coronagraph with the 217 mas diameter occulting spot. 
Satellite spots, attenuated copies of the stellar PSF, were generated by placing
a checkerboard pattern on the deformable mirror with a 50 nm amplitude and alternating its
phase between $0\arcdeg$ and $180\arcdeg$~\citep{Jovanovic2015b}. These spots were used for image
registration and spectrophotometric
calibration; their intensity relative to the star\footnote{The spot intensity calibration
changed following the observations described in this paper, so this equation may
not match what is provided elsewhere.} was given by
\begin{equation}
I_{spots}/I_{\star} = 4 \times 10^{-3} (\lambda/1.55 \; \micron)^{-2} .
\label{eqn:satspots}
\end{equation}
Exposures consisted of 86 co-added 60 $s$
frames (82 science frames, 4 sky frames) obtained over 92 minutes and covering a total parallactic angle rotation of $26.7\arcdeg$. Conditions were excellent; seeing was 0\farcs{}35-0\farcs{}40 at 0.5 $\micron$ and the wind speed was 3 m s$^{-1}$.
Although we did not obtain a real-time estimate of the Strehl ratio, the raw contrasts at $r$ $\sim$ 0\farcs{}2--0\farcs{}75 later estimated from spectrophotometrically calibrated data were characteristic of those 
obtained with $H$-band Strehls of 70--80\% \citep{Currie2018a}.  

 \begin{figure}[!ht]
 \begin{centering}
 \includegraphics[width=14cm]{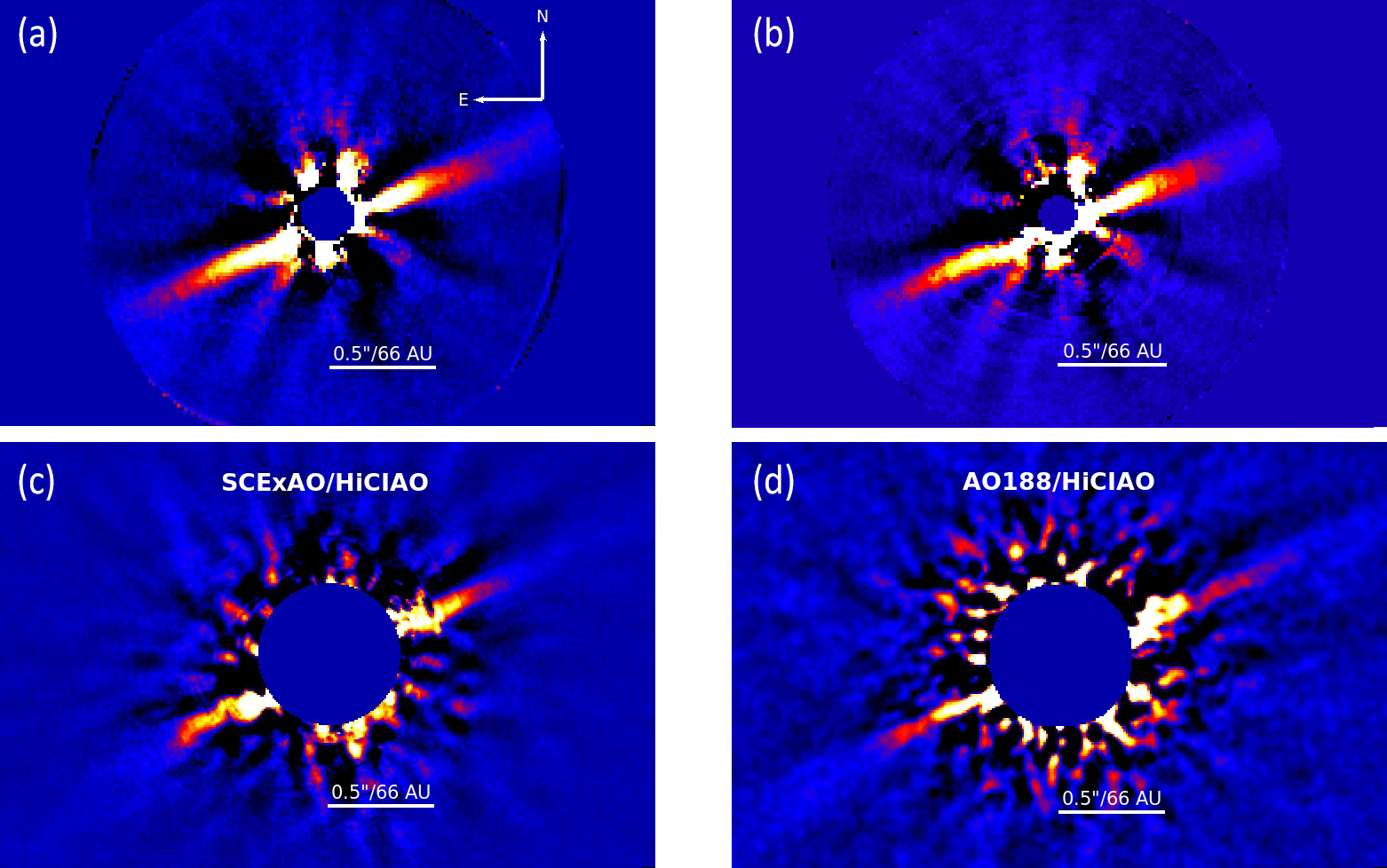}
 \caption
 [Improvement in observations of HIP79977 with time]
 {Illustrated here are the three different NIR datasets for HIP 79977.  The upper panels (a and b) are our 
 paper's main focus and
 show wavelength-collapsed images produced by two different KLIP-ADI reductions of the SCExAO/CHARIS data.
 Figure~\ref{fig:diskcomparison}c shows July 2016 $H$ band data from SCExAO + HiCIAO reduced using A-LOCI with local masking~\citep{Currie2012} 
 and has stronger residuals exterior to 0\farcs{}3--0\farcs{}4. 
 Finally, Figure~\ref{fig:diskcomparison}d shows the data published by~\citet{Thalmann2013},
 which were produced using the (non-extreme) AO188 and HiCIAO at $H$ band.  The data were processed using the ACORNS-ADI reduction package~\citep{Brandt2013}.
 The four images have the same intensity scaling. The circular region in the bottom
 two plots denotes the field of view of the CHARIS data.
 }
 \label{fig:diskcomparison}
 \end{centering}
 \end{figure}

To convert raw CHARIS files into data cubes, we employed the CHARIS Data Reduction Pipeline \citep[CHARIS DRP,][]{Brandt2017}.  After generating a wavelength solution from monochromatic ($\lambda_{0}$ = 1.550 $\micron$) lenslet flats, the pipeline extracted data cubes using the least squares method described by \citet{Brandt2017}, yielding a nominal spaxel scale of 0\farcs{}0164 and $\sim$ 1\farcs{}05 radius field of view.  
Subsequent processing steps -- e.g. image registration and spectrophotoemtric calibration -- followed those from \citet{Currie2018a}.

For PSF subtraction, we utilized the KLIP algorithm \citep{Soummer2012} in ADI-only mode as employed in 
\citet{Currie2014hd100546,Currie2017b}, where PSF subtraction is performed in annular regions with a 
rotation gap to limit signal loss from self-subtraction of astrophysical sources.   Key algorithm 
parameters -- the width of annulus over which PSF subtraction is performed ($\Delta$r), the rotation gap ($\delta$), the number of principal components ($N_{\rm pc}$) -- were varied to explore which combination maximized the total SNR of the disk in sequence-combined, wavelength-collapsed images.  While the detection of the HIP 79977 debris disk was robust across the entire range of parameter space, the signal to noise of the spine of the disk was maximized with a setting with $\Delta$r = 2 pixels, $N_{\rm pc} =2$, and $\delta = 1.0$ full width half maxima (FWHM) and then merging
the wavelength channels using a robust mean with outlier rejection instead of a median combination.  As described later, for computational efficiency and simplicity, we performed a second reduction with a larger annular width of $\Delta$r = 6 pixels ($\sim 2.5 \lambda$/D at 1.55 $\micron$).  Reductions retaining a slightly different number of principal components or value for the rotation gap yielded comparable results.  
 
\subsection{Detection of the HIP 79977 Debris Disk}
Figures~\ref{fig:diskcomparison}a and~\ref{fig:diskcomparison}b show the results of these two reductions of the
CHARIS data. Figures~\ref{fig:diskcomparison}c and~\ref{fig:diskcomparison}d contextualize the performance gain of SCExAO/CHARIS compared to earlier
observations. The disk is plainly visible down to an inner working angle of 
0\farcs{}11 in~\ref{fig:diskcomparison}a and~\ref{fig:diskcomparison}b.
Figure~\ref{fig:diskcomparison}c shows data collected on UT 17 July 2016 (Program UH-12B, PI K. Hodapp) using SCExAO (suboptimally tuned
and providing lower Strehl than that of the recent data) and the HiCIAO instrument at $H$ band.
Although the July 2016 SCExAO/HiCIAO image has a larger
field of view than the SCExAO/CHARIS image, it exhibits far stronger residuals interior to about 0\farcs{}3--0\farcs{}5.  
Figure~\ref{fig:diskcomparison}d shows the AO188~\citep[Subaru's facility AO system,][]{minowa10} + HiCIAO data previously published
by~\citet{Thalmann2013}, and this has even stronger residuals, particularly at smaller angular separations,
due to its much poorer AO correction.

\begin{figure}[!ht]
 \begin{centering}
 \includegraphics[width=1.0\textwidth]{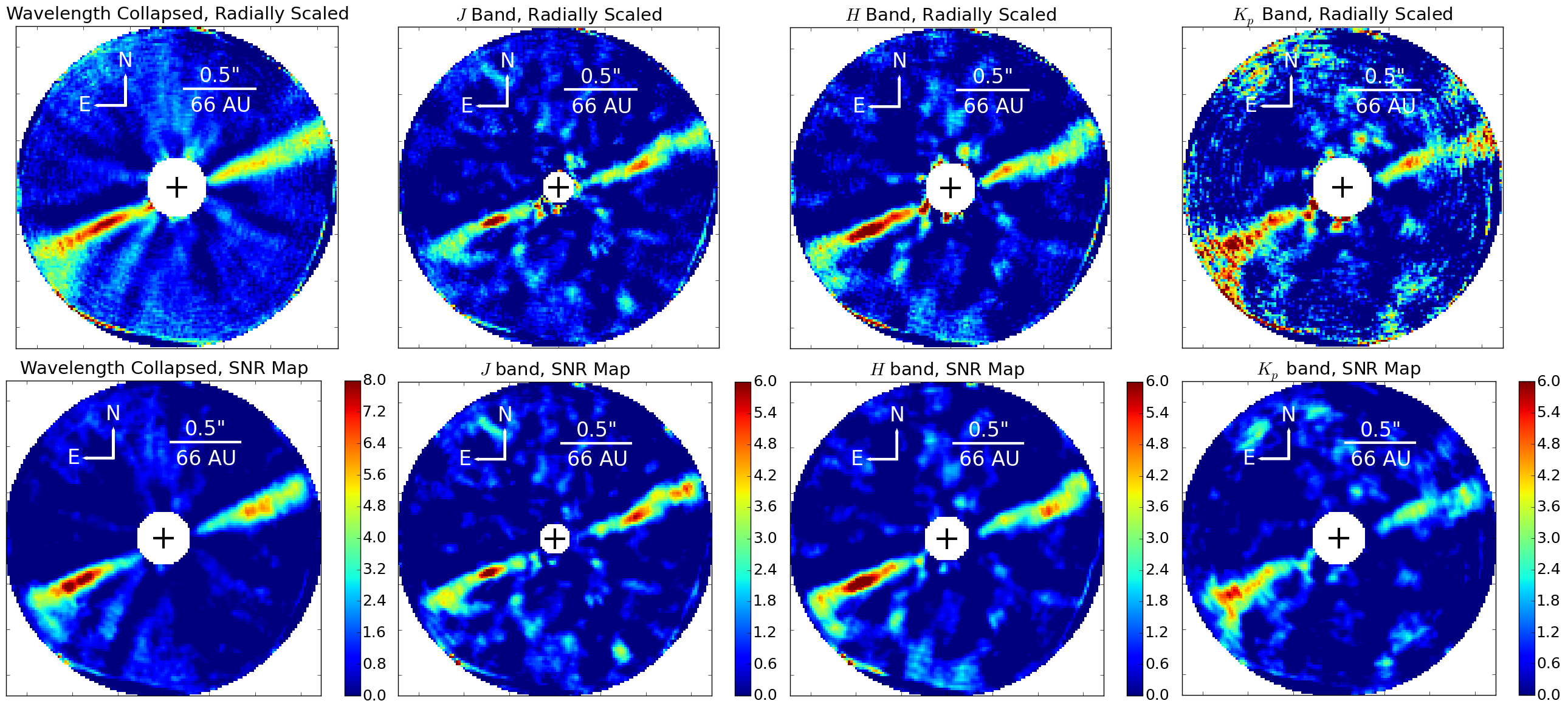}
 \caption
 [$J$-, $H$-, and $K_p$-band images and SNR maps of the HIP 79977 disk]
 {Shown here are flux images following KLIP PSF-subtraction (top) and the corresponding 
 signal-to-noise per resolution element
 maps (bottom). The CHARIS low-resolution mode produces data cubes with 22 spectral layers. We coadded all the layers (left) and the bands corresponding to (proceeding rightward) $J$, $H$, and $K_p$ bands. The flux images have
 arbitrary units and have been multiplied by an $r^2$ map in order to reveal structure away from the star.   
 The images presented here are rotated relative to those in Figure~\ref{fig:diskcomparison}.}
 \label{fig:combineddiskimages}
 \end{centering}
 \end{figure}

Figure~\ref{fig:combineddiskimages} shows the sequence-combined, wavelength-collapsed disk image scaled by the stellocentric distance squared, and analogous images obtained from combining channels covering the $J$ (channels 1--5; 1.16--1.33 $\mu$m), $H$ (channels 8--14; 1.47--1.80 $\mu$m), and $K_{p}$ (channels 16--21; 1.93--2.29$\mu$m) passbands. This image used the first set of KLIP parameters described above.
The disk is plainly visible in each image.  We computed the signal-to-noise per resolution element using the standard practice of replacing each pixel with the sum within a FWHM-sized aperture, computing the radial profile of the robust standard deviation of this summed image in the wavelength-collapsed image, dividing the two images, and correcting for small sample statistics \citep{Currie2011a}.   The disk is detected at a SNR/resolution element (SNRE) $>$ 3 exterior to 0\farcs{}25 and peaks at SNRE $\sim$ 9.1, 8, 9.1, and 5.8 in the broadband, $J$, $H$, and $K_{p}$ images, respectively\footnote{We achieved comparable results using a different algorithm, A-LOCI, using local masking as implemented in \citet{Currie2012,Currie2017a}.}.   These estimates are conservative as we do not mask the disk signal when computing the noise profile.  For our second reduction the SNRE values along the disk spine are slightly smaller at small angles but otherwise comparable, peaking at 9.6, 9, 8.4, and 5.6 in the broadband, $J$, $H$, and $K_{p}$ images, respectively.
 
For both reductions, the final images and SNR maps may reveal some evidence for a wavelength dependent brightness asymmetry between the eastern and western sides.  In the wavelength-collapsed image, the eastern
side of the disk appears about 50\% brighter and is detected at a higher significance ($\sim 8-9$ $\sigma$ vs. 
$5.5-6.5$ $\sigma$ along the disk spine beyond 0\farcs{}5).   From comparing images obtained over different passbands,  $H$ and $K_{p}$ band seem to be responsible for most of this brightness asymmetry.   

\section{Geometry of the HIP 79977 Debris Disk}\label{sec:geometry}
Our images clearly trace the major axis of the HIP 79977 debris disk.   To estimate the disk's position angle, we follow previous analysis performed for HD 36546 \citep{Currie2017a} and for $\beta$ Pic \citep{Lagrange2012}, determining the trace of the disk spine from the peak brightness as a function of separation (``maximum spine'' fitting) and from fitting a Lorentzian profile.   Our procedure used the \textit{mpfitellipse} package to estimate the disk spine from disk regions between 0\farcs{}15 and 0\farcs{}75, where the pixels are weighted by their SNRE, and explored a range of thresholds in SNRE (0--3) to define the spine.   

Precise astrometric calibration for CHARIS is ongoing and preliminary results will be described in full in a separate early-science paper focused on $\kappa$ Andromedae b (Currie et al. 2018, in prep.).   Briefly, we obtained near-infrared data for HD 1160 from SCExAO/CHARIS in September 2017 and Keck/NIRC2 in December 2017.  At a projected separation of $r$ $\sim$ 80 au, the low-mass companion HD 1160 B should not experience significant orbital motion \citep{Nielsen2012, Garcia2017}; Keck/NIRC2 is precisely calibrated, with a north position angle uncertainty of 0.02$^{o}$ and post-distortion corrected astrometric uncertainty of 0.5 mas \citep{Service2016}.   Thus, we pinned the SCExAO/CHARIS astrometry for HD 1160 B to that for Keck/NIRC2 to calibrate CHARIS's pixel scale and north position angle offset.   These steps yielded a north PA offset of $\sim -2.2\arcdeg$ east of north and a revised pixel scale of $\sim$ 0\farcs{}0162.  While the differences between the default and revised pixel scale lead to astrometric offsets are inconsequential for this paper (10 mas near the edge of CHARIS's field of view), the north position angle (PA) offset for CHARIS is necessary for an accurate estimate of the position angle for the disk's major axis.

After considering CHARIS's north PA offset, 
Lorentzian profile fitting yields a position angle of $114.59\arcdeg \pm 0.40\arcdeg$.   ``Maximum spine" fitting yields nearly identical results but with larger error bars:
$114.74\arcdeg \pm 1.88\arcdeg$.   These values are consistent with previous estimates from \citet{Engler2017} and \citet{Thalmann2013}.   For the rest of the paper, we adopt a position angle of $114.6\arcdeg$. 

\section{Modeling of the HIP 79977 Debris Disk}\label{sec:forwardmodeling}
\subsection{Methodology}\label{sec:methodology}
\subsubsection{Forward-Modeling of the Annealed Disk Due to PSF Subtraction}
To assess the morphology of the HIP 79977 debris disk, we forward-modeled synthetic disk images spanning a range of properties through empty data cubes, using the same eigenvalues and eigenvectors used in the reduction of our on-sky data \citep[e.g.][]{Soummer2012,Pueyo2016}.  Our specific implementation, following the formalism in \citet{Pueyo2016}, is described and justified in detail below.

The residual signal of a planet or disk in a target image with spatial dimensions x and an intrinsic signal \textbf{A(x)} after KLIP processing is nominally equal to the astrophysical signal in the target image minus its projection on the KLIP basis set constructed from references images from up to k = 1 $\cdots$ $K_{\rm klip}$ principal components, \textit{$Z_{k}$}:
\begin{equation}
P_{\rm residual,n} = A(x_{n}) - \Big(\sum_{k=1}^{K_{\rm klip}}<A(x_{n}),Z_{k}^{KL}> Z_{k}^{KL}(n)\Big)
\end{equation}

Here, \textbf{Z$_{k}$$^{KL}$} is the Karhunen-Loe\'ve transform of the reference image library \textbf{R} 
with eigenvalues \textbf{$\Lambda_{k}$} and eigenvectors
\textbf{$\nu_{k}$}:

\begin{equation}
Z_{k}^{KL}(x) = \frac{1}{\sqrt{\Lambda_{k}}}\sum_{m=1}^{K_{\rm klip}}\nu_{k}R_{m}(x).
\end{equation} 
When the astrophysical signal in a given image is not contained in reference images used for subtraction, then annealing is entirely due to \textit{oversubtraction}: confusion of the astrophysical signal with speckles.   
As described in \citet{Pueyo2016}, however, the presence of an astrophysical signal in the reference image library itself causes \textit{self-subtraction} of the source in the target image and perturbs \textbf{Z$_{k}$$^{KL}$}.  

In performing forward-modeling, we consider both oversubtraction and direct self-subtraction.
\citet{Pueyo2016} delineate regimes where oversubtraction and two different regimes of self-subtraction (direct and indirect) dominate.  For small $K_{\rm klip}$ values and  an astrophysical signal is small compared to the speckles over the region where principal component analysis is performed, oversubtraction dominates and a straightforward application of Equation 2 describes annealing.   For intermediate $K_{\rm klip}$ values, \textit{\underline{direct} self-subtraction} dominates, scaling linearly with the astrophysical signal and inversely with the square-root of the unperturbed eigenvalues: $\epsilon$/$\sqrt{\Lambda_{k}}$.  For large $K_{\rm klip}$, closer to a full-rank covariance matrix, \textit{\underline{indirect} self-subtraction} dominates and is inversely proportional to the eigenvalues: $\epsilon$/${\Lambda_{k}}$.   Other algorithm settings (e.g. using a larger rotation gap) can remove more astrophysical signal from the reference library, reducing the influence of self-subtraction.

These arguments and previous measurements of the HIP 79977 disk help identify the important biases/sources of annealing for our HIP 79977 data set.  In our reductions, the number of removed KL modes is small compared to the size of the reference library 
(K$_{\rm klip} \ll$ $N_{\rm images/channel}$).  In most channels, the disk is $\approx$ 5\% of the brightness of the local speckles.  Furthermore, we perform PSF subtraction in annular regions.  Over the angular separations modeled (0\farcs{}16--0\farcs{}75), results from \citet[][see their Fig. 6b]{Engler2017} imply that the nearly edge-on disk is present in no more than 10--20\% of the pixels at each angular separation.    Our rotation gap criterion (1 PSF footprint) further reduces self-subtraction.   As a result, the perturbed KL modes $\Delta$KL are far smaller than the unperturbed ones dominated by signal from the speckles: the indirect self-subtraction term is negligible.
Thus, in performing forward-modeling we consider oversubtraction (the dominant contribution) and direct self-subtraction (a secondary contribution) only.   

\subsubsection{Scattered Light Disk Models}
\begin{figure}[!ht]
 \begin{centering}
 \includegraphics[width=0.6\textwidth]{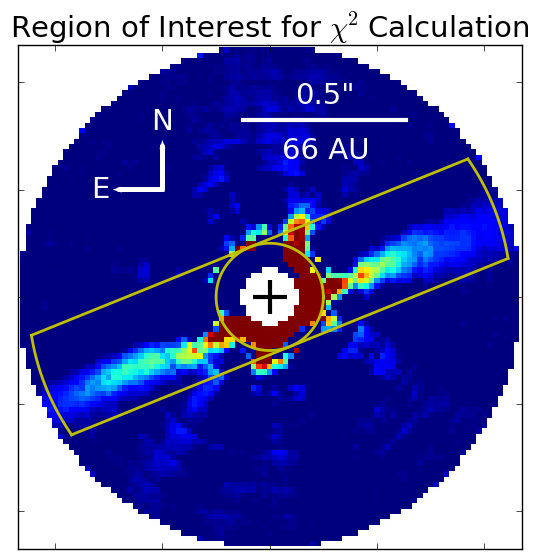}
 \caption
 [Region of interest used in forward modeling $\chi^2$ calculation]
 {The region bounded by the yellow lines was used for scaling the PSF-subtracted
 synthetic model disks and then computing their $\chi^2$ residuals relative to the
 on-sky data. The outer boundary is defined by the intersection of a rectangular box that
 is 100 pixels by 20 pixels where the major axis is rotated $22\arcdeg$ north of west and 
 a circle of radius $r=45$ pixels. The inner region is a circle of radius $r=10$ pixels.
 The disk in this figure is plotted from the same data as that used in 
 Figure~\ref{fig:combineddiskimages}, but it 
 has not been multiplied by an $r^2$ map.}
 \label{fig:roi}
 \end{centering}
 \end{figure}

Synthetic scattered light disk models were drawn from the GRaTeR code developed in \citet{Augereau1999}, convolved with the SCExAO/CHARIS instrumental PSF, and inserted into empty data cubes with the same position angles as the real data.  We then forward-modeled the annealing of each model disk due to KLIP PSF subtraction as described above and compared the wavelength-collapsed image of the residual disk model to the real data.  The fidelity of each model disk to the data is determined in the subtraction residuals in binned (by the instrument PSF size) pixels ($\sim$ 0\farcs{}04) over a region of interest defining the trace of the disk and any self-subtraction footprints (see Figure~\ref{fig:roi}).
This evaluation region encloses 237 binned pixels (N$_{\rm data}$). 
 
The set of acceptably-fitting solutions have chi-squared values of 
$\chi^{2} \le \chi^{2}_{\rm min} + \sqrt{2 N_{\rm data}}$ \citep[see][]{Thalmann2013}.  
At the 95\% confidence limit, this criterion equals $\chi_{\nu}^{2} \lesssim 1.092$. 

Because we performed KLIP PSF subtraction in annuli (not the entire field at once) and in 22 wavelength channels (not single-channel camera data), exploring 10$^{6}$ models covering a large parameter space as in \citet{Engler2017} would be extremely computationally expensive.  Rather, we leverage on inspection of the SCExAO/CHARIS wavelength-collapsed final image, our disk geometry modeling, and previous results from \citet{Engler2017} to focus on a smaller parameter space range.   

Inspection of the final CHARIS image shows that the disk is detected only on the near side, out to an angular 
separation of 0\farcs{}5--0\farcs{}6 before gradually fading in brightness at wider separations.   
Our fitting to the geometry of the disk reaffirms a position angle of 114.6$^{o}$.  Thus, our parameter space 
generally explores disks with moderate to strong forward-scattering, a sharp inner cutoff to the belt, and a shallower decay in dust density beyond the fiducial radius.

We varied six parameters in our search for the disk that best reproduced the on-sky data.
First, the Henyey-Greenstein parameter~\citep{Henyey1941} probes the visible extent of the dust's phase scattering function.
While it lacks a pure physical motivation and is known to fail at very small scattering angles for at least some debris disks \citep{Milli2017}\footnote{These angles correspond to the semimajor axis of the HIP 79977 debris disk and are inaccessible with our data.}, it is widely adopted in debris disk modeling literature and thus helps cast our results within the context established by other debris disks.   The H-G parameter ranges from $-1$ to $1$;
$g=0$ corresponds to neutral scattering, $g=-1$ indicates perfect backward scattering, and $g=1$ 
indicates dust that scatters light solely forward.

Second, we varied values of the fiducial radius $r_0$ of the disk, inside of which $\alpha_{in}$ ($\alpha_{in} > 0$)
describes the power law for the increase in dust particle number density and outside
of which $\alpha_{out}$ ($\alpha_{out} < 0$) describes the power law for its decrease.
These three variables, which were the second through fourth fitted parameters,
combine to give the radial distribution profile $R(r)$:
\begin{equation}
R(r) = \left[ \Big( \frac{r}{r_0} \Big) ^{-2 \alpha_{in}} + 
  \Big( \frac{r}{r_0} \Big)^{-2 \alpha_{out}} \right]^{-1/2}
\end{equation}
where $r$ is the distance from the center of the disk. The vertical profile $Z(h)$ is given by
\begin{equation}
Z(h) = \textrm{exp} \left( \frac{-| h |}{H(r)} \right) ,
\end{equation}
where
$h$ is the distance above the disk midplane. $H(r)$ is the scale height at radius $r$ and is given by
\begin{equation}
H(r) = \xi \left( \frac{r}{r_0} \right)^\beta ,
\end{equation}
where $\xi$ is the scale height at $r_0$ and $\beta$ is the disk's flare index. $\xi$ and $\beta$ were
the fifth and sixth parameters in our grid search.

We tested models with $g=0.3-0.8$, corresponding to moderate to strong forward scattering. 
Based on visual estimates of the disk images,
we produced model disks with fiducial radii of $r_0 = 43-91$ au. The parameters $\alpha_{in}$ and $\alpha_{out}$ determine
the power laws for the inner and outer radial emission profiles, respectively, and we selected values that produced
disks with relatively sharp inner cutoffs and slow radial decays. We sampled disks with a scale height at
the fiducial radius in the range of $\xi=0.5-3.2$ au; values outside this range would not be consistent with the
self-subtracted images. 
We adopt our value for the disk position angle determined in Section~\ref{sec:geometry}.
and 
used our available computing resources to probe a greater variety of the other parameters.
Values outside these parameter ranges produced synthetic disks whose morphology differed greatly 
from the on-sky results.
The left two columns of Table~\ref{table:syndisks} list each parameter and the associated range in 
parameter space explored.  Our nominal search considered only circular disks and consisted of 20,480 
models.

\clearpage
\begin{landscape}
\begin{deluxetable}{lccc}
\tablecaption
{The grid of synthetic model disks used in our forward modeling. \label{table:syndisks}}
\tablecolumns{4}
\tablewidth{0pt}
\tablehead{
\colhead{Parameter} & \colhead{Values tested} & \colhead{Value for} & \colhead{Acceptably-fitting} 
\\
\colhead{} & \colhead{} & \colhead{best model} & \colhead{values} 
}
\startdata
 Radius of belt $r_0$ (au) & [43, 53, 64, 69, 75, 80, 86, 91]  & 53 & [53, 64, 69, 75] \\
 Inner radial index $\alpha_{in}$ & [3, 4, 5, 6] & 6 & [3, 4, 5, 6]   \\
 Outer radial index $\alpha_{out}$ & [-1, -1.5, -2, -2.5, -3, -3.5, -4.5, -5.5] & -1.5 & [-1, -1.5, -2, -2.5, -3]  \\
 Vertical scale height $\xi$ (au) & [0.5, 1.1, 1.6, 2.1, 3.2] & 3.2 & [0.5, 1.1, 1.6, 2.1, 3.2] \\
 Flare index $\beta$ & [1, 2] & 1 & [1, 2]   \\
 H-G parameter $g_{sca}$ & [0.3, 0.4, 0.5, 0.55, 0.6, 0.65, 0.7, 0.8] & 0.6 & [0.5, 0.55, 0.6, 0.65] \\
\enddata
\tablecomments{We assumed inclination $i=84.5\arcdeg$, eccentricity $e=0$, and position angle $
\theta=114.6\arcdeg$. $\xi$ and $r_0$ are not round numbers because they were initially chosen based
on the distance to HIP 79977 provided by~\citet{vanLeeuwen2007}, which was refined by~\citet{Gaia2018}, 
causing the scale to change by $\sim 7 \%$. If one value of a parameter fell below the
acceptably-fitting $\chi_{\nu}^{2}$ threshold for at least one model, it was included here.
Figure~\ref{fig:histograms} shows which parameters values most frequently produced acceptably-fitting
models.}
\end{deluxetable} 
\clearpage
\end{landscape}

\subsection{Results}\label{sec:results}
\begin{figure}
 \begin{centering}
 \includegraphics[width=0.54\textwidth]{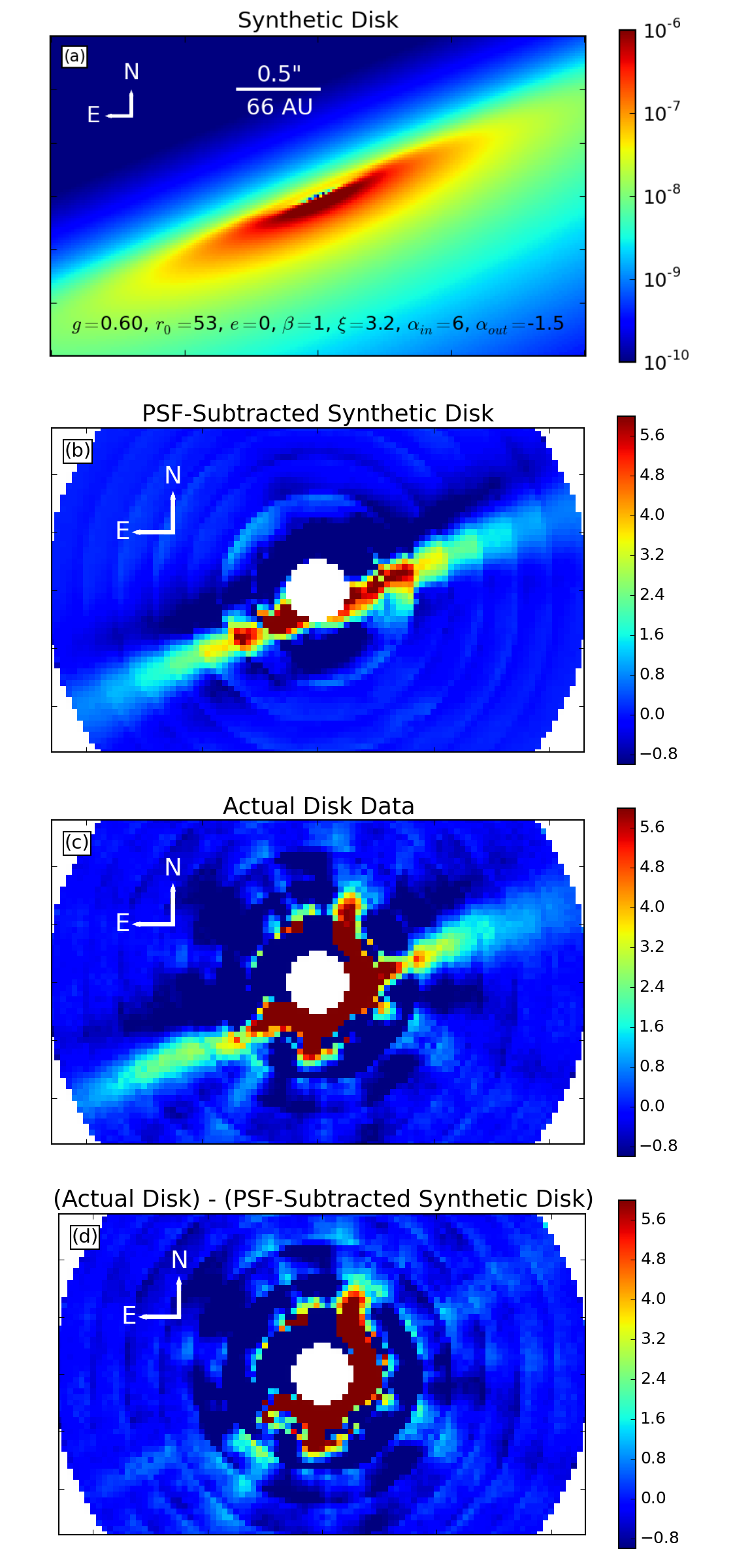}
 \caption
 [The best-fitting synthetic disk model]
 {From top to bottom are (a) the best-fitting synthetic disk; (b) that disk after
 it was convolved with the SCExAO PSF and then propagated through the KLIP PSF-subtraction
 using the same eigenvalues and eigenvectors as the on-sky data; (c) the wavelength-collapsed
 disk image (same as Figure~\ref{fig:diskcomparison}b) used in the $\chi^2$ comparison with 
 the synthetic model; and (d) the difference between panels (c) and (b). Minimal structure 
 remains in panel (d), indicating that the synthetic disk closely matches the actual data.
 The units are arbitrary. The distance scale is the same in all four panels.}
 \label{fig:bestsyndisk}
 \end{centering}
\end{figure}

Of the 20,480 synthetic disks, 132 produced residuals of $\chi^{2}_{\nu} \lesssim 1.092$ and 
therefore were acceptably fitting. 
The best model, which we defined as the model yielding the smallest 
$\chi^2_{\nu}$, produced $\chi^{2}_{\nu} = 1.000$.
The three panels of Figure~\ref{fig:bestsyndisk} show the best-fitting synthetic disk before 
and after PSF subtraction and the resulting residuals after it was subtracted from the on-sky 
data. This disk model
had $g=0.6$, indicating moderately strong forward scattering, a fiducial radius of $r=53$ au, a flare
index of $\beta=1$, a disk scale height at 
the fiducial radius of $\xi=3.2$ au, and dust emission with an inner power law of $\alpha_{in}=6$ 
and outer power law of $\alpha_{out}=-1.5$.

\begin{figure}[!ht]
 \begin{centering}
 \includegraphics[width=0.9\textwidth]{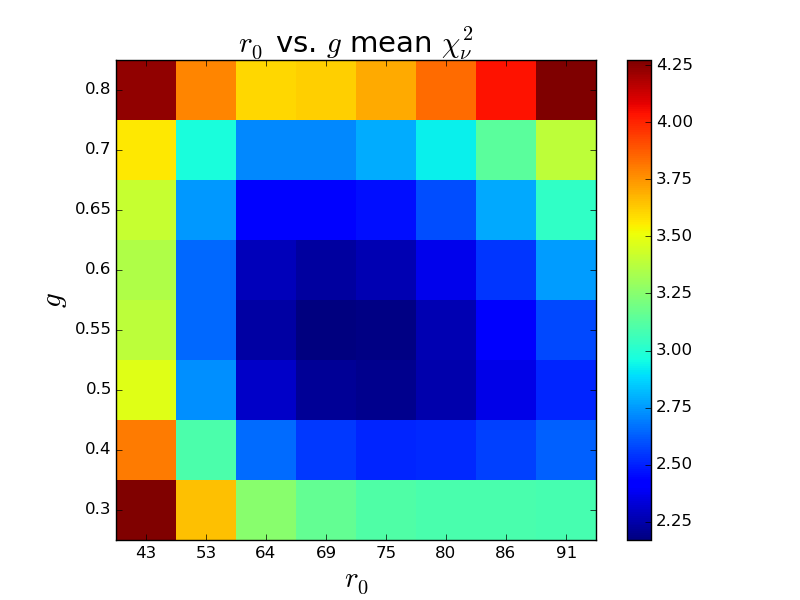}
 \caption
 [An example $\chi^2$ map]
 {Shown here is the mean $\chi^2_{\nu}$ for each value of $r_0$ and $g$. All values
 for the other parameters were included in the mean when calculating the value of each pixel. 
 We produced
 these maps for every variable against every other variable; this map is illustrative of the
 results. We used these maps to verify that our tested values adequately spanned the
 parameter space.   The region of parameter space minimizing $\chi^{2}$ is clear and well behaved.}
 \label{fig:contourmap}
 \end{centering}
\end{figure}

\begin{figure}[!ht]
 \begin{centering} 
 \includegraphics[width=0.75\textwidth]{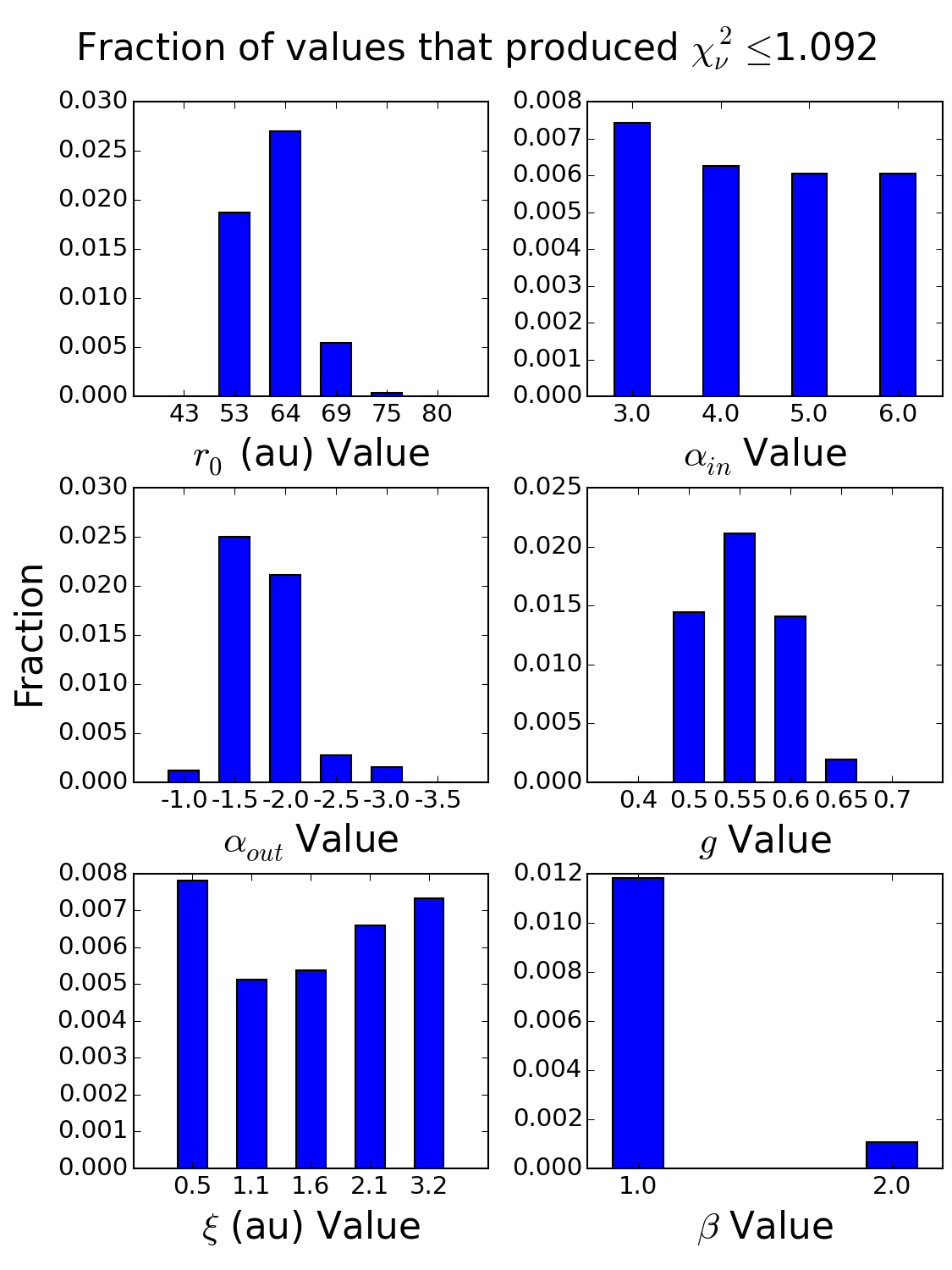}
 \caption
 [Histograms of well-fitting parameters.]
 {Each histogram bin contains the ratio of all models with that parameter value that
 produced an acceptably fitting $\chi_{\nu}^2$ compared to the number of models with that parameter
 value. The average of the bin heights in each plot is $132/20480 \approx 0.0064$. Some values with zero well-fitting disks have not been plotted in order to improve readability.}
 \label{fig:histograms}
 \end{centering}
\end{figure}

The range of parameters covered by 
the acceptably-fitting models is summarized in the fourth column of Table~\ref{table:syndisks}.  We 
produced contour maps of the average fit quality for every value of every parameter against every value of
every other parameter. An example map, showing the average $\chi^{2}_{\nu}$ for each value of $r_0$ and
$g$ averaged across the other parameters, is shown in Figure~\ref{fig:contourmap}. These maps helped
us ensure that we were sampling a reasonable range of values for each parameter.
Additionally, histograms of the parameter values that produced these acceptably-fitting models are shown
in Figure~\ref{fig:histograms}.  


Our modeling yielded improved constraints on the disk's radius and its scattering properties. As
shown in Figure~\ref{fig:contourmap}, there is a clear minimum in $\chi^2_{\nu}$ around
$g \approx 0.55$ and $r_0 \approx 70$ au. As shown in Figure~\ref{fig:histograms}, the family
of acceptably-fitting solutions has a small spread around these values. Our contour plots showed a 
strong preference for $\beta = 1$, indicating that the disk has low flaring.

On the other hand, the acceptably-fitting models covered the full range of considered
values of $\alpha_{in}$, indicating that $\alpha_{in}$ is not further constrained by our model fitting
beyond what was done in \citep{Engler2017}. This is likely because there was minimal disk available
between the inner working angle and the fiducial radius for the $\alpha_{in}$ fitting to occur.

We find numerically a good match between the wavelength-collapsed image and forward-modeled non-eccentric 
disk models, which show no brightness asymmetry. However, as evidenced by 
Figure~\ref{fig:combineddiskimages}, the HIP 79977 disk appears to exhibit asymmetrical
brightness. The east side of the disk is clearly brighter than the west side in $H$ band, 
and less clearly so in others. 
This brightness asymmetry may also be present
in SCExAO/HiCIAO $H$ band data from 2016 (Figure~\ref{fig:diskcomparison}c). 
This suggests that it may not be an artifact of the
data or processing. Plausible causes of the disk asymmetry are discussed in 
Section~\ref{sec:discussion}.

\section{HIP 79977 Disk Surface Brightness Profile and Colors}\label{sec:colors}
Next, we computed the surface brightness profile of the HIP 79977 disk in the $J$, $H$, and $K_p$ bands. We
began by using the satellite speckles (the PSF core was hidden by a coronagraph, but 
the flux of the satellite speckles was given by Equation~\ref{eqn:satspots}) and knowledge of the star's 
spectral type to spectrophotometrically calibrate the data cube. Second, we rotated the 
image so that the disk's spine was approximately horizontal and then fitted modified Gaussian
functions along the disk in order to find the spine's location with greater precision. We
then fit a fourth-order polynomial to these positions in order to smooth them and used
this fit as the trace of the disk in the subsequent steps.
Next, we merged the appropriate spectral channels to produce images equivalent to $J$, $H$, and $K_p$ bands 
and calculated a nominal surface brightness in each band along the disk's spine at radial intervals
of one PSF footprint. Uncertainties were calculated using the technique described in Section~\ref{sec:obs}. We 
divided the post-PSF-subtraction best-fitting synthetic model 
disk by the pre-PSF-subtraction version in order to produce a map of the attenuation that occurred
during the PSF subtraction. The PSF subtraction attenuated the disk spine by typically $15-20\%$,
and the attenuation increased with vertical displacement from the disk. Finally, we scaled the 
nominal surface brightnesses by to the attenuation map. 

\begin{figure}[!ht]
 \begin{centering}
 \includegraphics[width=0.7\textwidth]{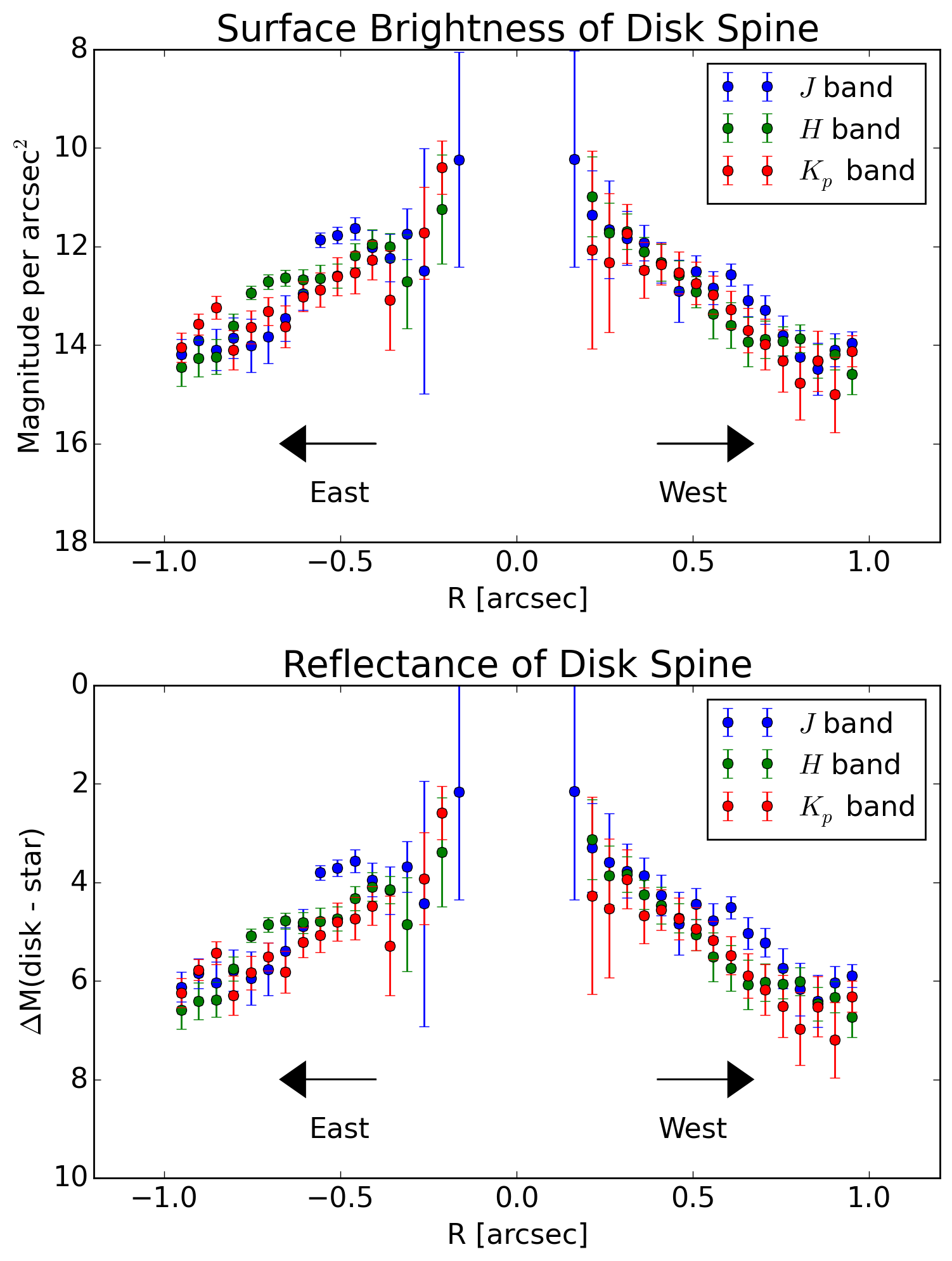}
 \caption
 [Surface brightness profiles]
 {The $J$, $H$, and $K_p$ band surface brightnesses along the disk spine are shown in
 the top plot.
 In the lower plot, we have subtracted the flux of the star ($J=8.062, H=7.854, K=7.800$)
 in order to see the disk's colors after removal of the stellar color.
 The disk is largely gray, though slightly blue at some radial separations.
 The three bands plotted individually are shown in Figure~\ref{fig:sb2}.}
 \label{fig:surfacebrightness}
 \end{centering}
 \end{figure}

\begin{figure}[!ht]
 \begin{centering}
 \includegraphics[width=0.7\textwidth]{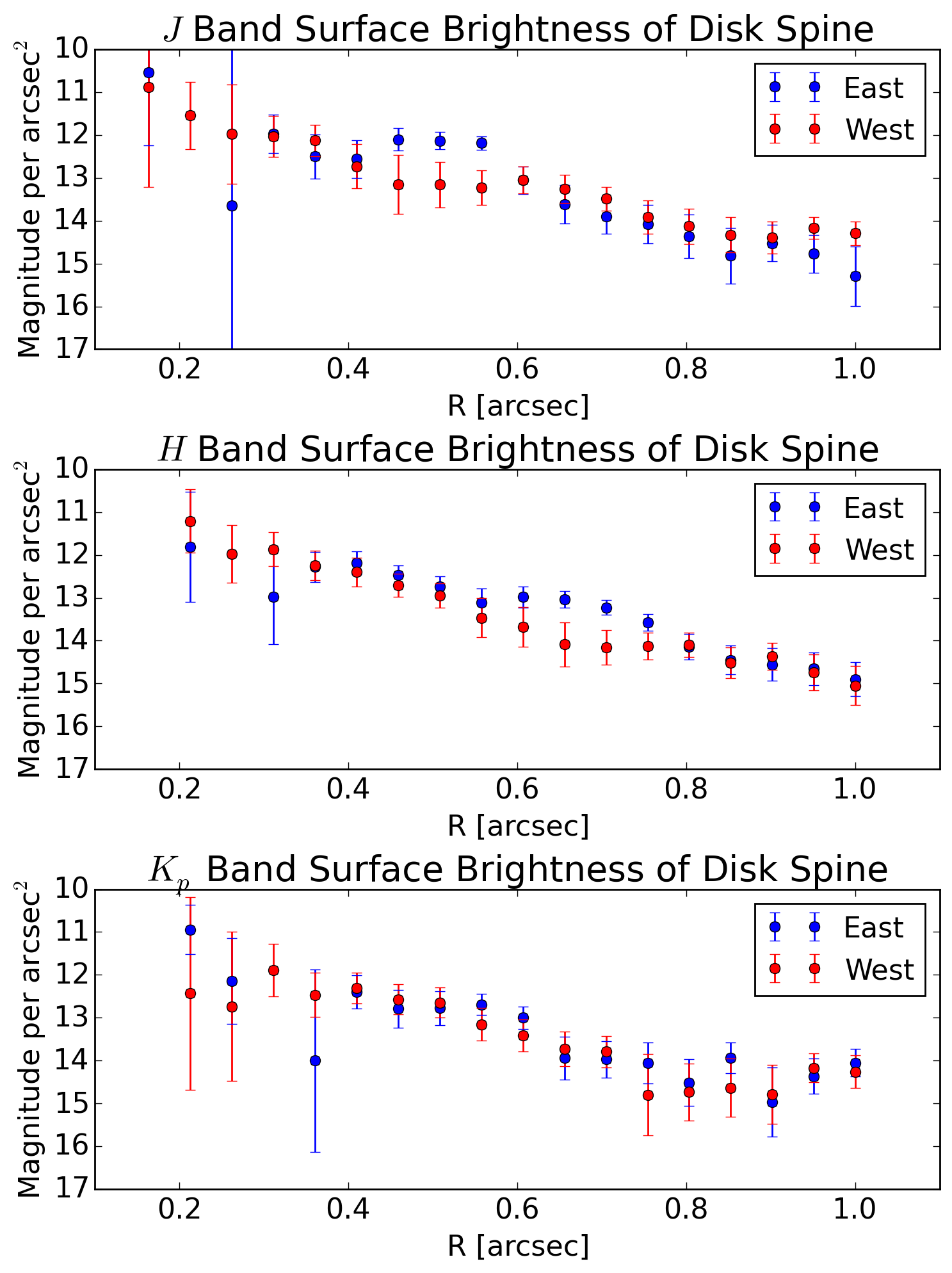}
 \caption
 [Surface brightness profiles corrected for star color]
 {From top to bottom are plots of the $J$, $H$, and $K_p$ band surface brightnesses
 of the disk. The brightness asymmetry of the east and west sides of the disk are visible
 in these plots, albeit at differing separations and significance.}
 \label{fig:sb2}
 \end{centering}
\end{figure}

Figures~\ref{fig:surfacebrightness} and~\ref{fig:sb2} show the surface brightnesses/reflectance on the
east and west sides of the disk for the three color bands. The uncertainties decrease
significantly at radial separations of $\gtrsim$ 0\farcs{}25. These measurements
extend the surface brightness measurements inward from those calculated by~\cite{Thalmann2013}.  
The reflectance of the disk is either comparable in all three bands, suggesting a grey color, or
is slightly blue. Figure~\ref{fig:sb2} clearly shows the excess $H$ band brightness of the east 
side of the disk compared to that of the west side. This asymmetry
appears present at $J$ band at a smaller inner separation and is marginal but
plausible at $K_p$ band at a larger separation.
The disk's surface brightness radial profile can be well fit with a power law with an exponential decay
term of $-4.1 \pm 0.4$.

\section{Discussion}\label{sec:discussion}
Our improved signal to noise and inner working angle compared to those of previous work
enabled us to better constrain HIP 79977's
disk parameters. Our fitted parameters agreed with those derived 
by~\citet{Engler2017} within $1 \sigma$ except for the fiducial radius, which differs by $1.4 \sigma$ (this
takes into account the different distance they assumed). While our picture of the disk qualitatively agrees with much of that from the discovery paper \citep{Thalmann2013}, we exclude some of the parameter space for dust scattering that they find (e.g. $g$ = 0.4) and find a larger disk radius than they adopted in their paper ($r_0=40$ au).

\begin{figure}[!ht]
 \begin{centering}
 \includegraphics[width=0.5\textwidth]{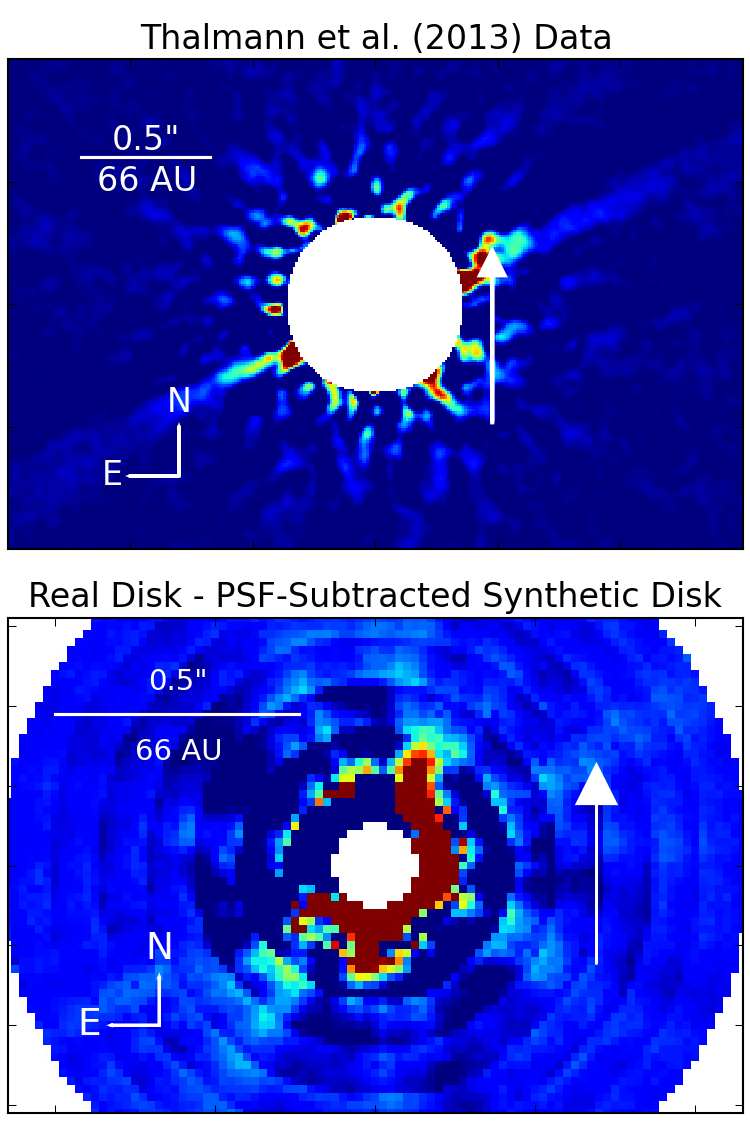}
 \caption
 [No planet found]
 {Top panel: reduction of the \citet{Thalmann2013}
 HIP 79977 data. An arrow points to the $4.6\sigma$ significance clump in their data.
 Bottom panel: our residuals after the forward-modeled best-fitting synthetic
 disk has been subtracted from the image. The same location is indicated with an arrow.}
 \label{fig:planet}
 \end{centering}
 \end{figure}

\citet{Thalmann2013} also note a candidate point source-like emission peak
located 0\farcs{}5 from the star, which appeared after subtracting their best-fit disk model.  They
posited that, if confirmed, this peak could be a localized clump of debris or
thermal emission from a $3-5 M_J$ planet\footnote{SCExAO
is a rapidly evolving platform that achieved a significant performance gain in the months after our
data were taken (O. Guyon, T. Currie, 2018 unpublished).   Thus, we defer discussion of limits on 
direct planet detections for a future HIP 79977 paper reporting new, substantially better data.}.
While our rereduction of the \citeauthor{Thalmann2013} data likewise show this emission, it does not
appear in the SCExAO/CHARIS data (Figure~\ref{fig:planet}) nor in the 2016 SCExAO/HiCIAO data.   
Given that both SCExAO data sets yield significantly deeper contrasts, we conclude that the emission
peak seen in AO188 data is likely residual speckle noise whose brightness highlights the stiff 
challenges in interpreting high-contrast imaging data where significant residual noise 
remains.
 
\clearpage
\begin{landscape}
\begin{deluxetable}{lcccccr}
\tablecaption{Scattered Light Resolved Debris Disks Around 5-30 Myr old Stars \label{table:diskscompared}}
\tablecolumns{7}
\tablewidth{0pt}
\tablehead{
\colhead{Star Name} & \colhead{Other} & \colhead{Age} & \colhead{$r_0$} & 
\colhead{H-G Para-} & \colhead{Inclina-} & \colhead{References} \\
\colhead{} & \colhead{Name} & \colhead{(Myr)} & \colhead{(au)} & 
\colhead{meter $g$} & \colhead{tion $i$ ($\arcdeg$)} & \colhead{}
}
\startdata
 HD 146897 & HIP 79977 & 11 & 53 & 0.6 & 84.5& \citet{Thalmann2013}, this work \\\\
 GSC 0739-00759 & - & 23 & 70 & 0.50 & 83& \citet{Sissa2018}\\
 HD 15115 & HIP 11360 & $<$100 & 90 & 0.25 & 86.2& \citet{Kalas2007,Mazoyer2014} \\
 %
 HD 36546 & HIP 26062 & 3-10 & 85 & 0.85 & 75 & \citet{Currie2017a}\\
 HD 39060 & $\beta$ Pic & 23 & 24--140 & 0.74 &85.2 & \citet{SmithTerrile1984}, \\
  & & & & & & \citet{MillarBlanchaer2015} \\
 HD 95086 & HIP 53524 & 17 &100--300& - & & \citet{Chauvin2018} \\
 %
 HD 106906 & HIP 59960 & 13 & 65 & 0.6 & 85.3& \citet{Lagrange2016}\\
 HD 109573 & HR 4796A & 10 & 77 & --\tablenotemark{a} & 76.5 & \citet{Schneider1999}, \citet{Milli2017}\\
 HD 110058 & HIP 61782 & 17 & 32 & - & $\sim$ 90? & \citet{Kasper2015}\\
 HD 111520 & HIP 62657 & 17 & 40-75 & - & 88?& \citet{Draper2016} \\
\enddata
\tablenotetext{a}{Note that a Henyey-Greenstein scattering function fails to reproduce this disk's scattering phase function \citep{Milli2017}.   See Discussion.  }
\tablecomments{
References are given for the first peer-reviewed publication of resolved optical/NIR imaging of the disk and the most recent paper that fitted for $r_0$ and $g$.
We report the age and best-fitting values of $g$ and $r_0$ from the second cited paper, unless there
has only been one publication, in which case we use its values.}
\end{deluxetable}

\begin{deluxetable}{lcccccr}
\tablecaption{Table~\ref{table:diskscompared} Continued.}
\tablecolumns{7}
\tablewidth{0pt}
\tablehead{
\colhead{Star Name} & \colhead{Other} & \colhead{Age} & \colhead{$r_0$} & 
\colhead{H-G Para-} & \colhead{Inclina-} & \colhead{References} \\
\colhead{} & \colhead{Name} & \colhead{(Myr)} & \colhead{(au)} & 
\colhead{meter $g$} & \colhead{tion $i$ ($\arcdeg$)} & \colhead{}
}
\startdata
 HD 114082 & HIP 64184 & 16 & 26-31\tablenotemark{b} & 0.07-0.23\tablenotemark{b} & 82.3 & \citet{Wahhaj2016} \\
 HD 115600 & HIP 64995 & 15 & 48 & 0 &79.5& \citet{Currie2015a}\\
 HD 120326 & HIP 67497 & 16 &59, 130\tablenotemark{c} & 0.82, -\tablenotemark{c} & 80 & \citet{Bonnefoy2017} \\
 HD 129590 & HIP 72070 & 10-16 & 59 & 0.43 & 75& \citet{Matthews2017} \\
 HD 131835 & HIP 73145 & 15 & 90 & 0.15 & 75.1 & \citet{Hung2015}, \citet{Feldt2017} \\
 HD 181327 & HIP 95270 & 23 & 88 & 0.3\tablenotemark{d} &31.7 & \citet{Schneider2006}, \\
  & & & & & & \citet{Schneider2014} \\
 HD 197481 & AU Mic & 23 & 40-50 & $>0.7$\tablenotemark{e} & $\sim$ 90? & \citet{Kalas2004}, \citet{Graham2007} \\ 
 TWA 7 & CE Ant & 10 & 25 & 0.63 & 13& \citet{Choquet2016}, \citet{Olofsson2018} \\
 TWA 25 & V1249 Cen & 7-13 & 78 & 0.7 & 75 & \citet{Choquet2016} \\
\enddata
\tablenotetext{b}{\citet{Wahhaj2016} reported values for three different data reductions, 
and we summarized their range of outcomes. Also,
instead of parameterizing the disk with $r_0$, inside and outside of which the disk drops off
in brightness, they assumed constant brightness between $r_{in}$ and $r_{in} + \Delta r$, with
falloff outside this range, and fit for both parameters. We reported their mean ring thickness 
$r_{in}+\frac{1}{2} \Delta r$.}
\tablenotetext{c}{\citet{Bonnefoy2017} detected two rings around HIP 67497 and modeled for both of them.}
\tablenotetext{d}{In their discovery paper,~\citet{Schneider2006} reported that HD 181327 had $r_0=86$
au and $g=0.3$. Later data modeled by \citet{Schneider2014} found $r_0=88$ au and surface brightness asymmetries that were not
well parameterized by a Henyey-Greenstein scattering function$g$ \citep[see also ][]{Stark2014}.}
\tablenotetext{e}{Au Mic has been extensively studied since~\cite{Graham2007}. However, publications
since then then have stopped fitting for $r_0$ and $g$ and have instead focused on characterization
of finer structures in the disk~\citep[e.g.][]{Boccaletti2018}.}
\end{deluxetable}
\clearpage
\end{landscape}

Table \ref{table:diskscompared} casts the derived dust scattering properties and radius for HIP 79977's debris disk within the context of other scattered light
resolved debris disks around young (5--30 Myr old) stars that have been observed at near-infrared wavelengths.   Our 
derived value of the Henyey-Greenstein parameter ($g=0.6$) 
falls in the middle to upper end of the range observed for other debris disks resolved in scattered 
light around 5--30 Myr old stars.
The fiducial radius of the HIP 79977 disk is fairly typical of values measured for other
debris disks.   Taking both parameters together, the location and dust scattering properties of the HIP 79977 disk appear most similar to that for HD 106906 \citep{Lagrange2016}, GSC 07396-00759 \citep{Sissa2018}, and TWA 25 \citep{Choquet2016}.   In particular, HD 106906's disk is likewise best modeled (within the Henyey-Greenstein formalism) by strongly forward-scattering dust and exhibits a clear east-west brightness asymmetry, similar to what our data hint at for HIP 79977.

However, HIP 79977's derived dust scattering parameter need not imply that its dust is \textit{intrinsically} more forward-scattering than that of other young, resolved debris disks. 
Early studies employing a single Henyey-Greenstein scattering function implied neutral dust grains \citep[$g$ $\lesssim$ 0.16][] {Schneider2009,Thalmann2011}.   However, more recent analysis based on extreme-AO observations probing small scattering angles showed that the disk's scattering function is not
well-fit by a single Henyey-Greenstein parameterization but by a weighted combination of a strongly forward-scattering and strongly backward-scattering H-G component \citep{Milli2017}.  Further improvements to scattering phase functions may require departures from standard Mie theory, e.g.  Distribution of Hollow Spheres \citep[e.g.][]{Milli2017}.   

Furthermore, as shown in \citet{Hughes2018}, the derived H-G $g$ value strongly correlates with the range of probed scattering angles: the closer to the forward-scattering peak probed by the data, the higher the derived $g$ value.  Indeed, all of the ostensibly strongly forward-scattering disks listed in Table \ref{table:diskscompared} are highly inclined, where such small angles are accessible.   If there is little intrinsic difference in the scattering properties of young debris disks, then a single scattering phase function \citep[e.g.][]{Hong1985} should be able to reproduce the available data. On the other hand, higher quality data for other ostensibly neutral scattering disks like HD 115600 \citep{Currie2015a} should likewise reveal a forward-scattering component inconsistent with the Henyey-Greenstein formalism.


The disk flux in our images is scattered primarily by dust grains that are micron-sized and larger.
Grains much smaller than our observing wavelengths scatter light isotropically, whereas larger
grains preferentially forward scatter light~\citep{Hughes2018}. Therefore, if the disk
was dominated by grains with sizes smaller than a micron, we would not have observed the 
forward scattering that we did.
Additionally, grains smaller than the observing wavelength
scatter light in the Rayleigh regime and thereby produce blue colors, and except for marginally
blue colors at certain radial separations, we do not observe this. The quality of our data
limits our ability to make further inferences about the dust properties; the disk's
dust properties could be better constrained by resolved spectra with higher signal to noise than 
our observations or multi-band polarimetric analysis.

A possible brightness asymmetry appears in our three bands (albeit at differing radii)
and seems plausible from the 2016 HiCIAO data
(Figure~\ref{fig:diskcomparison}c) but will require confirmation with additional data sets of comparable or greater depth.
If confirmed, there are several plausible physical explanations for this emission asymmetry. 
For example, the disk could be eccentric, and the preferential forward scattering of the dust
causes the side of the disk closer to us to appear brighter. Alternatively, collisions of the debris
in the disk could produce lumpiness and anisotrophies of brightness, and these would fade away on the
dynamical timescale of the disk. 
While ~\cite{Engler2017}
did not identify this brightness asymmetry, their data were at optical wavelengths and in polarized intensity.

The surface brightness power law measured in Section~\ref{sec:colors} is consistent with the
disk model proposed by~\citet{Strubbe2006}. They suggest that at the fiducial radius $r=r_0$,
micron-size grains are produced by the collisions of parent bodies with circular orbits. Outward
of this radius lie grains large enough to remain gravitationally bound to the star but having orbits that
have become eccentric due to stellar winds and radiation pressure from the star. This model
produces a surface brightness profile that drops off beyond the fiducial radius as 
$r^{-\alpha}$, where $\alpha \approx 4-5$. This agrees with our measured value of $-4.1 \pm 0.4$.



Since the acquisition of the data presented in this paper, SCExAO has achieved significant performance improvements, reaching in excess of 90\% Strehl at 1.6 $\micron$ for bright stars (Currie et al. 2018, in prep.).   Thus, future, deeper SCExAO/CHARIS observations of HIP 79977 will enable a more robust characterization of the HIP 79977 disk's morphology and access the inner 0\farcs{}25 with higher signal to noise.   Multi-wavelength photometry  obtained from these data can identify color gradients in the disk possibly traceable to different dust properties \citep[e.g.][]{Debes2008}.   These photometric points, complementary $L_{p}$ imaging, and spatially-resolved spectra can provide crucial insights into how the morphology and composition of HIP 79977's debris disk compare to the Kuiper belt and other debris disks probing the epoch of icy planet formation \citep[e.g.][]{Currie2015a,Rodigas2015,Milli2017}.

\section*{Acknowledgments}
We thank the anonymous referee for helpful suggestions that improved the quality of this work.   We also thank Laurent Pueyo for helpful conversations about KLIP forward-modeling. TC is supported by a NASA Senior Postdoctoral Fellowship.  MT is partly supported by the JSPS Grant-in-Aid
(15H02063). SG is supported from NSF award AST 1106391 and NASA Roses APRA award NNX 13AC14G.
The development of SCExAO was supported by the JSPS (Grant-in-Aid for Research \#23340051, \#26220704, \#23103002), the Astrobiology Center (ABC) of the National Institutes of Natural Sciences, Japan, the Mt Cuba Foundation and the directors contingency fund at Subaru Telescope.  
CHARIS was built at Princeton University in collaboration with the
National Astronomical Observatory of Japan under a Grant-in-Aid for
Scientific Research on Innovative Areas from MEXT of the Japanese
government (\#23103002).
We wish to emphasize the pivotal cultural role and reverence that the summit of Maunakea has always had within the indigenous Hawaiian community.  We are most fortunate to have the privilege to conduct scientific observations from this mountain. 

\bibliographystyle{apj}
\bibliography{ch5bib.bib}

%
\oneappendix{The Codes used to Produce the Preceding Results}\label{app:codes}
A repository of the codes Sean produced during his graduate school career can be viewed
at \texttt{https://github.com/seangoebelgradschool?tab=repositories}. Most of the codes
contain some self-documentation. Sean's codes were primarily written in (decreasing order
of prevalence) Python 2.7, IDL 7.1, or Bash. 
This section contains brief descriptions of a selection of his codes.

\vspace{5mm}
\noindent
Below are some of the codes used to produce the content of Chapter~\ref{chapter:overview}
(``Overview of SAPHIRA for AO Applications").
\begin{itemize}
\item \texttt{scexao/rrrfail.py} produced Figure~\ref{fig:rrr}.
\item \texttt{scexao/badbias.py} and \texttt{scexao/badbias\_analysis.py} were
used to produce Figure~\ref{fig:settling}.
\item \texttt{scexao/rrreduce.py} was widely used to form median dark/bias frames and subtract them from the data frames in order to reduce read-reset data.
\item \texttt{scexao/powspecproper.py} took into account how the detector spent its time
during clocking (including the periods at ends of rows and frames) in order to string together
frames and compute the power spectrum of noise common to all 32 outputs over frequencies from 1 Hz
to 500 kHz. An example result of this process was shown in Figure~\ref{fig:rfi1}.
\item \texttt{diodecapacitanceplot.py} produced Figure~\ref{fig:capacitance}. The data
plotted were measured by Ian Baker of Leonardo and sent by private correspondence. 
\end{itemize}

\vspace{5mm}
\noindent
Below are some of the codes used to produce the content of Chapter~\ref{chapter:preamps}
(``Commissioning of Cryogenic Preamplifiers for SAPHIRA").
\begin{itemize}
\item \texttt{scexao/powspecproper.py} took into account how the detector spent its time
during clocking (including the periods at ends of rows and frames) in order to string together
frames and compute the power spectrum of noise common to all 32 outputs over frequencies from 1 Hz
to 500 kHz. An example result of this process was shown in Figure~\ref{fig:rfi2}.
\item \texttt{scexao/voltgain.py} and \texttt{laptop/detectors/saphira/2018spie/voltgain\_complicated.py} measured volt gains at SCExAO and in the lab, respectively.
\texttt{laptop/detectors/saphira/2018spie/vgprettyplot.py} took their results and plotted them in Figures~\ref{fig:vglab} and~\ref{fig:vgscexao}.
\end{itemize}

\vspace{5mm}
\noindent
Below are some of the codes used to produce the content of Chapter~\ref{chapter:specklelives}
(``Measurements of Speckle Lifetimes in Near-Infrared Extreme Adaptive Optics Images for Optimizing Focal Plane Wavefront Control").
\begin{itemize}
\item \texttt{scexao/rrreduce.py} was widely used to form median dark/bias frames and subtract them from the data frames in order to reduce read-reset data.
\item \texttt{scexao/tiptiltastrometric.py} calculated offsets between frames in a user-specified
image cube by locating the artificial astrometric speckles. It then shifted the images in order to remove tips/tilts and saved them in a new file. In my testing, this was slightly more accurate
than using the cross-correlation method, but also significantly slower.
\item \texttt{scexao/tiptilt4.py} calculated offsets between frames in a user-specified image cube
by cross-correlating frames. It then shifted the images in order to remove tips/tilts and saved them in a new file.
\item \texttt{scexao/alignmulti.py} computed the shifts from one frame to the next using
cross-correlations and saved them in text files. Because this was computationally intensive and
involved millions of images, it multi-threaded it for improved performance.
\item \texttt{scexao/tiptilt5.py} read in the offsets produced by \texttt{alignmulti.py}, performed some smoothing, shifted the images in order to remove the tip/tilt, and saved the aligned images.
\item \texttt{scexao/specklelives5.py} produced the plots shown in Section~\ref{sec:sl1}
\item \texttt{scexao/corrcoeffs.py} was initially written to compute the Pearson's correlation 
coefficients discussed in Section~\ref{sec:sl2}.
\item \texttt{scexao/corrcoeffs2.py} computed the Pearson's correlation coefficients
discussed in Section~\ref{sec:sl2}. This was forked from \texttt{scexao/corrcoeffs.py} and 
optimized for speed, but it is less human-understandable. When compared to 
\texttt{scexao/corrcoeffs.py}, this uses less RAM, computes quantities ahead of time and
stores them in order to not re-compute them, and has the ability to multi-thread its
calculations across many CPUs.
\item \texttt{scexao/specklelivesfig2.py} produced Figure~\ref{fig:pupilfocalplanes}.
\end{itemize}

\vspace{5mm}
\noindent
Below are some of the codes used to produce the content of Chapter~\ref{chapter:hip79977}
(``SCExAO/CHARIS Near-IR High-Contrast Imaging and Integral Field Spectroscopy of the HIP 79977 Debris Disk").
\begin{itemize}
\item \texttt{scexaodisks/partialsub/charis\_klip\_disk\_fwdmod.pro} performs the forward
modeling process. It reads in a selection of synthetic disk models and propagates them
through the KLIP PSF-subtraction process (\texttt{charis\_subsklip.pro})
using the eigenvalues and eigenvectors saved from the processing of the actual data.
\item \texttt{scexaodisks/psfsub/charis\_subsklip.pro} performs the KLIP PSF-subtraction of
a user-specified image cube using a user-specified PSF.
\item \texttt{scexaodisks/sean/klipgrid.pro} calls \texttt{charis\_subsklip.pro} for
multiple models in order to check the effects of different reduction settings.
\item \texttt{scexaodisks/sean/scrapefitsinfo.pro} reads the header information of all
the forward-modeled synthetic models in order to compare them to and outputs it to a text file in order to compare them to the actual data and decide which synthetic disks are well-fitting.
\item \texttt{scexaodisks/specphotcal/calc\_disk\_sb.pro} calculates the surface brightness along the
disk spine. This is used in conjunction with \texttt{sbplot.py} and \texttt{sbplot2.py} to generate
Figures~\ref{fig:surfacebrightness} and~\ref{fig:sb2}.
\item \texttt{scexaodisks/syndisks/call\_grater.pro}, \texttt{scexaodisks/syndisks/scimage.pro}, and 
\texttt{scexaodisks/syndisks/density.pro} produce a synthetic model disk using user-specified parameters.
\item \texttt{scexaodisks/syndisks/grater\_grid.pro} calls \texttt{call\_grater.pro} in order to
produce synthetic model disks for a range of user-specified parameters. This has a deeply scandalous
11 nested for loops, so viewer discretion is advised.
\item \texttt{planet.py} produced Figure~\ref{fig:planet}.
\item \texttt{sbplot.py} used the data from \texttt{calc\_disk\_sb.pro} to produce Figure~\ref{fig:surfacebrightness}.
\item \texttt{spinefit.py} performed a fit to the spine of the disk in order to find its trace.
This was then used in the surface brightness calculations.
\item \texttt{contours.py} plotted the average $\chi^2$ of all disk parameters against
all other disk parameters, producing a number of interesting images. One of them was shown in Figure~\ref{fig:contourmap}.
\item \texttt{fwmodeling.py} produced Figure~\ref{fig:bestsyndisk}.
\item \texttt{plthists.py} produced Figure~\ref{fig:histograms}.
\item \texttt{showroi.py} produced Figure~\ref{fig:roi}.
\item \texttt{diskcomparison.py} produced an alternative version of Figure~\ref{fig:diskcomparison}
\item \texttt{makediskpics.py} produced Figure~\ref{fig:combineddiskimages}.
\item \texttt{sbplot2.py} used the data from \texttt{calc\_disk\_sb.pro} to produce Figure~\ref{fig:sb2}.
\item \texttt{snrcompare.py} compares the per-channel SNR of PSF-subtracted disks produced using various KLIP parameters in order to find the best set of parameters.
\end{itemize}

\vspace{5mm}
\noindent
Below are other useful codes.
\begin{itemize}
\item \texttt{scexao/pbserver/gui\_pizzabox} provides an easy user interface for operating the Rev.~3
Pizza Box, which is what was used at SCExAO deployments.
\item \texttt{scexao/pbserver/gui\_saphira} provides an easy user interface for starting and stopping
image saving at SCExAO, viewing the live images, and viewing data statistics.
\item \texttt{scexao/pbserver/temp.py} queries the Lakeshore temperature controller for SAPHIRA's temperature and prints it to the screen. 
\item \texttt{scexao/pbserver/temp\_notify2.py} is meant to be called from a cronjob. It queries the Lakeshore temperature controller for SAPHIRA's temperature. If the temperature is out of range, it
emails the user a warning. This proved infinitely useful during the many power surges at Subaru which
regularly caused the detector to start to warm. 
\end{itemize}

\end{document}